\documentclass[a4paper,12pt]{article}

\usepackage[a4paper]{geometry}
\usepackage[dvipsnames]{xcolor}
\usepackage[utf8]{inputenc} 
\usepackage[hidelinks]{hyperref}
\usepackage{float}
\usepackage{amssymb,amsmath,amsthm}
\usepackage{nicefrac}
\usepackage{amsfonts}

\pdfoutput=1
\usepackage{graphicx}
\usepackage{graphicx, float, tabularx, booktabs}
\usepackage{color}
\usepackage{enumerate}
\usepackage[american]{babel}
\usepackage{stackrel}
\usepackage[compress]{cite}
\usepackage[numbers, compress]{natbib}
\usepackage{subcaption}

\usepackage{amsfonts}
\usepackage{epsfig}

\usepackage{graphicx, tabularx}
\usepackage[toc,page]{appendix}

\newcommand{\beq}{\begin{equation}}
\newcommand{\eeq}{\end{equation}}
\newcommand{\bea}{\begin{eqnarray}}
\newcommand{\eea}{\end{eqnarray}}

\newcommand{\Mpl}{M_\mathrm{Pl}}
\newcommand{\ellp}{L_\mathrm{Pl}}

\newcommand{\diff}{\mathrm{d}}
\renewcommand{\d}{\diff}

\vspace{-0.1cm}
\renewcommand{\thefootnote}{\fnsymbol{footnote}}

\author{
Marco Knipfer$^{a,b}$\thanks{knipfer@fias.uni-frankfurt.de}, 
Sven K\"{o}ppel$^{a,b}$\thanks{koeppel@fias.uni-frankfurt.de},
Jonas Mureika$^{c}$\thanks{jmureika@lmu.edu} \ and
\and
Piero Nicolini$^{a,b}$\thanks{nicolini@fias.uni-frankfurt.de}
 \\[1ex]
\small $^a$ Frankfurt Institute for Advanced Studies (FIAS)\\[-0.5ex]
\small Ruth-Moufang-Str.~1, D-60438 Frankfurt am Main, Germany\\[1ex]
\small $^b$ Institut f\"{u}r Theoretische Physik, Johann Wolfgang Goethe-Universit\"{a}t Frankfurt\\[-0.5ex]
\small Max-von-Laue-Str.~1, D-60438 Frankfurt am Main, Germany\\[1ex]
\small $^c$ Department of Physics, Loyola Marymount University\\[-0.5ex]
\small Los Angeles, CA 90045, USA
}
\date{}
\title{Generalized Uncertainty Principle and Black Holes in Higher Dimensional Self-Complete Gravity}

\begin{document}
\maketitle

\vspace{-0.5cm}




\begin{abstract}
\noindent 
{\small 
In this paper we consider generalized uncertainty principle (GUP) effects in higher dimensional black hole spacetimes via a nonlocal gravity approach.  We study three possible modifications of momentum space measure emerging from GUP, including the original Kempf-Mangano-Mann (KMM) proposal. 
By following the KMM  model we derive a family of black hole spacetimes. The case of five spacetime dimensions is a special one.  We found an exact black hole solution with a Barriola-Vilenkin monopole at the origin. This object turns out to be the end point of the black hole evaporation. Interestingly for smaller masses, we found a ``naked monopole'' rather than a generic naked singularity.  We also show that the Carr-Lake-Casadio-Scardigli proposal leads to mild modifications of spacetime metrics with respect to the Schwarzschild-Tangherlini solution. Finally,  by demanding the same degree of convergence in the ultraviolet regime for any spacetime dimension, we derive a family of black hole solutions that fulfill the gravity self-completeness paradigm. The evaporation of such black holes is characterized by a fluctuating luminosity, which we dub a {\it lighthouse effect}.
    }
\noindent
    
\end{abstract}




\renewcommand{\thefootnote}{\arabic{footnote}}
\setcounter{footnote}{0}
\thispagestyle{empty}
\clearpage
\tableofcontents

\section{Introduction}

Black holes are perhaps one of the most intriguing objects in Physics, as they play a major role in a wide variety of both classical and quantum phenomena.  The gravitational wave events recently observed by LIGO were the product of binary black hole mergers \cite{LIGO15}, while the Event Horizon Telescope has produced the first image of the shadow of a supermassive black hole \cite{EHTM87}. 
Quantum mechanically, black holes show even more surprising features. Much in the same way as black bodies, they can emit thermal radiation at a temperature proportional to their surface gravity~\cite{Haw75}. 

More importantly, black holes bring into question our understanding of quantum mechanics \cite{western}. The Compton wavelength is conventionally believed to assume arbitrarily small values, provided one raises particles to high enough energies. This reasoning, however, breaks down at the Planck scale, where a black hole is expected to form due to the collapse of particles at such extreme energies~\cite{Adl10}. This is equivalent to saying that gravity is ultraviolet self-complete, \textit{i.e.}, there are no propagating quantum degrees of freedom in the trans-Planckian regime and length scales below the Planck length are inaccessible \cite{DvG10,DFG11,DGG11,SpA11,DFG12,DvG12,NiS12,MuN12,AuS13,Car14,DGI15,CMN15,DGL16,FKN16,Dva17,
Nic2018,CDRN18}. Interestingly, such features are effectively captured by a modification of commutation relations known as generalized uncertainty principle (GUP)~\cite{Ven86,ACV89,ACV93,Mag93,KMM95}, namely
 \begin{equation}
 [x^i, p_j] = i\, \hbar\, \delta^i_{\, j} \, (1+f( \vec{p}^2)) \,,
 \label{eq:commrel}
 \end{equation}
where the function $f$ customarily assumes the form $f(\vec{ p}^{\,\,2})\simeq \beta \vec{ p}^{\,\,2}+\dots$ at first order. From \eqref{eq:commrel}, one obtains that spatial resolution better than $\sqrt{\beta}$ is no longer possible, since the uncertainty relations reads 
\begin{equation}
\Delta x \Delta p \geq \frac \hbar2 (1+\beta(\Delta p)^2)\quad,
\label{eq:gup}
\end{equation}
where $\sqrt{\beta}\sim L_\mathrm{Pl}$ and $L_\mathrm{Pl}$ the Planck length.
For $\Delta p\gg 1/\sqrt{\beta}$, length scales become proportional to $\Delta p$ as expected from the presence of a black hole in the trans-Planckian regime. For reviews see \cite{SNB12,Hos13,TaM15}. 

The GUP has been invoked to improve the scenario of black hole evaporation, that is customarily affected by a divergent profile of the Hawking temperature $T$ in the terminal phase. If one employs \eqref{eq:gup}, with $\Delta p\sim T$ and $\Delta x\sim G_\mathrm{N}M$, the temperature profile is no longer divergent and  a Planckian black hole remnant forms as an evaporation endpoint \cite{AdS99,APS01}. Such a remnant has also been considered as a candidate for cold dark matter \cite{ChA03}. There are, however, potential problems that stem from such results. Planckian remnants have Planckian temperatures.
The surface gravity description of the temperature no longer holds.

To amend the above limitations, a new approach has been proposed in order to implement GUP effects in gravitational systems \cite{IMN13}. As a start, one can notice that the GUP introduces nonlocality by preventing infinitesimal resolution.  One might therefore be led to consider a nonlocal version of Einstein's equations \cite{Kra87,Tom97,Bar03,Mod12a}
\beq
\mathbb{G}^{-1}\left(L^2\Box\right)G_{\mu\nu}=8\pi T_{\mu\nu}\,,
\label{eq:nleinstein}
\eeq
where the gravitational constant, $G_\mathrm{N}$, becomes an invertible differential operator $\mathbb{G}$.  The term $\Box$ is the covariant d'Alembertian and $L$ is a length scale. Equation~\eqref{eq:nleinstein} can be either used to described large scale degravitating effects \cite{ADD02,DHK07,Bar05,Bar12} or short scale modified gravity theories \cite{GHS10,MMN11,Nic12,CMN14,FKN16}. In fact, one can select a specific profile of $\mathbb{G}^{-1}\left(L^2\Box\right)$ to reproduce the GUP momentum space deformation 
\begin{equation}
    \diff^3 p \to \frac{\diff^3 p}{1+\beta \vec{p}^2}\ 
    \label{eq:kempf}
\end{equation}
for the static potential due to virtual particle exchange by setting $L=\sqrt{\beta}$. The resulting non-rotating black hole metric allows for horizon extremization with consequent formation of a zero temperature remnant at the end of the evaporation~\cite{IMN13}. Such a black hole solution not only supersedes the aforementioned limitations of the scenario proposed in \cite{AdS99,APS01}, but offers additional interesting properties.  It removes the scale ambiguity of the Schwarzschild metric and fulfills the gravity ultraviolet self-completeness by preventing black hole radii smaller than the Planck length. It also allows for a semiclassical description of the whole evaporation process for the absence of relevant quantum back reaction during the SCRAM phase\footnote{The black hole SCRAM is a cooling down phase during the final stages of the evaporation. The term SCRAM has been introduced in \cite{Nic09} by borrowing it from nuclear reactor technology. SCRAM is a backronym for ``Safety control rod axe man'', introduced by Enrico Fermi in 1942 during the Manhattan Project at Chicago Pile-1. It still indicates an emergency shutdown of a nuclear reactor.} preceding the remnant formation \cite{NiW11}.

Higher dimensional black hole solutions play an important role in theoretical research for a variety of reasons. On the formal side, they are a key element of proposals aimed at a unified description of fundamental interactions, \textit{e.g.}, superstring theory and related paradigms, like the gauge/gravity duality.
On the phenomenological side, microscopic higher dimensional black holes would be the ``smoking  gun'' for quantum gravity well below the Planck scale \cite{BaF99,DiL01,GiT02}  and a viable resolution of the hierarchy problem \cite{AAD98,ADD98,ADD99,RaS99a,RaS99b,ACD01} (for reviews see \cite{Lan02,Cav03,Kan04,Hos04,CaS06,Win07,BlN10,Cal10a,Par12,NMS13,BlN14,KaW15,WNB17,WNB18}).

Given the above background, it is natural to consider the case of GUP effects in higher dimensional black hole metrics. It should be noted that there is no unique prescription for the GUP in the presence of extra dimensions \cite{Koppel:2017rsf}. As a result, before proceeding with this study, we will provide an analysis of the existing proposals for the GUP in $d$-dimensions, with $d=n+1$ and $n$ the number of spatial dimensions \cite{ScC03,Maz13,Car13,Car14,LaC15,LaC16,Carr:2017grh,Maz12,DMS15,Maz15,LaC18}. 
We will consider hyper-spherical black hole spacetimes that can fit both in the large extra dimension scenario (ADD \cite{ADD98,ADD99} or  AADD model \cite{AAD98})  as well as in the universal extra dimension scenario \cite{ACD01}. 
 For the sake of simplicity, we assume a new fundamental scale $M_* = C_n \Mpl^{2/(n-1)} R_c^{-(n-3)/(n-1)}$ according to the large extra dimensional model only, where $M_\text{Pl}$ is  the $4$-dimensional Planck mass, $R_c$ is the compactification scale, and the dimensionless prefactor is $C_n=\mathcal{O}(1)$.
Unless differently specified, all quantities are expressed in units of the fundamental  mass $M_*$, or of the fundamental length $L_*=1/M_*$. According to this notation, the effective gravitational coupling constant reads $G_*=1/M_*^{n-1}$. If the compactification radius drops to the Planck scale $R_c\sim 1/M_\text{Pl}$, one finds for consistency $M_*\sim M_\text{Pl}$ and $G_*\sim 1/M_\text{Pl}^{n-1}$.

The paper is organized as follows. In Section~\ref{sec:ambiguitygup} we summarize the existing proposals for a GUP in higher dimensional spacetimes. In Section~\ref{sec:KMM-GUP}, we consider the general set up to deform Einstein's equations in the presence of nonlocal effects and we implement the GUP proposed by the Kempf-Mangano-Mann (KMM) proposal~\cite{KMM95}. In Section~\ref{sec:HighDKMM} we present the higher dimensional extension of the KMM model. In Section~\ref{sec:Cas-Scar} we consider the proposal of 
Carr and Lake~\cite{Car13,Car14,LaC15,LaC16,Carr:2017grh,LaC18}, and
Casadio and Scardigli~\cite{ScC03} (CLCS).
In  Section~\ref{sec:ourGUP} we present an alternative proposal inspired by the work of Maziashvili~\cite{Maz12,DMS15,Maz15}. Finally in Section~\ref{sec:Conclusions}, we draw our conclusions.

\section{Review of GUP in higher dimensional spacetimes}
\label{sec:ambiguitygup}
According to the KMM model \cite{KMM95}, the GUP manifests itself via a deformation of the integration measure in momentum space. Following \eqref{eq:commrel} the Hilbert space representation of the identity becomes 
\begin{equation}
 \mathbb{I}=  \int \frac{\diff^{d-1}  p}{1+\beta {\vec p}^2}\ \left|p \right>\left< p\right|\,,
 \label{GUPmeasure}
  \end{equation}
where $\vec{p}$ is a ($d-1$)-dimensional spatial vector. While momentum operators preserve their feature as in quantum mechanics, position operators no longer admit physical eigenstates, as one should expect in the presence of a minimal resolution length $\sqrt{\beta}$. A closer inspection of (\ref{GUPmeasure}) shows that the measure is suppressed in the ultraviolet regime
\begin{equation}
\mathrm{d}{\cal V}_p \equiv \frac{\mathrm{d}^{d-1} p}{1 + \beta {\vec p}^2} \underset{\beta {\vec p}^2\gg 1}{\overset{}{\approx}}
|{\vec p}|^{\,d-4}\ \mathrm{d}|{\vec p}|\,.
\label{eq:intvol}
\end{equation}
We note that for $d=4$ one recovers \eqref{eq:kempf}, and the momentum term on the right-hand side
in \eqref{eq:intvol} disappears. Conversely for $d>4$, the measure diverges in the ultraviolet
regime. As $d$ increases, the effect of the GUP becomes increasingly weaker.

At this point, we remind the reader that the uncertainty relation \eqref{eq:gup} arises from a revision the standard reasoning of the Heisenberg microscope, a well known experiment in quantum mechanics. Specifically, one can introduce an additional displacement $\Delta x_\mathrm{g}$ for the electron due to the gravitational interaction with the incoming photon. By assuming a Newtonian description, ones finds that 
\begin{equation}
\Delta x_\mathrm{g}\sim G_\mathrm{N}\frac{M_{\rm eff}}{r^{2}} \left(\frac{r^2}{c^2}\right) 
\sim G_\mathrm{N}  \Delta p
\sim (\sqrt{\beta})^{2} \Delta p
\end{equation}
 for $d=4$, where $M_{\rm eff}=h/(\lambda c)$ is the effective photon mass and  $\sqrt{\beta}\sim L_\mathrm{Pl}$. The total uncertainty is obtained by adding $\Delta x_\mathrm{g}$ to the standard quantum uncertainty, namely $\Delta x=\Delta x_\mathrm{C}+\Delta x_\mathrm{g}$ with $\Delta x_\mathrm{C}\sim 1/\Delta p$. Following the reasoning of Scardigli and Casadio \cite{ScC03}, as well as Carr and Lake \cite{Car13,Car14,LaC15,LaC16,Carr:2017grh,LaC18}, the extension of the above calculation to the case $d>4$ leads to
\begin{equation}
    \Delta x_\mathrm{g}
    \sim G_\ast \frac{M_{\rm eff}}{r^{d-2}} \left(\frac{r^2}{c^2}\right)
    \sim G_\ast \frac{\Delta p}{r^{d-4}}\leadsto \Delta x_\mathrm{g}^{d-3}
    \sim L_\ast^{d-2} \Delta p\,,
    \label{eq:CLCSgup}
\end{equation}
having assumed $r\sim\Delta x_\mathrm{g}<R_\mathrm{c}$.
The uncertainty relation then reads
\begin{equation}\label{eq:modifGUP}
    \Delta x \Delta p \geq \frac{\hbar}{2} \left( 1 + \left( \sqrt{\beta}\, \Delta p \right)^{\frac{d-2}{d-3}} \right)\,,
\end{equation}
where $\sqrt{\beta}\sim L_\ast$. On  the other hand, for $r\sim\Delta x_\mathrm{g}>R_\mathrm{c}$, the uncertainty relation is assumed to be that for $d=4$ displayed in \eqref{eq:gup}. From \eqref{eq:modifGUP} one can show that the momentum space measure can be expressed as
\begin{equation}
\mathrm{d}{\cal V}_p \equiv \frac{\mathrm{d}^{d-1} p}{1 + (\beta {\vec p}^2)^{\frac{1}{2}\frac{d-2}{d-3}}} \underset{\beta {\vec p}^2\gg 1}{\overset{}{\approx}}
|{\vec p}|^{d-3}\ \mathrm{d}|{\vec p}|,
\label{eq:intvol2}
\end{equation}
for $d>4$. 
This means that in such a scenario the GUP corrections are even milder than those of the KMM model. 

Given the ambiguity in results described above, we proposed another revision to the Heisenberg microscope. From \eqref{eq:CLCSgup} one obtains that $\Delta x_\mathrm{g}\geq L_\ast$ only for $\Delta p \geq M_\ast$. As a result, for any $\Delta p \leq M_\ast$ the gravitational uncertainty is negligible, $\Delta x_\mathrm{g}<\Delta x_\mathrm{C}< R_\mathrm{c}$. In such a regime the typical interaction distance is controlled by the Compton wavelength, $r\sim \Delta x_\mathrm{C}\sim 1/\Delta p$. This implies that
\begin{equation}
\Delta x_\mathrm{g}
    \sim G_\ast \frac{M_{\rm eff}}{r^{d-2}} \left(\frac{r^2}{c^2}\right)
    \sim G_\ast \frac{\Delta p}{r^{d-4}}
    \sim L_\ast^{d-2} \Delta p^{d-3}\,.
\end{equation}
The above relation relaxes the proportionality condition between  
$\Delta x_\mathrm{g}$ and the radius of the Tangherlini-Schwarzschild black hole \cite{Koppel:2017rsf}. As a byproduct, however, one obtains a stronger correction in momentum space since gravity will begin to probe extra dimensions at scales
$r< R_\mathrm{c}$. This can be inferred from the condition
\begin{equation}
    \mathrm{d}{\cal V}_p \equiv \frac{\mathrm{d}^{d-1} p}{1 + (\beta {\vec p}^2)^{\frac{d-2}{2}}} \underset{\beta {\vec p}^2\gg 1}{\overset{}{\approx}}
    \ \mathrm{d}|{\vec p}|\,,
    \label{eq:intvol3}
\end{equation}
namely it is uniformly suppressed irrespective of the number of dimension $d$. 

There are three arguments in support of the above reasoning. First, the requirement that a  quantum gravity correction $\Delta x_\mathrm{g}$ is proportional to a classical quantity like the radius of the Tangherlini-Schwarzschild black hole is not fully consistent. It makes sense only for $r > R_\mathrm{c}$, namely at length scales at which gravity becomes classical and only four dimensions are visible. 
Second, investigations of string scattering showed that the position-momentum uncertainty relation is of the form \eqref{eq:gup} \cite{ACV89,ACV93}. Such a result has, however, been obtained in the eikonal limit and higher order corrections for the ultraviolet regime of momentum space are expected. Third, the above corrections are consistent with the algebra proposed in \cite{Maz12,DMS15,Maz15}.

In the remainder of this paper, we present higher dimensional black hole solutions that emerge from the non-local field equations \eqref{eq:nleinstein} that account for GUP effects following from \eqref{eq:intvol}, \eqref{eq:intvol2} and \eqref{eq:intvol3} according to the method proposed in \cite{IMN13}. Table \ref{table:gup-comparison} offers an overview
about the selected models and their abilities to tame the ultraviolet
divergences in higher dimensional space.

\begin{table}[h!]
	\begin{tabularx}{\linewidth}{rlll}
		\firsthline
		Eq. & Short name  & Momentum space volume element & Large $p$ limit \\
		\hline 
		\hline 
		& No GUP   & $\d^{d-1} p$           & $|\vec{p}|^{d-2} \d |\vec{p}|$ \\
		\eqref{eq:intvol2} & CLCS   & $\d^{d-1} p \ \left[ 1 + (\sqrt \beta~|\vec{p}|)^{\frac{(d-2)}{(d-3)}}\right]^{-1}$ & $|\vec{p}|^{d-3} \d |\vec{p}|$ \\
		\eqref{eq:kempf} & KMM      & $\d^{d-1} p \ \left[1 + (\sqrt \beta~|\vec{p}|)^2   \right]^{-1}$ & $|\vec{p}|^{d-4} \d |\vec{p}|$  \\
		\eqref{eq:intvol3} & Revised GUP & $\d^{d-1} p  \ \left[1 + (\sqrt \beta~|\vec{p}|)^{d-2}\right]^{-1}$ & $\d |\vec{p}|$ \\
		\hline
	\end{tabularx}
	\caption{
	  Overview of the different GUP models studied in the present work.
	  The ordering follows the large momentum limit of the modified momentum
	  space volume element. ``No GUP'' denotes the
	  ordinary Heisenberg uncertainty principle.
	  \\	  
	  The Carr-Lake-Casadio-Scardigli
	  model (CLCS) provides a weaker regularization
	  than the Kempf-Mangano-Mann model (KMM), while our ``improved'' GUP in Section~\ref{sec:ourGUP} 
	  provides a dimension-independent measure.
   }
	\label{table:gup-comparison}
\end{table}

\section{GUP black holes from the KMM momentum measure}
\label{sec:KMM-GUP}
To begin, we review the calculation of the GUP modified Schwarzschild solution~\cite{IMN13} in $(3+1)$-dimensions. 
We chose a specific profile of the operator $\mathbb{G}^{-1}(L^2\Box)$
such that effectively the momentum measure is modified as in the KMM model~\cite{KMM95}.  Equation~\eqref{eq:nleinstein} can be cast in the form
\beq
    \label{eq:nlee}
    R_{\mu\nu} - \frac{1}{2} g_{\mu\nu} R = 8 \pi \mathbb{G}(L^2\Box) T_{\mu\nu}\,,
\eeq
that is equivalent to coupling Einstein gravity to a non-standard energy momentum tensor. In case of a static, spherically symmetric spacetime, one has  
\beq 
    \label{eq:cltoo}
    T^{\ 0}_0=-M \delta^{(3)}(\vec{x})\,, 
\eeq
corresponding to a vanishing mass distribution, apart from the origin where a curvature singularity is present \cite{BaN93,BaN94,DeB08}. The action of the operator $\mathbb{G}(L^2\Box)$ determines a smearing of the source term that reads
\begin{equation}
  \mathbb{T}^{\ 0}_0(\vec{x})\equiv\frac{1}{G_\mathrm{N}} \ \mathbb{G}(L^2\Box) T^{\ 0}_0  = -\rho(\vec{x})\,.
\end{equation}
Since the Dirac delta can be represented as 
\begin{equation}
    \delta^{(3)}(\vec{x}) = \frac{1}{(2\pi)^{3}}\int \diff^3 p\ e^{i \vec{x}\cdot\vec{p}}\quad,
    \label{eq:delta}
\end{equation}
one obtains 
\begin{equation}
    \rho(\vec{x}) = \frac{M}{(2\pi)^{3}}\int \frac{\diff^3 p}{1+\beta \vec{p}^2}  e^{i \vec{x}\cdot\vec{p}}= M\
    \frac{e^{\frac{-|\vec{x}|}{\sqrt{\beta}}}}{4\pi |\vec{x}| \beta}\,,
    \label{eq:smeared}
\end{equation}
provided the operator is chosen as
\beq
\label{eq:goperator}
    \mathbb{G}(L^2\Box) = \frac{G_\mathrm{N}}{1 - L^2 \Box}\,,
\eeq
with $L^2=\beta$. From \eqref{eq:goperator} it can be seen that in the low energy regime  
$-\Box\ll L^{-2}$, \eqref{eq:nlee} match Einstein's equations and $\mathbb{G}(L^2\Box) \to G_\mathrm{N}$. Conversely for $-\Box\sim L^{-2}$, strong non-local corrections enter the game, gravity becomes increasingly weaker ($\mathbb{G}(L^2\Box) \ll G_\mathrm{N}$), and the source can no longer be compressed as in \eqref{eq:cltoo}. We note that the profile in \eqref{eq:goperator} modifies the momentum measure in the same way as in \eqref{eq:kempf}, namely the KMM model \cite{KMM95}.

For a generic static, spherically symmetric metric
\begin{align}
    \diff s^2 &= -\left( 1-\frac{2G_\mathrm{N}\mathcal{M}(r)}{r}\right)\diff t^2 
    + \left( 1-\frac{2G_\mathrm{N}\mathcal{M}(r)}{r}\right)^{-1}\diff r^2 + r^2 \diff \Omega^2\ 
    \label{eq:gupschwarzschild}
\end{align}
the solution of \eqref{eq:nlee} corresponds to determining the unknown function  
\begin{align}
    \mathcal{M}(r) &= \int_{B_r}\diff^3 x\,\rho(\vec{x}) 
\end{align}
that represents the cumulative mass distribution, where $B_r$ is the $3$-ball of radius $r$ with $r\equiv|\vec{x}|$. Given \eqref{eq:gupschwarzschild}, the conservation of the energy momentum tensor implies its  form, namely $\mathbb{T}_\mu^\nu=\mathrm{diag} \left(-\rho, p_r, p_\perp, p_\perp\right)$ with $p_r=-\rho$ and $p_\perp=-\rho-\frac{1}{2}r (\diff \rho/\diff r)$.

\begin{figure}[t]
    \centering
    \includegraphics[width=0.72\textwidth]{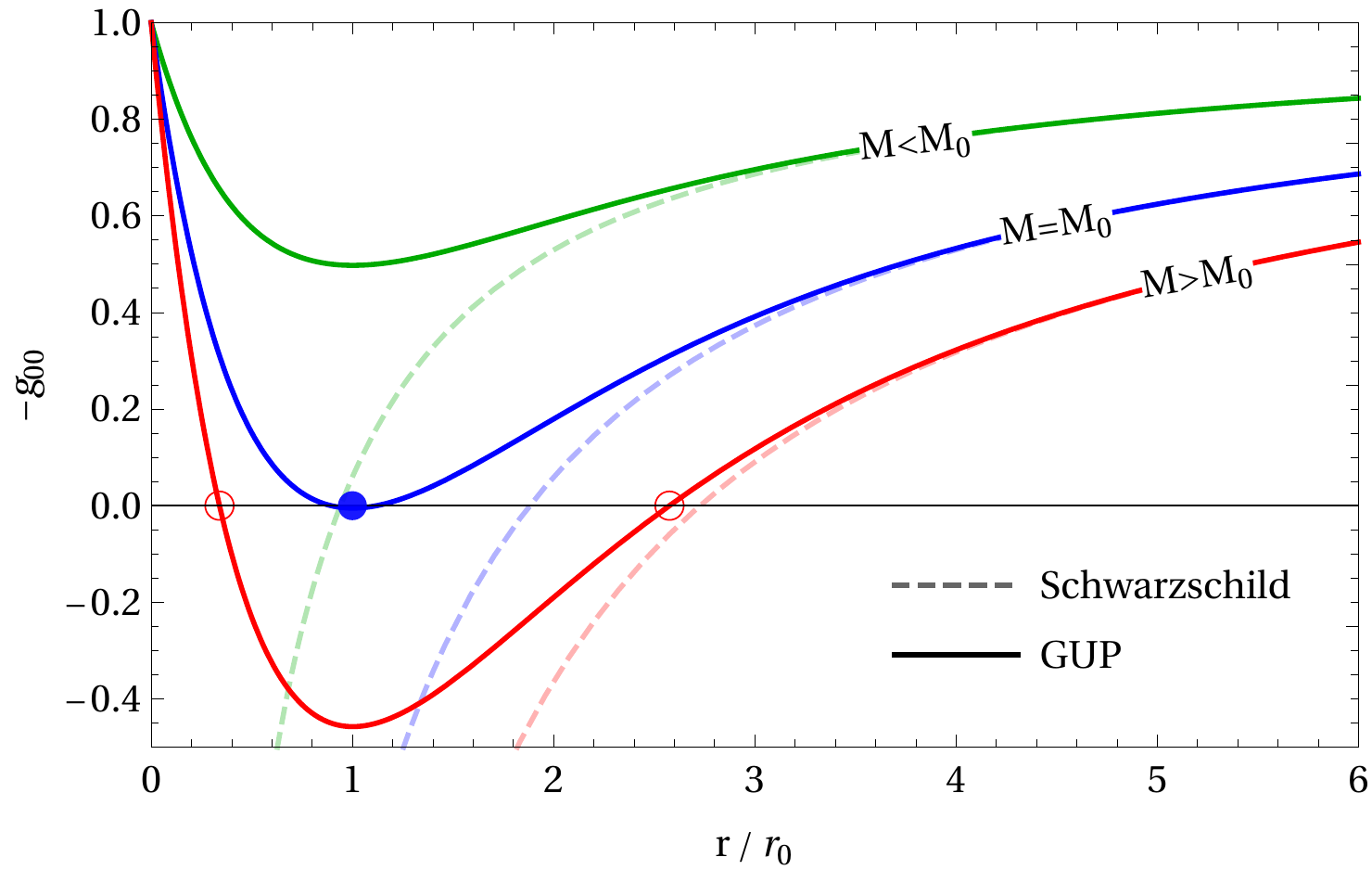}
    \caption{The horizon structure of the KMM GUP
        metric~(\ref{eq:gupschwarzschild}) in $(3+1)$-dimensions. Here $-g_{00}(r)$ is shown
        in units
        of the extremal radius $r_{0} \approx 1.793 \sqrt{\beta}$ for the three different cases
        discussed in the main text, namely for $M > M_0$ (two horizons), $M = M_0$ (one extremal horizon) and $M < M_0$ (no horizon), where $M_0=1.68\sqrt{\beta}/G_\mathrm{N}$ is the critical (threshold) mass.  }
    \label{fig:g00kempf}
\end{figure}

 By assuming $\rho(\vec{x})$ given in \eqref{eq:smeared} one finds
\begin{align}
    \mathcal{M}(r) &= M\gamma\left(2;\frac{r}{\sqrt{\beta}}\right)\ = M\left[ 1 - e^{-\frac{r}{\sqrt{\beta}}} - \frac{r}{\sqrt{\beta}}
    e^{-\frac{r}{\sqrt{\beta}}}\right]\,.
    \label{eq:mass}
\end{align}
The behaviour of the metric can best be seen when plotting the metric coefficient $g_{00}(r)$ as
in Figure~\ref{fig:g00kempf}. The horizon structure resembles that of the Reissner-Norstr\"{o}m solution: there exist an outer event horizon  $r_+$ and an inner Cauchy horizon $r_-$. 
The two eventually merge at the critical mass parameter $M=M_0= 1.68 \sqrt{\beta}/G_\mathrm{N}$, corresponding to an extremal configuration. The curvature still diverges at the origin, $R\to 1/4\pi \beta r$, but less brutally than in the Schwarzschild case. 
This can be seen from the fact that, in contrast to the Schwarzschild case, the metric is no longer divergent at the origin. That is, only the first and higher derivatives of the metric are singular at this point, which implies a softer singularity. 

Such a property of the GUP inspired black holes is similar to that of the recently proposed holographic metric~\cite{NiS12,FKN16}. For other quantum corrected black hole solutions, however, the metric and all its derivatives are regular at the origin implying a removal of the curvature singularities~\cite{NSS06b,Nic09,NiS10,Nic12,NSW19}.
Despite the singular behaviour of the spacetime \eqref{eq:gupschwarzschild} the gravitational field, $\vec{g}=\frac{1}{2}\vec{\nabla} g_{00}$, can be computed in a neighborhood of the origin: it turns out to be constant and repulsive.  Much in the same way as the aforementioned regular geometries, the quantum fluctuations of the manifold provide an outer pressure that prevents the energy density to collapse in a Dirac delta profile.

\begin{figure}[t]
    \centering
      \includegraphics[width =0.7 \textwidth]{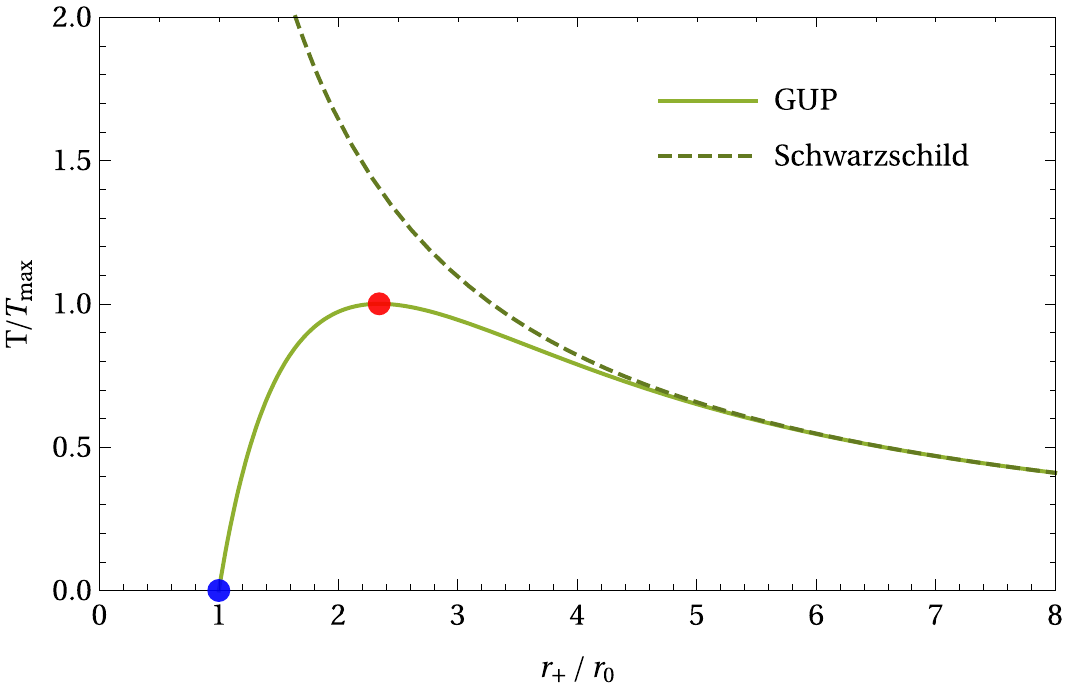}
    \caption{Temperature of the GUP black hole in $n=3$ spatial dimensions as a
        function of the outer horizon radius $r_+$.
        For comparison, the Hawking temperature of
        a traditional Schwarzschild black hole displayed.
        The blue dot at $r_{0} \approx 2.59 \ellp$ corresponds to the cold remnant ($T=0$), while the red dot marks the maximum temperature $T_{\max}\approx 9.34\times 10^{-3}\Mpl$ at $r_{+, \max} \approx 6.09 L_\mathrm{Pl}$. 
        }
    \label{fig:kempftemp}
\end{figure}

The existence of an extremal configuration has an impact on the thermodynamics. The Hawking temperature does not diverge as the black hole evaporates, but rather reaches a maximum before the SCRAM phase, {\it i.e.}, an asymptotic cooling towards a zero temperature black hole remnant. The plot in Fig.~\ref{fig:kempftemp} shows the temperature of the metric \eqref{eq:gupschwarzschild}, namely
\begin{equation}
    T(r_+) = \frac{\kappa}{2\pi} = \frac{1}{4\pi} \frac{\diff g_{00}}{\diff 
    r}\bigg\rvert_{r=r_+}
    =
    \frac{1}{4\pi\, r_+}
    \left(
    1 - \frac{r_+^2}{\beta}
    \frac{e^{-r_+ / \sqrt{\beta}}}{\gamma(2;\, r_+/\sqrt{\beta})}
    \right)\,.
\end{equation}
Such a temperature resembles the behaviour of the Reissner-Nordstr\"{o}m, Kerr and Kerr-Newman metrics. 
One has to note, however, that despite the similar profile, the evaporation of Reissner-Nordstr\"{o}m, Kerr and Kerr-Newman metrics is drastically different. Indeed the SCRAM phase never takes place in such cases. Rather than cooling down, such charged, rotating, charged-rotating metrics reach a Schwarzschild configuration at the end of the balding and spin-down phases, that fatally occur in the presence of emissions like Hawking evaporation and  superradiance. On the other hand, the metric in \eqref{eq:gupschwarzschild} extends the thermodynamics of the Schwarzschild phase by properly taking into account the quantum backreaction. To see this one can consider the ratio $T/M<T_{\max}/M_0\approx 8.1 \times 10^{-3} G_\mathrm{N}/\beta$, being the maximum temperature $T_{\max}\approx 1.35\times 10^{-2}/\sqrt{\beta}$ and the extremal mass $M_0=1.68 \sqrt{\beta}/G_\mathrm{N}$. 

At this point, one can use the argument of the gravity self-completeness to set the value of the parameter $\beta$. By requiring that the evaporation remnant fulfills the particle-black hole condition,
\begin{equation}
\frac{2\pi\hbar}{M_0}\equiv r_0\,,
\label{eq:selfcomplcond}
\end{equation}
one obtains $\sqrt{\beta}=1.45\ellp$. This value allows for a further estimate of the back reaction. During the evaporation process the ratio $T/M$ is always small, namely
\begin{equation}
T/M<T_{\max}/M_0\approx 4.0\times 10^{-3}\,,
\end{equation}
being now $T_{\max}\simeq 9.31\times 10^{-3}\ \Mpl$, $M_0\simeq 2.42\Mpl$ and $r_0\simeq 2.59\ellp$.

The self-completeness condition plays an important role to protect the curvature singularity. For $M<M_0$ one finds a horizonless geometry with a  gravitational field $\vec{g}=\frac{1}{2}\vec{\nabla} g_{00}$ that is constant and repulsive in a neighborhood of the origin and vanishing at infinity.  Being the spacetime still singular one should speak of a naked singularity. The self-completeness is, however, an argument to rule out this case. Either by matter compression or Hawking decay, there is no possibility to land on such a horizonless spacetime. In other words, in such a regime, the typical length scale associated to the mass $M$ is its Compton wavelength, $\sim 1/M$. The naked singularity can never be probed.

On similar grounds, the self-completeness allows to circumvent the problem of the Cauchy instability of the inner horizon \cite{BaN10,BrM11}. Since $r_0$ actually acts as a genuine quantum gravity ultraviolet cutoff, it is no longer meaningful to consider classical perturbations at length scales $r_-<r_0$.

\section{Higher dimensional KMM Black Holes}
\label{sec:HighDKMM}
\begin{figure}[h]
	\centering
    \includegraphics[width=0.75\linewidth]{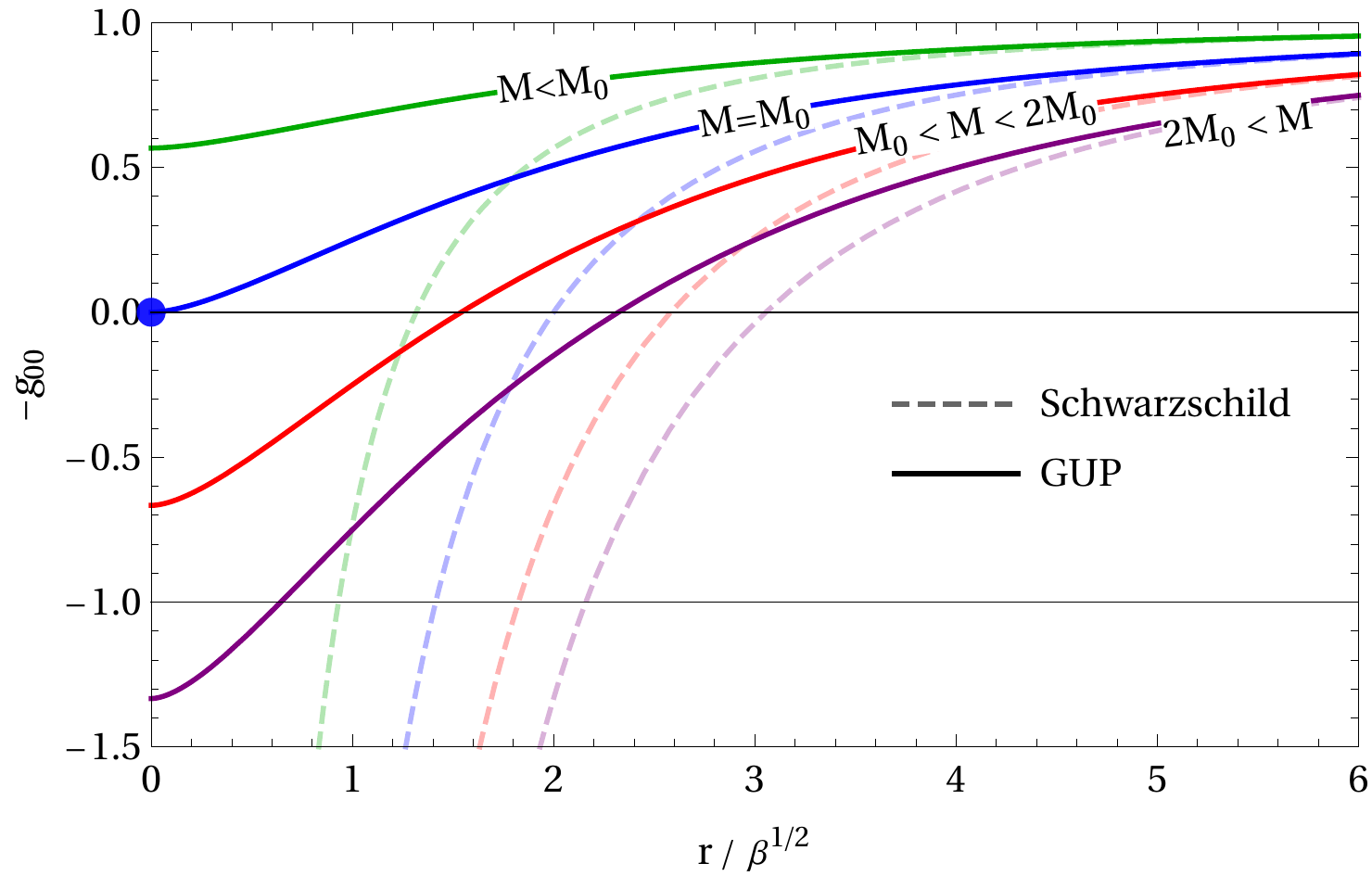}
	\caption{
		The GUP metric component $g_{00}$ according to the KMM measure in $4+1$  
		dimensional spacetime, shown in units of $\sqrt{\beta}$.
		At the origin the spacetime has  a gravitational monopole. Depending on the value of the mass one has one or no horizon. The blue curve represents the final   configuration of the horizon evaporation for a mass $M=M_0=3\pi\beta/2G_\ast$ corresponding to a deficit angle $2\pi$. This configuration is represented as a blue dot in Figure \ref{fig:kempfextramdimTemp} where temperatures as a function of the horizon radius for different spacetime dimensions are displayed. The red and green lines represent configurations with deficit angle smaller than $2\pi$, and one and no horizon respectively. The purple line represents a configuration with an excess angle.	
	}\label{fig:kempfextradimg00}
\end{figure}

In this section we consider the extension of the KMM model to the higher dimensional scenario along the lines of Section \ref{sec:KMM-GUP}. This corresponds to having an integration measure of the kind in \eqref{eq:intvol}. Following the previous exposition, one has to start by determining the energy momentum tensor $\mathbb{T}_{\mu\nu}$. Apart from the higher dimensional gravitational constant $G_\ast$ in place of $G_\mathrm{N}$, the profile of the operator $\mathbb{G}\left(L^2\Box\right)$ remains the same as in the $(3+1)$-dimensional case,  \eqref{eq:goperator}. As a result, one has
\begin{align}
\mathbb{T}^0_0(\vec{x}) = -\rho(\vec{x})=-
\frac{M}{(2\pi)^{d-1}}\int\diff^{d-1}p\, \frac{e^{i\vec{x}\cdot\vec{p}}}{1 + 
	\beta \vec{p}^2}\,.
\end{align}
The integration includes $d-4$ additional spatial dimensions and leads to the following result~\cite{Knipfer2014}: 
\begin{equation}
\rho(\vec{x}) = 
 \frac{M}{(2\pi)^{n/2}}
\left( \frac{|\vec{x}|}{\sqrt{\beta}} \right)^{1 - \nicefrac n2}
K_{\frac{n}{2}-1}\left( \frac{|\vec{x}|}{\sqrt{\beta}} \right)\,,
\label{eq:ndimrho}
\end{equation}
where $n= d-1$ is the number of spatial dimensions and $K_\alpha$ is the modified Bessel function of the second kind.
Integrating equation~\eqref{eq:ndimrho} over an $n$-ball~$B_r$ of radius~$r$ yields the cumulative mass distribution 
\begin{align}\label{eq:mass-ndim}
\mathcal{M}(r) &= \int_{B_r}\diff^n x~\rho(\vec{x}) 
= \frac{(2\pi)^{\nicefrac n2}}{\Gamma(\nicefrac n2)} \int_0^r \diff |\vec{x}|~ |\vec{x}|^{n-1} 
~\rho(|\vec{x}|)\nonumber
\\&
=
M \left[ 1 - \frac{2^{1-\nicefrac{n}{2}}}
{\Gamma\left(\nicefrac{n}{2}\right)}
\left( \frac r{\sqrt{\beta}} \right)^{\nicefrac{n}{2}}
K_{\nicefrac{n}{2}}\left( \frac{r}{\sqrt{\beta}} 
\right)
\right]
\,.
\end{align}

The metric can be written as
\begin{equation}
\diff s^2 = - f_n(r)\ \diff t^2 + 
f_n^{-1}(r)\ \diff r^2 + r^2 \diff\Omega^2_{n-1}\,,
\label{eq:fen}
\end{equation}
where $\diff\Omega^2_{n-1}$ is the $(n-1)$-dimensional spherical surface element
 and the metric function $f_n(r)$ is given by
\begin{equation}
f_n(r) =1- \frac{8 G_\ast  \, \Gamma(\nicefrac{n}{2})}{(n-1)\pi^{\nicefrac{n}{2}-1}}
\frac{\mathcal{M}(r)}{r^{n-2}}
\,.
\label{eq:tangherlini} 
\end{equation}
The Ansatz for the metric \eqref{eq:fen} requires a conserved energy momentum tensor of the form
$\mathbb{T}_\mu^\nu=\mathrm{diag} \left(-\rho, p_r, p_\perp, p_\perp, ...\right)$ with $p_r=-\rho$ and $p_\perp=-\rho-\frac{r}{n-1} (\diff \rho/\diff r)$.

The metric coefficient can be cast in a  more compact form as
\begin{equation}
f_n(r) =1- \frac{2 G_\ast m\ \mu(r)}{r^{n-2}}
\,,
\end{equation}
where
\begin{equation}
\mu(r)\equiv  1 - \frac{2^{1-\nicefrac{n}{2}}}
{\Gamma\left(\nicefrac{n}{2}\right)}
\left( \frac r{\sqrt{\beta}} \right)^{\nicefrac{n}{2}}
K_{\nicefrac{n}{2}}\left( \frac{r}{\sqrt{\beta}} 
\right)
\,,
\end{equation}
and
\begin{equation}
m\equiv M \frac{4  \, \Gamma(\nicefrac{n}{2})}{(n-1)\pi^{\nicefrac{n}{2}-1}}
\,.
\end{equation}
By considering the following system:
\begin{equation}
\begin{cases} 
f_n(r)=0\\ \frac{\mathrm{d}f_n(r)}{\mathrm{d}r}=0
\,,
\\ 
\end{cases} 
\end{equation}
one can look for a solution $r=r_0$ representing the vanishing minimum of $f_n(r)$. If the solution exists for a specific value $m=m_0$, one has determined what is physically known as an extremal black hole configuration. For $n=3$ one finds $m_0=M_0=1.68 \sqrt{\beta}/G_\mathrm{N}$ and $r_0=1.79\sqrt{\beta}$,
as was shown in the previous section. For $n>4$, the system has no positive defined solution. This can easily be seen by considering the expansion of the function $\mu(r)$ for small arguments:
\begin{equation}
\mu(r) \xrightarrow[r \ll \sqrt{\beta}]{}
(2n-4)^{-1}
\left( \frac r{\sqrt{\beta}} \right)^{2}\,.
\label{eq:muexpansion}
\end{equation}
For small radii, one finds $\mu\sim r^2$ and the function $f_n$ diverges negatively at the origin for $n>4$.  Conversely, it asymptotes to the Minkowski space at large distances where $\mu\approx 1$. This behaviour suggests that $f_n$ is a monotonic increasing function having a single zero, \textit{i.e.}, the event horizon.

For $n=4$, one finds a surprising case. The solution of the system is $m_0 = 2 \beta/G_*$ and $r_0=0$. Evidently this does not represent an extremal configuration, but instead reveals the presence of a gravitational object of a different nature (see Fig. \ref{fig:kempfextradimg00}). By using \eqref{eq:muexpansion}, one can write the metric in a region near the origin for $n=4$ as
\begin{equation}
\diff s^2 \approx - \left(1-\frac{2G_\ast M}{3\pi\beta}\right)\ \diff t^2 + 
\left(1-\frac{2G_\ast M}{3\pi\beta}\right)^{-1}\ \diff r^2 + r^2 \diff\Omega^2_{3}\,.
\end{equation}
We note the Newtonian potential is constant at short scales. This implies that the mass does not produce any gravitational field near the origin. One can see this by rescaling the $r$ and $t$ variables and by expressing the above metric in the form
\begin{equation}
\diff s^2 = -  \diff t^2 + 
\diff r^2 + \left(1-\frac{2G_\ast M}{3\pi\beta}\right) r^2\left( \diff\theta^2_2+\sin^2\theta_2\left(\diff\theta_1^2+\sin^2\theta_1\diff \phi^2\right)\right)\,,
\end{equation}
which introduces a deficit angle. 
Indeed by considering the surface $t=\mathrm{const.}$, $\theta_1=\theta_2=\pi/2$, one finds the geometry of a cone, whose conical singularity is a curvature singularity.
The behaviour of the energy density for $n=4$ at short scales, $\rho(\vec{x})\sim |\vec{x}|^{-2}$, confirms this pathology of the manifold. The above scenario reveals that for $n=4$, the gravitational object at the origin is a Barriola-Vilenkin global monopole \cite{BaV89}, \textit{i.e.}, a spacetime object that resembles a cosmic string \cite{FIU89}. 

This is just a first glimpse of the interesting properties of the manifold. The topology of the spacetime changes by varying the parameter $M$. For $M>3\pi\beta/2G_\ast$, the coefficient $g_{00}$ is positive and larger than $1$. The aforementioned surface then reads
\begin{equation}
\diff s^2 = - 
\diff r^2 + r^2\diff \phi^2\,,
\label{eq:innermonopole}
\end{equation}
with $0\leq \phi<2\pi \sqrt{\left|1-\frac{2G_\ast M}{3\pi\beta}\right|}$. The above line element describes the short scale spacetime inside the event horizon, in the presence of an excess angle.

For $\frac{3\pi\beta}{2G_\ast}<M<\frac{3\pi\beta}{G_\ast}$, the line element also describes the short scale spacetime inside the event horizon, but in this case with a deficit angle. By decreasing $M$, \textit{e.g.}, during the horizon evaporation, the mass parameter tends to $\frac{3\pi\beta}{2G_\ast}$. This limit implies a deficit angle $\mathrm{def}(\phi)=2\pi$, corresponding to the closure of the cone and to a complete evaporation of the horizon. Indeed in such a limit the event horizon tends to zero, \textit{i.e.}, $r_+\to 0$. The evaporation end point is a degenerate monopole whose geometry is no longer a cone but a straight line (see discussion below). It is interesting to note that our metric is an exact solution that interpolates the geometry of the monopole at the origin with that of the Schwarzschild black hole at large scales. Contrary to Barriola and Vilenkin we do not employ any perturbation theory.

For lighter objects, $M<\frac{3\pi\beta}{2G_\ast}$ no horizon forms and $g_{00}$ is negative and larger than $-1$.
By the usual coordinate rescaling the aforementioned surface at the origin can be written as
\begin{equation}
\diff s^2 =  
\diff r^2 + r^2\diff \phi^2\,,
\label{eq:nakedmonopole}
\end{equation}
with $0\leq \phi<2\pi \sqrt{1-\frac{2G_\ast M}{3\pi\beta}}$.
When $M=0$ there is no deficit angle, and the spacetime is the regular Minkoswki spacetime. For any mass in the interval $0<M<\frac{3\pi\beta}{2G_\ast}$, the deficit angle is non vanishing and the conical singularity develops. As a result one finds the geometry of a  ``naked monopole''. The left limit $M\to \frac{3\pi\beta}{2G_\ast}$ implies $\mathrm{def}(\phi)\to 2\pi$, namely the degenerate cone. Note that the right and the left limits $M\to \left(\frac{3\pi\beta}{2G_\ast}\right)^{\pm}$ are characterized by a different sign of the $g_{rr}$ coefficient in \eqref{eq:innermonopole} and \eqref{eq:nakedmonopole}. This corresponds to the peculiar situation in which the naked monopole and the monopole inside the black hole coalesce to form a straight line geometry while the event horizon dissolves.

\begin{figure}[t]
	\centering
	\includegraphics[width=0.75\linewidth]{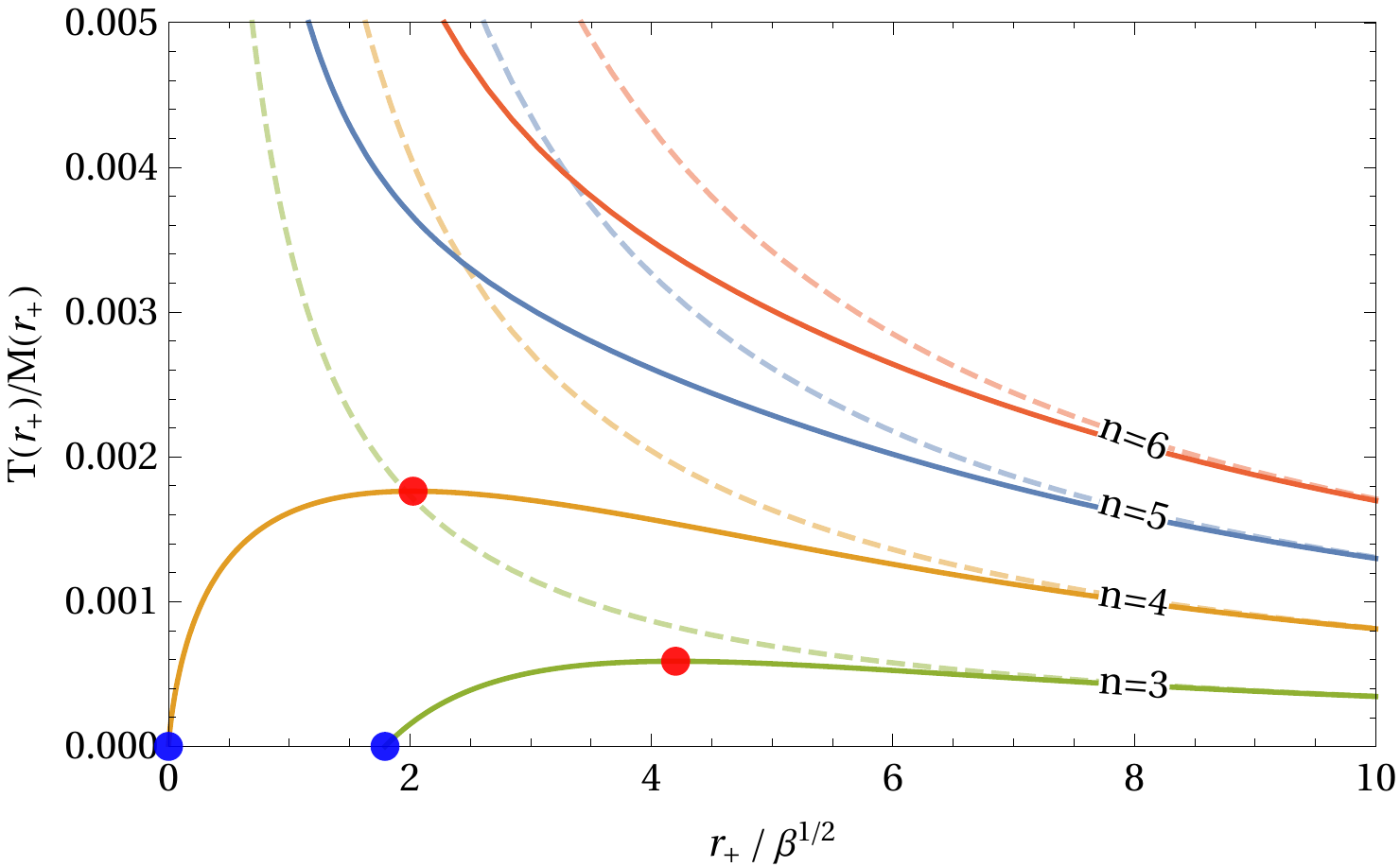}
	\caption{
			Hawking temperatures as a function of the horizon radius for different spacetime dimensions according to the GUP with the KMM measure. For $n=3$ there is a SCRAM phase ending in an extremal configuration. For $n=4$ there is a SCRAM phase ending in a gravitational monopole. For $n>4$, the GUP corrections are clearly negligible and the temperature diverges in a similar way to the Schwarzschild-Tangherlini case (dashed lines).
	}\label{fig:kempfextramdimTemp}
\end{figure}

A study of the related thermodynamics can be done by considering the Hawking temperature:
\begin{equation}
T=\frac{n-2}{4\pi r_+}\left[1-\frac{r_+}{n-2}\frac{\mu^\prime(r_+)}{\mu(r_+)}\right]\,.
\end{equation}
Before displaying the exact expression for  $T$, we consider its asymptotic nature. Since the function $\mu\to 1$ for $r\gg \sqrt{\beta}$ , $T$ approaches the standard semiclassical result at large distances. Conversely at short scales, \eqref{eq:muexpansion} leads to 
\begin{equation}
T\approx\frac{n-2}{4\pi r_+}\left[1-\frac{2}{n-2}\right], \quad \mathrm{for}\ r\ll \sqrt{\beta}\,.
\end{equation}
For $n>4$ the temperature has a divergent behaviour as in the semiclassical case. On the other hand, for $n=4$ the temperature vanishes in the limit $r_+\to 0$. This means the  temperature admits a maximum and undergoes a SCRAM phase. Following what is discussed above, the horizon structure prevents the formation of an extremal configuration. As a result, this analysis confirms our previous predictions: the final state of the evaporation is nothing but a global monopole with mass 
$M=M_0=3\pi\beta/2G_\ast$, corresponding to a degenerate cone with deficit angle $2\pi$.

The full expression of the temperature for any number of spatial dimensions $n$ is~\cite{Knipfer2014}
\begin{equation}
T=
\frac{n-2}{4\pi r_+}
\,\left(
1-
\frac{r_+^{n/2 + 1}}
{
 (n-2)
 \Gamma\left(\frac n2\right) 2^{\frac n2} \beta^{\frac n4}
}
\frac{
\beta^{-\frac 12}
\left(
K_{n/2-1} + K_{n/2+1}
\right)
-\frac n{r_+} K_{n/2}
}{
1 -
\frac{2^{1-n/2} r_+^{n/2}}{
\Gamma( n/2 ) \beta^{n/4}}
K_{n/2}
}
\right)
\,,
\end{equation}
with $K_i = K_i(r_+ / \sqrt{\beta})$. From Figure \ref{fig:kempfextramdimTemp} it is clear that for $n>4$ the back reaction cannot be neglected in the final stages of the evaporation. On the other hand, for $n=4$ there is a maximum temperature, $T_{\max} \approx 1.93\times 10^{-2}/\sqrt\beta$. From the same Figure one learns that the ratio $T(r_+)/M(r_+)<2\times 10^{-3}$ always. 
This means that the geometry describing a monopole inside a black hole is not only an exact solution but back reaction free.

By applying the argument of gravity self-completeness to the above monopole parameters for $n=4$ one would be tempted to fix the value of the parameter $\beta$. This is problematic, since the monopole at the end of the evaporation is not an extremal black hole configuration, since $r_+\to 0$. As a result, one cannot invoke \eqref{eq:selfcomplcond} to determine $\beta$. There is, however, a possible way out. The black hole undergoes a SCRAM phase. This means that, in contrast to the Schwarzschild-Tangherlini case, the final stages of the evaporation are characterized by thermodynamic stability. The heat capacity of the system is actually positive defined, namely
\begin{equation}
C(r_+)=\frac{\partial M}{\partial r_+}\left(\frac{\partial T}{\partial r_+}\right)^{-1}>0 \quad \mathrm{for}\quad 0<r_+<r_{+, \max}\,.
\end{equation}
This means that the black hole can reach the thermal equilibrium if immersed in a background  at temperature $T_\mathrm{b}$, with $0<T_\mathrm{b}< T_{\max}$. As a result such an equilibrium point, $(r_{+, \mathrm{b}},\ M_\mathrm{b}$), can represent the transition between the particle and black hole phases. Accordingly the self-complete scenario is safe and one can invoke the condition
\begin{equation}
\frac{2\pi\hbar}{M_\mathrm{b}}=r_{+,\mathrm{b}}\,.
\end{equation}
The parameter $\beta$ can be fixed and turns out to be a function of the background temperature,  $\beta=\beta(T_\mathrm{b})$.
On the other hand, for $n>4$, the self-completeness scenario is lost much in the same way as in the Schwarzschild-Tangherlini case. 

\section{Higher dimensional  black holes from the CLCS momentum measure}
\label{sec:Cas-Scar}

\begin{figure}[t]
    \centering
    \includegraphics[width = 0.7 \textwidth]{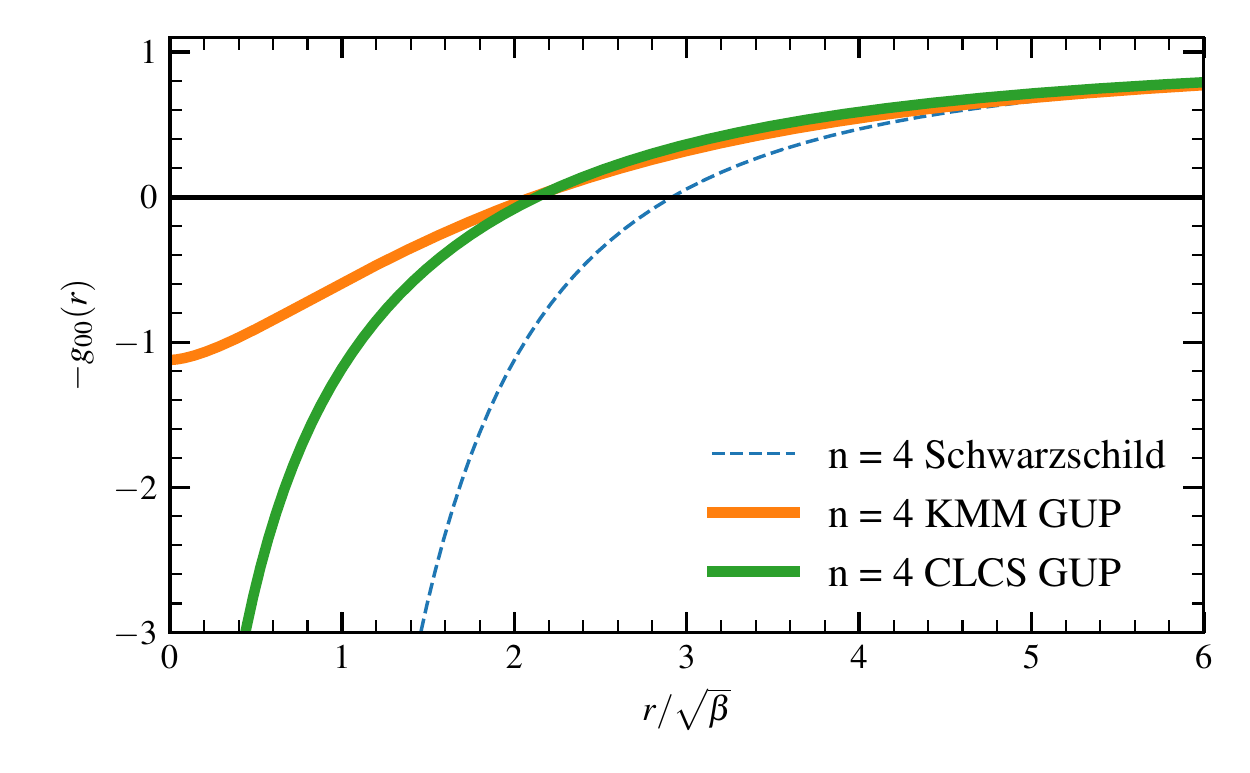}
    \caption{
      Comparison of the metric component $-g_{00}(r)$ in the different GUP models, in $n=4$ spatial dimensions. Shown
      are the Schwarzschild-Tangherlini solution, the Kempf-Mangano Mann model \eqref{eq:kempf} and the Carr-Lake-Casadio-Scardigli model
      \eqref{eq:intvol2}. Here, all metric components are shown for 
      $M=10 \sqrt{\beta}/G_\ast$.
    }
    \label{fig:N4-comparison}
\end{figure}

In this section we consider a modification of the momentum measure according to the proposal 
by Carr-Lake-Casadio-Scardigli (CLCS) in Equation~\ref{eq:intvol2}.
From the large momentum limit, one expects a minimal deviation with respect to the Schwarzschild black hole 
(see Table~\ref{table:gup-comparison}).

As a start, one has to determine the corresponding operator $\mathbb{G}\left(L^2\Box\right)$, namely
\begin{equation}
\mathbb{G}\left(L^2\Box\right)=\frac{G_\ast}{1-\left(L^2\Box\right)^{\frac{1}{2}\frac{n-1}{n-2}}}\equiv G_\ast \sum_{N=0}^\infty \left[\left(L^2\Box\right)^{\frac{1}{2}\frac{n-1}{n-2}}\right]^N\,,
\end{equation}
where, in case of non-integer exponents, $\frac{n-1}{n-2}=\frac{3}{2}, \frac{5}{2}, \frac{7}{2}, \dots$, the following Schwinger representation can be used to express  powers of an arbitrary operator, $\hat{O}$:
\begin{equation}
\hat{O}^{\alpha}=\frac{1}{\Gamma(-\alpha)}\int_0^\infty ds \ s^{-\alpha - 1} \ e^{-s\hat{O}}, \quad \alpha\in \mathbb{R}\setminus\mathbb{N}\,.
\end{equation}
The resulting energy momentum tensor has the same form of that presented in the previous Section. The energy density and the other components, however, have a different profile.

In order to determine $\mathbb{T}_0^0$, one has to evaluate the integral 
\begin{align}
\mathbb{T}^0_0(\vec{x}) = -\rho(\vec{x})=-
\frac{M}{(2\pi)^{n}} \int \frac{\diff^n p}{1 + 
(\sqrt{\beta} |\vec{p}|)^{(n-1)/(n-2)} }e^{i \vec{x}\cdot\vec{p}}\,,
\label{eq:t00newgup}
\end{align}
where $L=\sqrt{\beta}$. Although the above integral cannot be solved analytically,  it is possible to do so numerically via the approach described in Appendix~\ref{apx:integration}. In a similar fashion,
the mass distribution is obtained by solving \eqref{eq:mass-ndim}.

The strongest corrections with respect the Schwarzschild geometry occur for $n=4$. As a result we focus our analysis on this case only. The metric component $g_{00}$ is shown in Figure~\ref{fig:N4-comparison}, in comparison with the KMM model presented in section \ref{sec:HighDKMM}. As expected, the KMM model is the only one with
a non-singular metric component, while the CLCS model is qualitatively
indistinguishable from the Schwarzschild
black hole. 

For $n>4$, the CLCS measure offers even milder corrections and negligible deviations with respect to the Schwarzschild black hole.
One can see this by performing an analytic approximation. Being $\rho(\vec{x})\approx 0$ for $|\vec{x}|\gg \sqrt{\beta}$, one finds that the short distance behaviour of the geometry is controlled by the ultraviolet limit of the above integral, namely
\begin{align}
\rho(\vec{x})\approx
\frac{M}{(2\pi)^{n}} \int \frac{\diff^n p}{
(\sqrt{\beta} |\vec{p}|)^{(n-1)/(n-2)} }e^{i \vec{x}\cdot\vec{p}}\sim \frac{M}{(\sqrt{\beta})^{\frac{n-1}{n-2}}|\vec{x}|^{n-\frac{n-1}{n-2}}}\quad \mathrm{for}\ |\vec{x}|\ll \sqrt{\beta}\,.
\end{align}
As a result the degree of divergence is less brutal than in the Schwarzschild case, but more severe than in the KMM model for which one finds $\rho(\vec{x})\sim |\vec{x}|^{-(n - 2)}$ for $|\vec{x}|\ll \sqrt{\beta}$. 

 Another drawback of the CLCS model is that, for $n>3$, it is never ultraviolet self-complete. The black hole can decay to sizes smaller than $L_\ast$ and expose the curvature singularity. These results are a motivation to consider a further model to improve the Schwarzschild geometry according to the GUP tenets.

\section{Revised GUP in higher dimensions}
\label{sec:ourGUP}

The discussion in Section \ref{sec:ambiguitygup} suggests that the KMM model might be ineffective to tame the ultraviolet divergences in higher dimensional space. This fact is translated in mild modifications of the metric and the thermodynamics of black hole geometries for $n>4$. Only for $n=3$ and $n=4$ does the KMM model provide a significant departure from Einstein gravity. In the former case the KMM model predicts an extremal configuration as an end point of the black hole evaporation. In the latter case, the end point is a gravitational monopole resulting from a full evaporation of the event horizon. Furthermore, the discussion in Section \ref{sec:Cas-Scar} shows that the GUP as proposed in \eqref{eq:intvol2} is not suitable for  improving the thermodynamics of the Schwarzschild-Tangherlini solution in $n>3$ spatial dimensions.

In this section, we consider the GUP model whose integration measure is of the form denoted in \eqref{eq:intvol3}.  
As a start, we present the corresponding operator $\mathbb{G}\left(L^2\Box\right)$, namely
\begin{equation}
\mathbb{G}\left(L^2\Box\right)=\frac{G_\ast}{1-\left(L^2\Box\right)^{\frac{n-1}{2}}}\equiv G_\ast \sum_{N=0}^\infty \left[\left(L^2\Box\right)^{\frac{n-1}{2}}\right]^N\,.
\end{equation}
To determine $\mathbb{T}_0^0$, one has to consider the integral 
\begin{align}
\mathbb{T}^0_0(\vec{x}) = -\rho(\vec{x})=-
\frac{M}{(2\pi)^{n}} \int \frac{\diff^n p}{1 + 
(\sqrt{\beta} |\vec{p}|)^{n-1} }e^{i \vec{x}\cdot\vec{p}}\,.
\label{eq:t00newgup}
\end{align}
Due to the complex structure of the integral measure in \eqref{eq:t00newgup}, an analytic solution is not viable. However, analytic approximations are still possible. At large distances one finds $\rho(\vec{x})\approx 0$. By expanding the integrand function for high momenta, one can estimate the behaviour of the energy density at short scales, namely
\begin{align}
\rho(\vec{x})\sim \frac{M}{(\sqrt{\beta})^{n-1}|\vec{x}|}\quad \mathrm{for}\ |\vec{x}|\ll \sqrt{\beta}\,.
\end{align}
The spacetime is still singular but the degree of divergence no longer increases with the number of dimensions. It is actually independent of $n$ and turns to be softer than any previous higher dimensional GUP models under consideration. This result descends from having a GUP model with a uniform ultraviolet behaviour of integrals in momentum space, as shown in Table~\ref{table:gup-comparison}.

A numerical approach allows one to obtain an energy density at any length scale, and subsequently a mass distribution, by solving \eqref{eq:mass-ndim}. To do so, the $n$-dimensional Fourier transformation in \eqref{eq:t00newgup} is rewritten as a Hankel transformation, and technical details are given in Appendix~\ref{apx:integration}. Figure~\ref{fig:adjustedGUP-MassDistr} displays the resulting mass distribution $\mathcal{M}(r)$ for $n=3-7$ spatial dimensions. Note that the case $n=3$ is the only one having $\mathcal{M}(r)$ described by a monotonic increasing function. 

For $n>3$ there is a surprising new behaviour: the function oscillates with an amplitude that increases with $n$ and decreases with $r$. A naive interpretation of these oscillations is the presence of negative contributions in the energy density for some regions close to the spatial origin. 
We recall that such negative density regions are not a remote possibility, at least during the early stages of the Universe, for the presence of strong quantum fluctuations of the spacetime manifold \cite{MTY88,Mann97}.

\begin{figure}[t]
	\centering
	\includegraphics[width=0.78\linewidth]{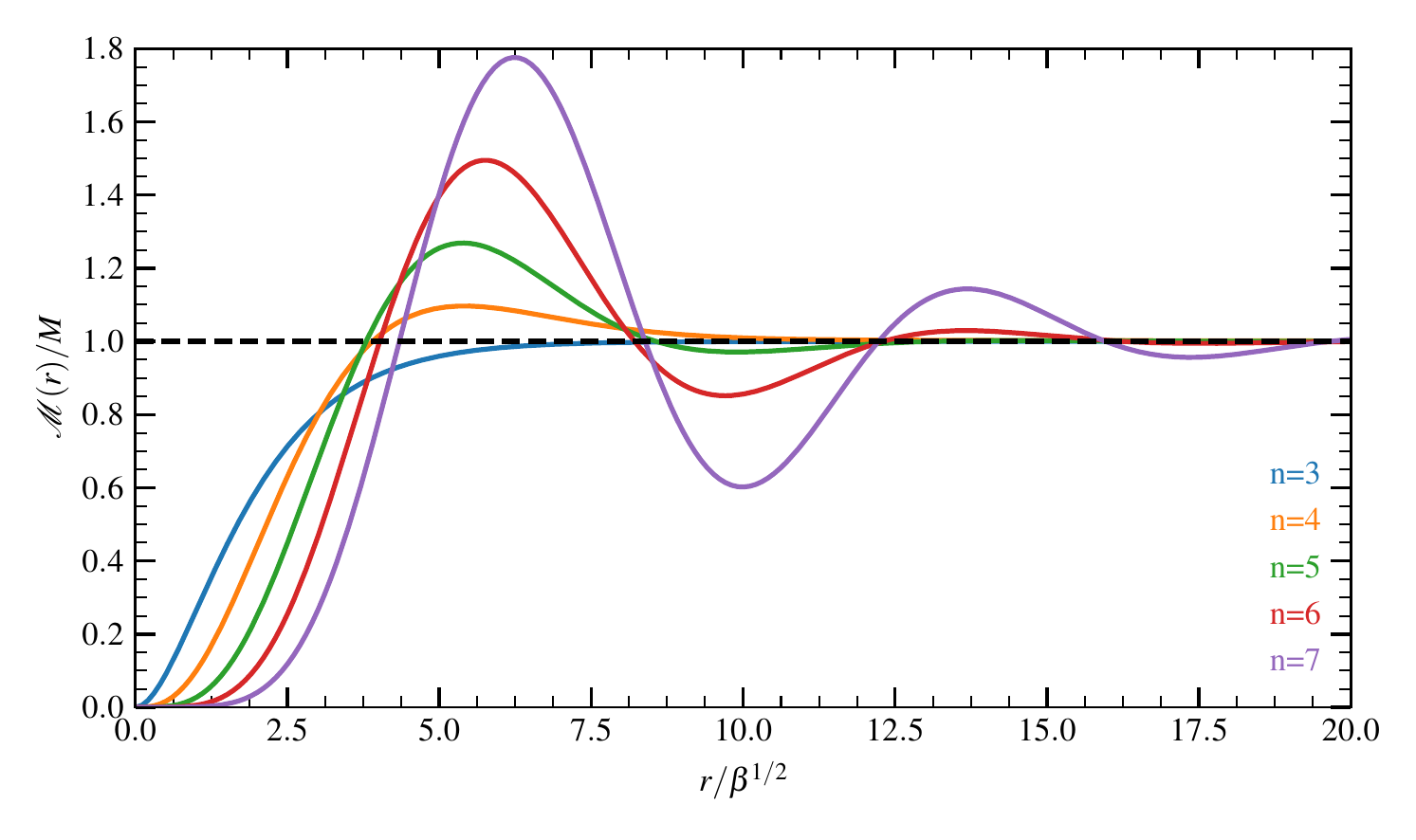}
	\caption{
		 Cumulative mass distributions induced by the revised GUP model
		in $n$ spatial dimensions. The dashed line corresponds to the classical matter distribution (Schwarzschild).
	}\label{fig:adjustedGUP-MassDistr}
\end{figure}

\begin{figure}[t]
	\centering
	\includegraphics[width=0.78\linewidth]{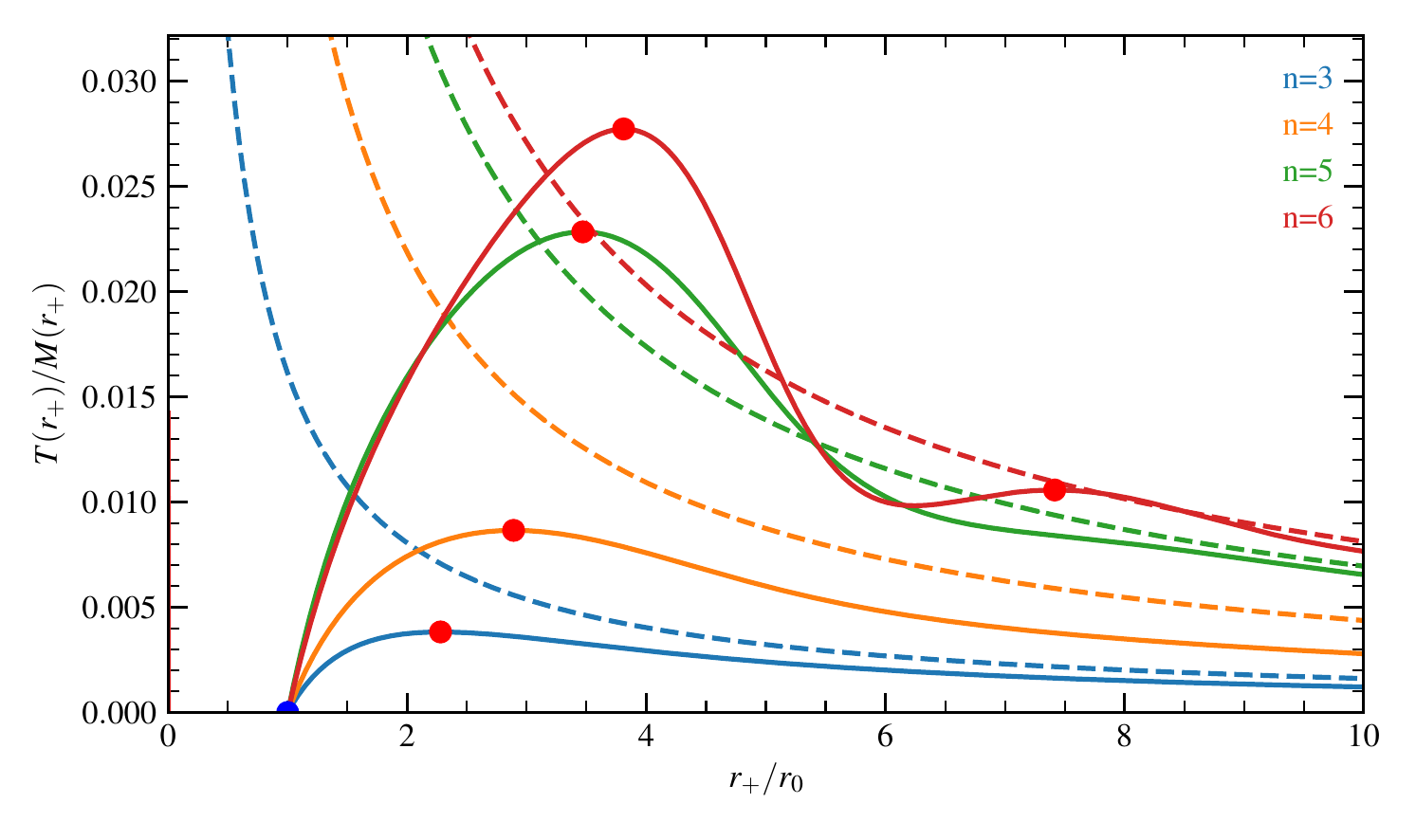}
	\caption{
		Temperatures of the $n$-dimensional revised GUP model for $n=3--6$, each compared with the Schwarzschild
		equivalent (dashed lines). The blue dot corresponds to the cold
		evaporation endpoint (note that all outer horizons $r_+$ are rescaled
		by the individual evaporation endpoint radius $r_0$), while the red dots
		correspond to local temperature maxima.
	}\label{fig:adjustedGUP-temps}
\end{figure}

One can also propose another interpretation based on the presence of $n-1$ tachyon states of mass, $i/\sqrt{\beta}\sim i M_\ast$, emerging from the poles of the integrand function in \eqref{eq:t00newgup}. As a consequence the energy density $\rho(\vec{x})$, despite being positive defined at the origin,  oscillates around zero for larger values of $r$. 
A possible explanation  can be found in the fact that the GUP captures only part of the non-perturbative corrections of quantum gravity. As a result, this interpretation is consistent with the fact that GUP corrections emerge from the  eikonal limit of string collisions at the Planck scale \cite{ACV89,ACV93,ACV87,ACV88,ACV90,GrM88,GrM87}. Interestingly, such oscillations of the Newtonian potential have been found in a variety of other formulations aiming to amend Einstein gravity. These include $f(R)$-gravity \cite{NoO2003,Olmo2005,Faraoni96,CST2007,NoO2007,CENOSZ2008,CDLF2010,BeG2011,Schell2016},  string induced, ghost free, non-local gravity \cite{EKMaz2016,FroZ2016} and other non-local formulations \cite{KeMa2014}.   
  On the other hand, in the low energy limit for which only the three spatial dimensions are visible, the oscillations disappear as expected in similar quantum gravity contexts \cite{HaH02,Perivo16}.

Even if the above effects are just a short-scale quantum mechanical property of the solution, they have important repercussions for the thermodynamics of the system. The profile of the  temperature  is presented in Figure~\ref{fig:adjustedGUP-temps}. One can see that the  oscillations of $\mathcal{M}(r)$  produce temperature oscillations corresponding to phase transitions from negative to positive heat capacity phases. The resulting variable luminosity of the black hole can be termed as a \textit{lighthouse effect}. Again such an effect increases with $n$.
For lower $n$  one obtains small amplitude oscillations of the temperature. 

\begin{figure}
	\centering
	\makebox[\textwidth][c]{ 
		\includegraphics[width=1.3\textwidth]{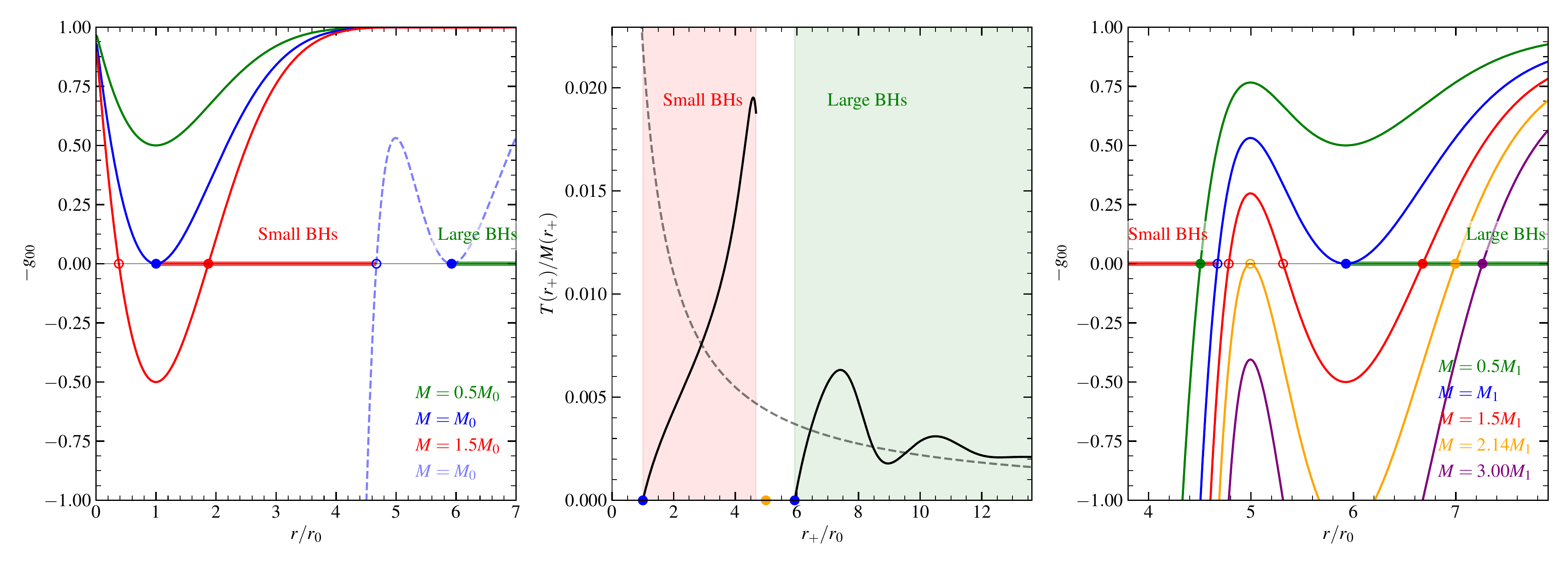}
	}
	\caption{
		Gravitational potentials and temperature 
		for the revised GUP black hole
		(Section~\ref{sec:ourGUP})
		in $n=8$ spatial dimensions.
		The left and right panels show the metric coefficient $g_{00}$
		for different masses while the center panel shows the
		temperature.
		The colouring of the metric function follows 
		Figure~\ref{fig:g00kempf}.
	}\label{fig:higher-dimensional-adjustedGUP}
\end{figure}

More importantly, further scenarios are possible. In Figure~\ref{fig:higher-dimensional-adjustedGUP},
the case of $d=9$ space time dimensions is depicted. For this number of dimensions, two regimes emerge that depend on the black hole mass. For large mass black holes, $M>M_1=3.73 \times 10^6 \sqrt{\beta}/G_*$, the temperature oscillation determines an anticipated shut-down of the Hawking emission with the formation of a zero temperature remnant. 
We refer to the mass of this zero temperature configuration as $M_1$ (blue curve, right panel, Figure \ref{fig:higher-dimensional-adjustedGUP}).
Such a large mass regime is characterized by a rich horizon structure. For $M>2.14 M_1$, there are just two horizons, an event horizon, $r_+$ and an inner Cauchy horizon, $r_-$ (violet curve in Figure \ref{fig:higher-dimensional-adjustedGUP} -- the inner horizon is very close to the origin and not visible in the aforementioned Figure.). For $M=2.14 M_1$, the function $g_{00}$ admits a double zero between the aforementioned two horizons (yellow curve in Figure \ref{fig:higher-dimensional-adjustedGUP}). For smaller masses  $M_1<M<2.14 M_1$, the function $g_{00}$ admits $4$ simple (positive) zeros, \textit{i.e.}, $r_-<r_2<r_3<r_+$ (red curve in Figure \ref{fig:higher-dimensional-adjustedGUP}). Finally for $M=M_1=3.73 \times 10^6 \sqrt{\beta}/G_*$, there is the merge between $r_3$ and $r_+$ that form a double zero, corresponding to the extremal configuration, \textit{i.e.}, the end stage of the evaporation (blue curve in Figure \ref{fig:higher-dimensional-adjustedGUP}). 

In terms of horizon radii, there is a regime that is not realized, {\it i.e.} $r_i < r < r_1$, where $r_1 =5.93r_0$ is the
size of black hole with mass $M_1$, and an intermediate radius $r_i=5.0r_0$. 
For even smaller masses, $M<M_1$ however, there are horizons again, $r_\pm$, that eventually merge in a new extremal configuration $r_0=r_-=r_+$ for $M=M_0=5.15 \times 10^3 \sqrt{\beta}/G_*$ (see left panel in Figure \ref{fig:higher-dimensional-adjustedGUP}). 

In conclusion we have a small mass and a large mass regime for black holes. The former can be thought to form due the high density fluctuation of the early Universe \cite{CaH74} or by de Sitter space decay, as predicted in terms of the instanton formalism \cite{MaR95,BoH96,MaN11}. 

We also note that the presence of two remnant masses posits an ambiguity of the scale at which gravity may be self-complete \cite{MuN12,DFG11,DFG12,SpA11,Nic2018}. In other words, gravity is able to mask singularities by covering them with an extremal configuration but the latter has no unique mass scale. In contrast to larger ones, however, small remnants can be the result of the particle compression phase during a scattering at energy $\sim M_\ast$. This fact promotes them as the transition point between the particle and black hole phases and realizes univocally the self-complete scenario for this model, irrespective of the number of extra dimensions. 

The self-complete nature of the geometry has an important repercussion.
For $M<M_0$, one finds a singular horizonless geometry for all $n$. Such a naked singularity can, however, never be probed. The self-complete characters guarantees that singularities are masked in any actual physical process. Finally, Figs.
\ref{fig:adjustedGUP-temps} and \ref{fig:higher-dimensional-adjustedGUP} show that the back reaction is negligible. The ratio $T(r_+)/M(r_+) \ll 1$ for any $n$.

\section{Conclusions}
\label{sec:Conclusions}
In this paper we have studied GUP effects via a nonlocal gravity approach with black holes in four- and higher-dimensions. Specifically, we discussed three possible deformation of Einstein gravity inspired by the GUP.

As a first step, we derived black hole spacetimes by adopting the KMM formulation in $d\geq 4$ dimensions.  In particular, the case of one extra dimension ($d=5$) yields a black hole solution possessing a number of interesting characteristics.  The temperature profile for the $d=5$ case shows a  SCRAM behaviour similar to that of the $d=4$ black hole, although instead of ending its evaporation as a remnant, the former reaches a configuration known as a Barriola-Vilenkin monopole. Interestingly no perturbation theory has been employed to obtain such an exact result. Furthermore, the SCRAM guarantees the absence of relevant back reaction during the whole evaporation.  The topology of the spacetime is found to be dependent on the ADM mass, such that the monopole is enveloped by either a single horizon for large masses, or no coordinate singularity for small masses ({\it i.e.} a naked monopole).
For $d>5$, the KMM solution shows limited to no influence from the GUP, and behaves in a similar fashion to higher-dimensional Einstein gravity.

As a second step, we have derived black hole spacetimes following the CLCS proposal. We showed that they exhibit minor deviations from the ordinary Schwarzschild -Tangherlini metric. As a result, they suffer from relevant back reaction and do not fulfill the requirement of gravitational self-completeness.

Lastly, we have derived a black hole spacetime based on a new GUP measure in momentum space that guarantees uniform ultraviolet convergence for any number of dimensions. Such a property allows us to overcome the limitations of the KMM model.  The most novel feature of this new measure is that the mass distribution ${\cal M}(r)$ shows an oscillatory behaviour as $r$ decreases.  This is possibly due to negative energy density contributions introduced by quantum fluctuations, or alternatively the presence of tachyonic states. A subsequent oscillation of the Hawking temperature also results, revealing a new feature we appropriately call a ``lighthouse effect''.  The effect becomes enhanced as the number of extra dimensions increases.  In all cases the endpoint of evaporation is an extremal black hole remnant, though for $n=8$ the oscillations determine two mass regimes admitting a zero temperature configuration.
Contrary to the previous two cases, these spacetimes are back reaction free and are thus gravitationally self-complete.

Before concluding we would like to comment about the observational consequences of our findings. The possibility of observing the effects presented in this paper and other quantum gravity phenomenology is related to the energy scale at which quantum gravity itself is expected to show up.  The energy scale $M_\ast$ can be Planckian or at an intermediate value between the electroweak scale and the Planck scale.
It is interesting to note that the value of $M_\ast$ does not affect the results of our paper since they are valid for a generic value of the fundamental scale. The presented phenomenology will show up at an energy scale $M_\ast$ whatever it is. Only the possibility of detecting such a phenomenology depends on the value of $M_\ast$. 

If the fundamental scale is Planckian, $M_\ast\sim M_\mathrm{Pl} \sim 10^{16}$ TeV, quantum gravity phenomenology is expected to be quite hard to expose. Direct observations of microscopic black hole formation in particle detectors would be virtually impossible with  current and near future technology. Indirect effects in cosmic messengers might offer a more promising testbed (see for instance \cite{SNB11a}). Being a more promising arena does not mean, however, that the observation of such new effects is easy. The detection of multi messengers  has required big technological efforts (\textit{e.g.} gravitational waves). Deviations from Einstein gravity encoded into them are even harder to detect since they affect the signal profile  at a subleading order and might be covered by the background noise\footnote{For sake of completeness we mention recent proposals claiming the possibility of observing extremely small Planckian effects at future gravitational wave experiments \cite{Maselli2017,Addazi2018}.}. 

On the other hand, if the scale $M_\ast$ is lower than the Planck scale, the scenario might be drastically different. As of today we know that for $M_\ast\lesssim 13$ TeV, there is no evidence of black hole formation in particle detectors, at least according to Einstein gravity \cite{Sirunyan:2018xwt}. This would suggest that the value of $M_\ast$ is somewhere between $13$ TeV and $10^{16}$ TeV. Our  results are compatible with such a conclusion but they offer also an alternative possibility. The scale 
$M_\ast$ might still be around the terascale, $M_\ast\lesssim 13$ TeV, but the formation of gravitational objects would be prevented by a threshold mass $M_0$, set by the end stage of the evaporation. In other words, as long as the accelerator working energy is $E<M_0$ there is no black hole production despite the fundamental scale has been reached, $E\sim M_\ast\lesssim 13$ TeV. This is a new feature since Einstein gravity does not set any threshold mass \cite{MNS12}.

Finally the light house effect might offer an alternative phenomenology. Provided we are able to observe an evaporating black hole we expect two main features. First, the spectrum of the emitted particle should greatly differ from that of a Tangherlini-Schwarzschild black hole with the same mass. In particular we expect emission of particles at varying energy. For photons this corresponds to an alternating frequency. Second, an oscillating temperature might lead to the formation of cold stable black hole remnants with mass $M\sim M_1$ where $M_\ast \ll M_0\ll M_1$. The latter might be an alternative candidate of dark matter component with masses not exceeding $10^{-2}$ kg for $M_*\sim 10^{16}$ TeV, or $10^{-17}$ kg for $M_*\sim 13$ TeV (see for instance the case $d=9$ in Section \ref{sec:ourGUP}). In both cases such mass regimes would  escape current constraints on primordial black holes that tend to exclude the interval $M>10^{19}$ kg. \cite{Niikura:2017zjd}.

We believe the current results open the door towards many future investigations. It would be crucial to understand, for example, why temperature oscillations show up only in the presence of extra dimensions.  We note that a similar behaviour arises in a GUP-modified Reissner-Nordstr\"om metric \cite{CMMN}.   A deeper understanding of ghosts in quantum gravity would be pertinent in this respect.

\subsection*{Acknowledgements}
JM would like to thank the generous hospitality of Frankfurt Institute 
for Advanced Studies (FIAS), at which this work was started.
The work of JM was supported by a Frank R. Seaver Research Fellowship from Loyola Marymount University. 
The work of PN has been supported by the German Research Foundation (DFG) grant NI 1282/2-1, partially by the Helmholtz International Center for FAIR within the framework of the LOEWE program (Landesoffensive zur Entwicklung Wissenschaftlich-\"{O}konomischer Exzellenz) launched by the State of Hesse and partially by GNFM, the Italian National Group for Mathematical Physics. The authors thank Maximiliano Isi for collaboration during the early stages of this work.


\begin{thebibliography}{127}%
\makeatletter
\providecommand \@ifxundefined [1]{%
 \@ifx{#1\undefined}
}%
\providecommand \@ifnum [1]{%
 \ifnum #1\expandafter \@firstoftwo
 \else \expandafter \@secondoftwo
 \fi
}%
\providecommand \@ifx [1]{%
 \ifx #1\expandafter \@firstoftwo
 \else \expandafter \@secondoftwo
 \fi
}%
\providecommand \natexlab [1]{#1}%
\providecommand \enquote  [1]{``#1''}%
\providecommand \bibnamefont  [1]{#1}%
\providecommand \bibfnamefont [1]{#1}%
\providecommand \citenamefont [1]{#1}%
\providecommand \href@noop [0]{\@secondoftwo}%
\providecommand \href [0]{\begingroup \@sanitize@url \@href}%
\providecommand \@href[1]{\@@startlink{#1}\@@href}%
\providecommand \@@href[1]{\endgroup#1\@@endlink}%
\providecommand \@sanitize@url [0]{\catcode `\\12\catcode `\$12\catcode
  `\&12\catcode `\#12\catcode `\^12\catcode `\_12\catcode `\%12\relax}%
\providecommand \@@startlink[1]{}%
\providecommand \@@endlink[0]{}%
\providecommand \url  [0]{\begingroup\@sanitize@url \@url }%
\providecommand \@url [1]{\endgroup\@href {#1}{\urlprefix }}%
\providecommand \urlprefix  [0]{URL }%
\providecommand \Eprint [0]{\href }%
\providecommand \doibase [0]{http://dx.doi.org/}%
\providecommand \selectlanguage [0]{\@gobble}%
\providecommand \bibinfo  [0]{\@secondoftwo}%
\providecommand \bibfield  [0]{\@secondoftwo}%
\providecommand \translation [1]{[#1]}%
\providecommand \BibitemOpen [0]{}%
\providecommand \bibitemStop [0]{}%
\providecommand \bibitemNoStop [0]{.\EOS\space}%
\providecommand \EOS [0]{\spacefactor3000\relax}%
\providecommand \BibitemShut  [1]{\csname bibitem#1\endcsname}%
\let\auto@bib@innerbib\@empty
\bibitem [{\citenamefont {Abbott}\ \emph {et~al.}(2016)\citenamefont {Abbott}
  \emph {et~al.}}]{LIGO15}%
  \BibitemOpen
  \bibfield  {author} {\bibinfo {author} {\bibfnamefont {B.~P.}\ \bibnamefont
  {Abbott}} \emph {et~al.} (\bibinfo {collaboration} {Virgo, LIGO
  Scientific}),\ }\href {\doibase 10.1103/PhysRevLett.116.061102} {\bibfield
  {journal} {\bibinfo  {journal} {Phys. Rev. Lett.}\ }\textbf {\bibinfo
  {volume} {116}},\ \bibinfo {pages} {061102} (\bibinfo {year} {2016})},\
  \Eprint {http://arxiv.org/abs/1602.03837}{arXiv:1602.03837
  [gr-qc]}\BibitemShut {NoStop}%
\bibitem [{\citenamefont {Akiyama}\ \emph {et~al.}(2019)\citenamefont {Akiyama}
  \emph {et~al.}}]{EHTM87}%
  \BibitemOpen
  \bibfield  {author} {\bibinfo {author} {\bibfnamefont {K.}~\bibnamefont
  {Akiyama}} \emph {et~al.} (\bibinfo {collaboration} {Event Horizon
  Telescope}),\ }\href {\doibase 10.3847/2041-8213/ab0ec7} {\bibfield
  {journal} {\bibinfo  {journal} {Astrophys. J.}\ }\textbf {\bibinfo {volume}
  {875}},\ \bibinfo {pages} {L1} (\bibinfo {year} {2019})}\BibitemShut
  {NoStop}%
\bibitem [{\citenamefont {Hawking}(1975)}]{Haw75}%
  \BibitemOpen
  \bibfield  {author} {\bibinfo {author} {\bibfnamefont {S.}~\bibnamefont
  {Hawking}},\ }\href {\doibase 10.1007/BF02345020, 10.1007/BF02345020}
  {\bibfield  {journal} {\bibinfo  {journal} {Commun. Math. Phys.}\ }\textbf
  {\bibinfo {volume} {43}},\ \bibinfo {pages} {199} (\bibinfo {year}
  {1975})}\BibitemShut {NoStop}%
\bibitem [{\citenamefont {'t~Hooft}\ \emph {et~al.}(2018)\citenamefont
  {'t~Hooft}, \citenamefont {Giddings}, \citenamefont {Rovelli}, \citenamefont
  {Nicolini}, \citenamefont {Mureika}, \citenamefont {Kaminski},\ and\
  \citenamefont {Bleicher}}]{western}%
  \BibitemOpen
  \bibfield  {author} {\bibinfo {author} {\bibfnamefont {G.}~\bibnamefont
  {'t~Hooft}}, \bibinfo {author} {\bibfnamefont {S.~B.}\ \bibnamefont
  {Giddings}}, \bibinfo {author} {\bibfnamefont {C.}~\bibnamefont {Rovelli}},
  \bibinfo {author} {\bibfnamefont {P.}~\bibnamefont {Nicolini}}, \bibinfo
  {author} {\bibfnamefont {J.}~\bibnamefont {Mureika}}, \bibinfo {author}
  {\bibfnamefont {M.}~\bibnamefont {Kaminski}}, \ and\ \bibinfo {author}
  {\bibfnamefont {M.}~\bibnamefont {Bleicher}},\ }\bibfield  {booktitle} {\emph
  {\bibinfo {booktitle} {{Proceedings, 2nd Karl Schwarzschild Meeting on
  Gravitational Physics (KSM 2015): Frankfurt am Main, Germany, July 20-24,
  2015}}},\ }\href {\doibase 10.1007/978-3-319-94256-8_2} {\bibfield  {journal}
  {\bibinfo  {journal} {Springer Proc. Phys.}\ }\textbf {\bibinfo {volume}
  {208}},\ \bibinfo {pages} {13} (\bibinfo {year} {2018})},\ \Eprint
  {http://arxiv.org/abs/1609.01725}{arXiv:1609.01725 [hep-th]}\BibitemShut
  {NoStop}%
\bibitem [{\citenamefont {Adler}(2010)}]{Adl10}%
  \BibitemOpen
  \bibfield  {author} {\bibinfo {author} {\bibfnamefont {R.~J.}\ \bibnamefont
  {Adler}},\ }\href {\doibase 10.1119/1.3439650} {\bibfield  {journal}
  {\bibinfo  {journal} {Am.J.Phys.}\ }\textbf {\bibinfo {volume} {78}},\
  \bibinfo {pages} {925} (\bibinfo {year} {2010})},\ \Eprint
  {http://arxiv.org/abs/1001.1205}{arXiv:1001.1205 [gr-qc]}\BibitemShut
  {NoStop}%
\bibitem [{\citenamefont {Dvali}\ and\ \citenamefont {Gomez}(2010)}]{DvG10}%
  \BibitemOpen
  \bibfield  {author} {\bibinfo {author} {\bibfnamefont {G.}~\bibnamefont
  {Dvali}}\ and\ \bibinfo {author} {\bibfnamefont {C.}~\bibnamefont {Gomez}},\
  }\href@noop {} {\  (\bibinfo {year} {2010})},\ \Eprint
  {http://arxiv.org/abs/1005.3497}{arXiv:1005.3497 [hep-th]}\BibitemShut
  {NoStop}%
\bibitem [{\citenamefont {Dvali}\ \emph
  {et~al.}(2011{\natexlab{a}})\citenamefont {Dvali}, \citenamefont {Folkerts},\
  and\ \citenamefont {Germani}}]{DFG11}%
  \BibitemOpen
  \bibfield  {author} {\bibinfo {author} {\bibfnamefont {G.}~\bibnamefont
  {Dvali}}, \bibinfo {author} {\bibfnamefont {S.}~\bibnamefont {Folkerts}}, \
  and\ \bibinfo {author} {\bibfnamefont {C.}~\bibnamefont {Germani}},\ }\href
  {\doibase 10.1103/PhysRevD.84.024039} {\bibfield  {journal} {\bibinfo
  {journal} {Phys.Rev.}\ }\textbf {\bibinfo {volume} {D84}},\ \bibinfo {pages}
  {024039} (\bibinfo {year} {2011}{\natexlab{a}})},\ \Eprint
  {http://arxiv.org/abs/1006.0984}{arXiv:1006.0984 [hep-th]}\BibitemShut
  {NoStop}%
\bibitem [{\citenamefont {Dvali}\ \emph
  {et~al.}(2011{\natexlab{b}})\citenamefont {Dvali}, \citenamefont {Giudice},
  \citenamefont {Gomez},\ and\ \citenamefont {Kehagias}}]{DGG11}%
  \BibitemOpen
  \bibfield  {author} {\bibinfo {author} {\bibfnamefont {G.}~\bibnamefont
  {Dvali}}, \bibinfo {author} {\bibfnamefont {G.~F.}\ \bibnamefont {Giudice}},
  \bibinfo {author} {\bibfnamefont {C.}~\bibnamefont {Gomez}}, \ and\ \bibinfo
  {author} {\bibfnamefont {A.}~\bibnamefont {Kehagias}},\ }\href {\doibase
  10.1007/JHEP08(2011)108} {\bibfield  {journal} {\bibinfo  {journal} {JHEP}\
  }\textbf {\bibinfo {volume} {1108}},\ \bibinfo {pages} {108} (\bibinfo {year}
  {2011}{\natexlab{b}})},\ \Eprint
  {http://arxiv.org/abs/1010.1415}{arXiv:1010.1415 [hep-ph]}\BibitemShut
  {NoStop}%
\bibitem [{\citenamefont {Spallucci}\ and\ \citenamefont
  {Ansoldi}(2011)}]{SpA11}%
  \BibitemOpen
  \bibfield  {author} {\bibinfo {author} {\bibfnamefont {E.}~\bibnamefont
  {Spallucci}}\ and\ \bibinfo {author} {\bibfnamefont {S.}~\bibnamefont
  {Ansoldi}},\ }\href@noop {} {\bibfield  {journal} {\bibinfo  {journal} {Phys.
  Lett.}\ }\textbf {\bibinfo {volume} {B701}},\ \bibinfo {pages} {471}
  (\bibinfo {year} {2011})},\ \Eprint
  {http://arxiv.org/abs/1101.2760}{arXiv:1101.2760 [hep-th]}\BibitemShut
  {NoStop}%
\bibitem [{\citenamefont {Dvali}\ \emph {et~al.}(2012)\citenamefont {Dvali},
  \citenamefont {Franca},\ and\ \citenamefont {Gomez}}]{DFG12}%
  \BibitemOpen
  \bibfield  {author} {\bibinfo {author} {\bibfnamefont {G.}~\bibnamefont
  {Dvali}}, \bibinfo {author} {\bibfnamefont {A.}~\bibnamefont {Franca}}, \
  and\ \bibinfo {author} {\bibfnamefont {C.}~\bibnamefont {Gomez}},\
  }\href@noop {} {\  (\bibinfo {year} {2012})},\ \Eprint
  {http://arxiv.org/abs/1204.6388}{arXiv:1204.6388 [hep-th]}\BibitemShut
  {NoStop}%
\bibitem [{\citenamefont {Dvali}\ and\ \citenamefont {Gomez}(2012)}]{DvG12}%
  \BibitemOpen
  \bibfield  {author} {\bibinfo {author} {\bibfnamefont {G.}~\bibnamefont
  {Dvali}}\ and\ \bibinfo {author} {\bibfnamefont {C.}~\bibnamefont {Gomez}},\
  }\href {\doibase 10.1088/1475-7516/2012/07/015} {\bibfield  {journal}
  {\bibinfo  {journal} {JCAP}\ }\textbf {\bibinfo {volume} {1207}},\ \bibinfo
  {pages} {015} (\bibinfo {year} {2012})},\ \Eprint
  {http://arxiv.org/abs/1205.2540}{arXiv:1205.2540 [hep-ph]}\BibitemShut
  {NoStop}%
\bibitem [{\citenamefont {Nicolini}\ and\ \citenamefont
  {Spallucci}(2014)}]{NiS12}%
  \BibitemOpen
  \bibfield  {author} {\bibinfo {author} {\bibfnamefont {P.}~\bibnamefont
  {Nicolini}}\ and\ \bibinfo {author} {\bibfnamefont {E.}~\bibnamefont
  {Spallucci}},\ }\href {\doibase 10.1155/2014/805684} {\bibfield  {journal}
  {\bibinfo  {journal} {Adv. High Energy Phys.}\ }\textbf {\bibinfo {volume}
  {2014}},\ \bibinfo {pages} {805684} (\bibinfo {year} {2014})},\ \Eprint
  {http://arxiv.org/abs/1210.0015}{arXiv:1210.0015 [hep-th]}\BibitemShut
  {NoStop}%
\bibitem [{\citenamefont {Mureika}\ and\ \citenamefont
  {Nicolini}(2013)}]{MuN12}%
  \BibitemOpen
  \bibfield  {author} {\bibinfo {author} {\bibfnamefont {J.}~\bibnamefont
  {Mureika}}\ and\ \bibinfo {author} {\bibfnamefont {P.}~\bibnamefont
  {Nicolini}},\ }\href {\doibase 10.1140/epjp/i2013-13078-0} {\bibfield
  {journal} {\bibinfo  {journal} {Eur.Phys.J.Plus}\ }\textbf {\bibinfo {volume}
  {128}},\ \bibinfo {pages} {78} (\bibinfo {year} {2013})},\ \Eprint
  {http://arxiv.org/abs/1206.4696}{arXiv:1206.4696 [hep-th]}\BibitemShut
  {NoStop}%
\bibitem [{\citenamefont {Aurilia}\ and\ \citenamefont
  {Spallucci}(2013)}]{AuS13}%
  \BibitemOpen
  \bibfield  {author} {\bibinfo {author} {\bibfnamefont {A.}~\bibnamefont
  {Aurilia}}\ and\ \bibinfo {author} {\bibfnamefont {E.}~\bibnamefont
  {Spallucci}},\ }\href@noop {} {\  (\bibinfo {year} {2013})},\ \Eprint
  {http://arxiv.org/abs/1309.7186}{arXiv:1309.7186 [gr-qc]}\BibitemShut
  {NoStop}%
\bibitem [{\citenamefont {Carr}(2016)}]{Car14}%
  \BibitemOpen
  \bibfield  {author} {\bibinfo {author} {\bibfnamefont {B.~J.}\ \bibnamefont
  {Carr}},\ }\bibfield  {booktitle} {\emph {\bibinfo {booktitle} {{Proceedings,
  1st Karl Schwarzschild Meeting on Gravitational Physics (KSM 2013): Frankfurt
  am Main, Germany, July 22-26, 2013}}},\ }\href {\doibase
  10.1007/978-3-319-20046-0_19} {\bibfield  {journal} {\bibinfo  {journal}
  {Springer Proc. Phys.}\ }\textbf {\bibinfo {volume} {170}},\ \bibinfo {pages}
  {159} (\bibinfo {year} {2016})},\ \Eprint
  {http://arxiv.org/abs/1402.1427}{arXiv:1402.1427 [gr-qc]}\BibitemShut
  {NoStop}%
\bibitem [{\citenamefont {Dvali}\ \emph {et~al.}(2015)\citenamefont {Dvali},
  \citenamefont {Gomez}, \citenamefont {Isermann}, \citenamefont {Lüst},\ and\
  \citenamefont {Stieberger}}]{DGI15}%
  \BibitemOpen
  \bibfield  {author} {\bibinfo {author} {\bibfnamefont {G.}~\bibnamefont
  {Dvali}}, \bibinfo {author} {\bibfnamefont {C.}~\bibnamefont {Gomez}},
  \bibinfo {author} {\bibfnamefont {R.~S.}\ \bibnamefont {Isermann}}, \bibinfo
  {author} {\bibfnamefont {D.}~\bibnamefont {Lüst}}, \ and\ \bibinfo {author}
  {\bibfnamefont {S.}~\bibnamefont {Stieberger}},\ }\href {\doibase
  10.1016/j.nuclphysb.2015.02.004} {\bibfield  {journal} {\bibinfo  {journal}
  {Nucl. Phys.}\ }\textbf {\bibinfo {volume} {B893}},\ \bibinfo {pages} {187}
  (\bibinfo {year} {2015})},\ \Eprint
  {http://arxiv.org/abs/1409.7405}{arXiv:1409.7405 [hep-th]}\BibitemShut
  {NoStop}%
\bibitem [{\citenamefont {Carr}\ \emph {et~al.}(2015)\citenamefont {Carr},
  \citenamefont {Mureika},\ and\ \citenamefont {Nicolini}}]{CMN15}%
  \BibitemOpen
  \bibfield  {author} {\bibinfo {author} {\bibfnamefont {B.~J.}\ \bibnamefont
  {Carr}}, \bibinfo {author} {\bibfnamefont {J.}~\bibnamefont {Mureika}}, \
  and\ \bibinfo {author} {\bibfnamefont {P.}~\bibnamefont {Nicolini}},\ }\href
  {\doibase 10.1007/JHEP07(2015)052} {\bibfield  {journal} {\bibinfo  {journal}
  {JHEP}\ }\textbf {\bibinfo {volume} {07}},\ \bibinfo {pages} {052} (\bibinfo
  {year} {2015})},\ \Eprint {http://arxiv.org/abs/1504.07637}{arXiv:1504.07637
  [gr-qc]}\BibitemShut {NoStop}%
\bibitem [{\citenamefont {Dvali}\ \emph {et~al.}(2016)\citenamefont {Dvali},
  \citenamefont {Gomez},\ and\ \citenamefont {Lüst}}]{DGL16}%
  \BibitemOpen
  \bibfield  {author} {\bibinfo {author} {\bibfnamefont {G.}~\bibnamefont
  {Dvali}}, \bibinfo {author} {\bibfnamefont {C.}~\bibnamefont {Gomez}}, \ and\
  \bibinfo {author} {\bibfnamefont {D.}~\bibnamefont {Lüst}},\ }\href
  {\doibase 10.1016/j.physletb.2015.11.073} {\bibfield  {journal} {\bibinfo
  {journal} {Phys. Lett.}\ }\textbf {\bibinfo {volume} {B753}},\ \bibinfo
  {pages} {173} (\bibinfo {year} {2016})},\ \Eprint
  {http://arxiv.org/abs/1509.02114}{arXiv:1509.02114 [hep-th]}\BibitemShut
  {NoStop}%
\bibitem [{\citenamefont {Frassino}\ \emph {et~al.}(2016)\citenamefont
  {Frassino}, \citenamefont {Köppel},\ and\ \citenamefont {Nicolini}}]{FKN16}%
  \BibitemOpen
  \bibfield  {author} {\bibinfo {author} {\bibfnamefont {A.~M.}\ \bibnamefont
  {Frassino}}, \bibinfo {author} {\bibfnamefont {S.}~\bibnamefont {Köppel}}, \
  and\ \bibinfo {author} {\bibfnamefont {P.}~\bibnamefont {Nicolini}},\ }\href
  {\doibase 10.3390/e18050181} {\bibfield  {journal} {\bibinfo  {journal}
  {Entropy}\ }\textbf {\bibinfo {volume} {18}},\ \bibinfo {pages} {181}
  (\bibinfo {year} {2016})},\ \Eprint
  {http://arxiv.org/abs/1604.03263}{arXiv:1604.03263 [gr-qc]}\BibitemShut
  {NoStop}%
\bibitem [{\citenamefont {Dvali}(2017)}]{Dva17}%
  \BibitemOpen
  \bibfield  {author} {\bibinfo {author} {\bibfnamefont {G.}~\bibnamefont
  {Dvali}},\ }\bibfield  {booktitle} {\emph {\bibinfo {booktitle}
  {{Proceedings, 53rd International School of Subnuclear Physics: The Future of
  our Physics Including New Frontiers (ISSP 2015): Erice, Italy, June 24-July
  3, 2015}}},\ }\href {\doibase 10.1142/9789813208292_0005} {\bibfield
  {journal} {\bibinfo  {journal} {Subnucl. Ser.}\ }\textbf {\bibinfo {volume}
  {53}},\ \bibinfo {pages} {189} (\bibinfo {year} {2017})},\ \Eprint
  {http://arxiv.org/abs/1607.07422}{arXiv:1607.07422 [hep-th]}\BibitemShut
  {NoStop}%
\bibitem [{\citenamefont {Nicolini}(2018)}]{Nic2018}%
  \BibitemOpen
  \bibfield  {author} {\bibinfo {author} {\bibfnamefont {P.}~\bibnamefont
  {Nicolini}},\ }\href {\doibase 10.1016/j.physletb.2018.01.013} {\bibfield
  {journal} {\bibinfo  {journal} {Phys. Lett.}\ }\textbf {\bibinfo {volume}
  {B778}},\ \bibinfo {pages} {88} (\bibinfo {year} {2018})},\ \Eprint
  {http://arxiv.org/abs/1712.05062}{arXiv:1712.05062 [gr-qc]}\BibitemShut
  {NoStop}%
\bibitem [{\citenamefont {Casadio}\ \emph {et~al.}(2018)\citenamefont
  {Casadio}, \citenamefont {Nicolini},\ and\ \citenamefont
  {da~Rocha}}]{CDRN18}%
  \BibitemOpen
  \bibfield  {author} {\bibinfo {author} {\bibfnamefont {R.}~\bibnamefont
  {Casadio}}, \bibinfo {author} {\bibfnamefont {P.}~\bibnamefont {Nicolini}}, \
  and\ \bibinfo {author} {\bibfnamefont {R.}~\bibnamefont {da~Rocha}},\ }\href
  {\doibase 10.1088/1361-6382/aad664} {\bibfield  {journal} {\bibinfo
  {journal} {Class. Quant. Grav.}\ }\textbf {\bibinfo {volume} {35}},\ \bibinfo
  {pages} {185001} (\bibinfo {year} {2018})},\ \Eprint
  {http://arxiv.org/abs/1709.09704}{arXiv:1709.09704 [hep-th]}\BibitemShut
  {NoStop}%
\bibitem [{\citenamefont {Veneziano}(1986)}]{Ven86}%
  \BibitemOpen
  \bibfield  {author} {\bibinfo {author} {\bibfnamefont {G.}~\bibnamefont
  {Veneziano}},\ }\href {\doibase 10.1209/0295-5075/2/3/006} {\bibfield
  {journal} {\bibinfo  {journal} {Europhys.Lett.}\ }\textbf {\bibinfo {volume}
  {2}},\ \bibinfo {pages} {199} (\bibinfo {year} {1986})}\BibitemShut {NoStop}%
\bibitem [{\citenamefont {Amati}\ \emph {et~al.}(1989)\citenamefont {Amati},
  \citenamefont {Ciafaloni},\ and\ \citenamefont {Veneziano}}]{ACV89}%
  \BibitemOpen
  \bibfield  {author} {\bibinfo {author} {\bibfnamefont {D.}~\bibnamefont
  {Amati}}, \bibinfo {author} {\bibfnamefont {M.}~\bibnamefont {Ciafaloni}}, \
  and\ \bibinfo {author} {\bibfnamefont {G.}~\bibnamefont {Veneziano}},\ }\href
  {\doibase 10.1016/0370-2693(89)91366-X} {\bibfield  {journal} {\bibinfo
  {journal} {Phys.Lett.}\ }\textbf {\bibinfo {volume} {B216}},\ \bibinfo
  {pages} {41} (\bibinfo {year} {1989})}\BibitemShut {NoStop}%
\bibitem [{\citenamefont {Amati}\ \emph {et~al.}(1993)\citenamefont {Amati},
  \citenamefont {Ciafaloni},\ and\ \citenamefont {Veneziano}}]{ACV93}%
  \BibitemOpen
  \bibfield  {author} {\bibinfo {author} {\bibfnamefont {D.}~\bibnamefont
  {Amati}}, \bibinfo {author} {\bibfnamefont {M.}~\bibnamefont {Ciafaloni}}, \
  and\ \bibinfo {author} {\bibfnamefont {G.}~\bibnamefont {Veneziano}},\ }\href
  {\doibase 10.1016/0550-3213(93)90367-X} {\bibfield  {journal} {\bibinfo
  {journal} {Nucl. Phys.}\ }\textbf {\bibinfo {volume} {B403}},\ \bibinfo
  {pages} {707} (\bibinfo {year} {1993})}\BibitemShut {NoStop}%
\bibitem [{\citenamefont {Maggiore}(1993)}]{Mag93}%
  \BibitemOpen
  \bibfield  {author} {\bibinfo {author} {\bibfnamefont {M.}~\bibnamefont
  {Maggiore}},\ }\href {\doibase 10.1016/0370-2693(93)91401-8} {\bibfield
  {journal} {\bibinfo  {journal} {Phys. Lett.}\ }\textbf {\bibinfo {volume}
  {B304}},\ \bibinfo {pages} {65} (\bibinfo {year} {1993})},\ \Eprint
  {http://arxiv.org/abs/hep-th/9301067}{arXiv:hep-th/9301067
  [hep-th]}\BibitemShut {NoStop}%
\bibitem [{\citenamefont {Kempf}\ \emph {et~al.}(1995)\citenamefont {Kempf},
  \citenamefont {Mangano},\ and\ \citenamefont {Mann}}]{KMM95}%
  \BibitemOpen
  \bibfield  {author} {\bibinfo {author} {\bibfnamefont {A.}~\bibnamefont
  {Kempf}}, \bibinfo {author} {\bibfnamefont {G.}~\bibnamefont {Mangano}}, \
  and\ \bibinfo {author} {\bibfnamefont {R.~B.}\ \bibnamefont {Mann}},\ }\href
  {\doibase 10.1103/PhysRevD.52.1108} {\bibfield  {journal} {\bibinfo
  {journal} {Phys.Rev.}\ }\textbf {\bibinfo {volume} {D52}},\ \bibinfo {pages}
  {1108} (\bibinfo {year} {1995})},\ \Eprint
  {http://arxiv.org/abs/hep-th/9412167}{arXiv:hep-th/9412167
  [hep-th]}\BibitemShut {NoStop}%
\bibitem [{\citenamefont {Sprenger}\ \emph {et~al.}(2012)\citenamefont
  {Sprenger}, \citenamefont {Nicolini},\ and\ \citenamefont
  {Bleicher}}]{SNB12}%
  \BibitemOpen
  \bibfield  {author} {\bibinfo {author} {\bibfnamefont {M.}~\bibnamefont
  {Sprenger}}, \bibinfo {author} {\bibfnamefont {P.}~\bibnamefont {Nicolini}},
  \ and\ \bibinfo {author} {\bibfnamefont {M.}~\bibnamefont {Bleicher}},\
  }\href@noop {} {\bibfield  {journal} {\bibinfo  {journal} {Eur.J.Phys.}\
  }\textbf {\bibinfo {volume} {33}},\ \bibinfo {pages} {853} (\bibinfo {year}
  {2012})},\ \Eprint {http://arxiv.org/abs/1202.1500}{arXiv:1202.1500
  [physics.ed-ph]}\BibitemShut {NoStop}%
\bibitem [{\citenamefont {Hossenfelder}(2013)}]{Hos13}%
  \BibitemOpen
  \bibfield  {author} {\bibinfo {author} {\bibfnamefont {S.}~\bibnamefont
  {Hossenfelder}},\ }\href {\doibase 10.12942/lrr-2013-2} {\bibfield  {journal}
  {\bibinfo  {journal} {Living Rev. Rel.}\ }\textbf {\bibinfo {volume} {16}},\
  \bibinfo {pages} {2} (\bibinfo {year} {2013})},\ \Eprint
  {http://arxiv.org/abs/1203.6191}{arXiv:1203.6191 [gr-qc]}\BibitemShut
  {NoStop}%
\bibitem [{\citenamefont {Tawfik}\ and\ \citenamefont {Diab}(2015)}]{TaM15}%
  \BibitemOpen
  \bibfield  {author} {\bibinfo {author} {\bibfnamefont {A.~N.}\ \bibnamefont
  {Tawfik}}\ and\ \bibinfo {author} {\bibfnamefont {A.~M.}\ \bibnamefont
  {Diab}},\ }\href {\doibase 10.1088/0034-4885/78/12/126001} {\bibfield
  {journal} {\bibinfo  {journal} {Rept. Prog. Phys.}\ }\textbf {\bibinfo
  {volume} {78}},\ \bibinfo {pages} {126001} (\bibinfo {year} {2015})},\
  \Eprint {http://arxiv.org/abs/1509.02436}{arXiv:1509.02436
  [physics.gen-ph]}\BibitemShut {NoStop}%
\bibitem [{\citenamefont {Adler}\ and\ \citenamefont {Santiago}(1999)}]{AdS99}%
  \BibitemOpen
  \bibfield  {author} {\bibinfo {author} {\bibfnamefont {R.~J.}\ \bibnamefont
  {Adler}}\ and\ \bibinfo {author} {\bibfnamefont {D.~I.}\ \bibnamefont
  {Santiago}},\ }\href {\doibase 10.1142/S0217732399001462} {\bibfield
  {journal} {\bibinfo  {journal} {Mod. Phys. Lett.}\ }\textbf {\bibinfo
  {volume} {A14}},\ \bibinfo {pages} {1371} (\bibinfo {year} {1999})},\ \Eprint
  {http://arxiv.org/abs/gr-qc/9904026}{arXiv:gr-qc/9904026 [gr-qc]}\BibitemShut
  {NoStop}%
\bibitem [{\citenamefont {Adler}\ \emph {et~al.}(2001)\citenamefont {Adler},
  \citenamefont {Chen},\ and\ \citenamefont {Santiago}}]{APS01}%
  \BibitemOpen
  \bibfield  {author} {\bibinfo {author} {\bibfnamefont {R.~J.}\ \bibnamefont
  {Adler}}, \bibinfo {author} {\bibfnamefont {P.}~\bibnamefont {Chen}}, \ and\
  \bibinfo {author} {\bibfnamefont {D.~I.}\ \bibnamefont {Santiago}},\ }\href
  {\doibase 10.1023/A:1015281430411} {\bibfield  {journal} {\bibinfo  {journal}
  {Gen.Rel.Grav.}\ }\textbf {\bibinfo {volume} {33}},\ \bibinfo {pages} {2101}
  (\bibinfo {year} {2001})},\ \Eprint
  {http://arxiv.org/abs/gr-qc/0106080}{arXiv:gr-qc/0106080 [gr-qc]}\BibitemShut
  {NoStop}%
\bibitem [{\citenamefont {Chen}\ and\ \citenamefont {Adler}(2003)}]{ChA03}%
  \BibitemOpen
  \bibfield  {author} {\bibinfo {author} {\bibfnamefont {P.}~\bibnamefont
  {Chen}}\ and\ \bibinfo {author} {\bibfnamefont {R.~J.}\ \bibnamefont
  {Adler}},\ }\href {\doibase 10.1016/S0920-5632(03)02088-7} {\bibfield
  {journal} {\bibinfo  {journal} {Nucl.Phys.Proc.Suppl.}\ }\textbf {\bibinfo
  {volume} {124}},\ \bibinfo {pages} {103} (\bibinfo {year} {2003})},\ \Eprint
  {http://arxiv.org/abs/gr-qc/0205106}{arXiv:gr-qc/0205106 [gr-qc]}\BibitemShut
  {NoStop}%
\bibitem [{\citenamefont {Isi}\ \emph {et~al.}(2013)\citenamefont {Isi},
  \citenamefont {Mureika},\ and\ \citenamefont {Nicolini}}]{IMN13}%
  \BibitemOpen
  \bibfield  {author} {\bibinfo {author} {\bibfnamefont {M.}~\bibnamefont
  {Isi}}, \bibinfo {author} {\bibfnamefont {J.}~\bibnamefont {Mureika}}, \ and\
  \bibinfo {author} {\bibfnamefont {P.}~\bibnamefont {Nicolini}},\ }\href
  {\doibase 10.1007/JHEP11(2013)139} {\bibfield  {journal} {\bibinfo  {journal}
  {JHEP}\ }\textbf {\bibinfo {volume} {1311}},\ \bibinfo {pages} {139}
  (\bibinfo {year} {2013})},\ \Eprint
  {http://arxiv.org/abs/1310.8153}{arXiv:1310.8153 [hep-th]}\BibitemShut
  {NoStop}%
\bibitem [{\citenamefont {Krasnikov}(1987)}]{Kra87}%
  \BibitemOpen
  \bibfield  {author} {\bibinfo {author} {\bibfnamefont {N.}~\bibnamefont
  {Krasnikov}},\ }\href {\doibase 10.1007/BF01017588} {\bibfield  {journal}
  {\bibinfo  {journal} {Theor.Math.Phys.}\ }\textbf {\bibinfo {volume} {73}},\
  \bibinfo {pages} {1184} (\bibinfo {year} {1987})}\BibitemShut {NoStop}%
\bibitem [{\citenamefont {Tomboulis}(1997)}]{Tom97}%
  \BibitemOpen
  \bibfield  {author} {\bibinfo {author} {\bibfnamefont {E.~T.}\ \bibnamefont
  {Tomboulis}},\ }\href@noop {} {\  (\bibinfo {year} {1997})},\ \Eprint
  {http://arxiv.org/abs/hep-th/9702146}{arXiv:hep-th/9702146
  [hep-th]}\BibitemShut {NoStop}%
\bibitem [{\citenamefont {Barvinsky}(2003)}]{Bar03}%
  \BibitemOpen
  \bibfield  {author} {\bibinfo {author} {\bibfnamefont {A.~O.}\ \bibnamefont
  {Barvinsky}},\ }\href {\doibase 10.1016/j.physletb.2003.08.055} {\bibfield
  {journal} {\bibinfo  {journal} {Phys. Lett.}\ }\textbf {\bibinfo {volume}
  {B572}},\ \bibinfo {pages} {109} (\bibinfo {year} {2003})},\ \Eprint
  {http://arxiv.org/abs/hep-th/0304229}{arXiv:hep-th/0304229
  [hep-th]}\BibitemShut {NoStop}%
\bibitem [{\citenamefont {Modesto}(2012)}]{Mod12a}%
  \BibitemOpen
  \bibfield  {author} {\bibinfo {author} {\bibfnamefont {L.}~\bibnamefont
  {Modesto}},\ }\href {\doibase 10.1103/PhysRevD.86.044005} {\bibfield
  {journal} {\bibinfo  {journal} {Phys.Rev.}\ }\textbf {\bibinfo {volume}
  {D86}},\ \bibinfo {pages} {044005} (\bibinfo {year} {2012})},\ \Eprint
  {http://arxiv.org/abs/1107.2403}{arXiv:1107.2403 [hep-th]}\BibitemShut
  {NoStop}%
\bibitem [{\citenamefont {Arkani-Hamed}\ \emph {et~al.}(2002)\citenamefont
  {Arkani-Hamed}, \citenamefont {Dimopoulos}, \citenamefont {Dvali},\ and\
  \citenamefont {Gabadadze}}]{ADD02}%
  \BibitemOpen
  \bibfield  {author} {\bibinfo {author} {\bibfnamefont {N.}~\bibnamefont
  {Arkani-Hamed}}, \bibinfo {author} {\bibfnamefont {S.}~\bibnamefont
  {Dimopoulos}}, \bibinfo {author} {\bibfnamefont {G.}~\bibnamefont {Dvali}}, \
  and\ \bibinfo {author} {\bibfnamefont {G.}~\bibnamefont {Gabadadze}},\
  }\href@noop {} {\  (\bibinfo {year} {2002})},\ \Eprint
  {http://arxiv.org/abs/hep-th/0209227}{arXiv:hep-th/0209227}\BibitemShut
  {NoStop}%
\bibitem [{\citenamefont {Dvali}\ \emph {et~al.}(2007)\citenamefont {Dvali},
  \citenamefont {Hofmann},\ and\ \citenamefont {Khoury}}]{DHK07}%
  \BibitemOpen
  \bibfield  {author} {\bibinfo {author} {\bibfnamefont {G.}~\bibnamefont
  {Dvali}}, \bibinfo {author} {\bibfnamefont {S.}~\bibnamefont {Hofmann}}, \
  and\ \bibinfo {author} {\bibfnamefont {J.}~\bibnamefont {Khoury}},\ }\href
  {\doibase 10.1103/PhysRevD.76.084006} {\bibfield  {journal} {\bibinfo
  {journal} {Phys. Rev.}\ }\textbf {\bibinfo {volume} {D76}},\ \bibinfo {pages}
  {084006} (\bibinfo {year} {2007})},\ \Eprint
  {http://arxiv.org/abs/hep-th/0703027}{arXiv:hep-th/0703027
  [HEP-TH]}\BibitemShut {NoStop}%
\bibitem [{\citenamefont {Barvinsky}(2005)}]{Bar05}%
  \BibitemOpen
  \bibfield  {author} {\bibinfo {author} {\bibfnamefont {A.~O.}\ \bibnamefont
  {Barvinsky}},\ }\href {\doibase 10.1103/PhysRevD.71.084007} {\bibfield
  {journal} {\bibinfo  {journal} {Phys. Rev.}\ }\textbf {\bibinfo {volume}
  {D71}},\ \bibinfo {pages} {084007} (\bibinfo {year} {2005})},\ \Eprint
  {http://arxiv.org/abs/hep-th/0501093}{arXiv:hep-th/0501093
  [hep-th]}\BibitemShut {NoStop}%
\bibitem [{\citenamefont {Barvinsky}(2012)}]{Bar12}%
  \BibitemOpen
  \bibfield  {author} {\bibinfo {author} {\bibfnamefont {A.~O.}\ \bibnamefont
  {Barvinsky}},\ }\href {\doibase 10.1016/j.physletb.2012.02.075} {\bibfield
  {journal} {\bibinfo  {journal} {Phys. Lett.}\ }\textbf {\bibinfo {volume}
  {B710}},\ \bibinfo {pages} {12} (\bibinfo {year} {2012})},\ \Eprint
  {http://arxiv.org/abs/1107.1463}{arXiv:1107.1463 [hep-th]}\BibitemShut
  {NoStop}%
\bibitem [{\citenamefont {Gaete}\ \emph {et~al.}(2010)\citenamefont {Gaete},
  \citenamefont {Helayel-Neto},\ and\ \citenamefont {Spallucci}}]{GHS10}%
  \BibitemOpen
  \bibfield  {author} {\bibinfo {author} {\bibfnamefont {P.}~\bibnamefont
  {Gaete}}, \bibinfo {author} {\bibfnamefont {J.~A.}\ \bibnamefont
  {Helayel-Neto}}, \ and\ \bibinfo {author} {\bibfnamefont {E.}~\bibnamefont
  {Spallucci}},\ }\href {\doibase 10.1016/j.physletb.2010.07.058} {\bibfield
  {journal} {\bibinfo  {journal} {Phys. Lett.}\ }\textbf {\bibinfo {volume}
  {B693}},\ \bibinfo {pages} {155} (\bibinfo {year} {2010})},\ \Eprint
  {http://arxiv.org/abs/1005.0234}{arXiv:1005.0234 [hep-ph]}\BibitemShut
  {NoStop}%
\bibitem [{\citenamefont {Modesto}\ \emph {et~al.}(2011)\citenamefont
  {Modesto}, \citenamefont {Moffat},\ and\ \citenamefont {Nicolini}}]{MMN11}%
  \BibitemOpen
  \bibfield  {author} {\bibinfo {author} {\bibfnamefont {L.}~\bibnamefont
  {Modesto}}, \bibinfo {author} {\bibfnamefont {J.~W.}\ \bibnamefont {Moffat}},
  \ and\ \bibinfo {author} {\bibfnamefont {P.}~\bibnamefont {Nicolini}},\
  }\href {\doibase 10.1016/j.physletb.2010.11.046} {\bibfield  {journal}
  {\bibinfo  {journal} {Phys. Lett.}\ }\textbf {\bibinfo {volume} {B695}},\
  \bibinfo {pages} {397} (\bibinfo {year} {2011})},\ \Eprint
  {http://arxiv.org/abs/1010.0680}{arXiv:1010.0680 [gr-qc]}\BibitemShut
  {NoStop}%
\bibitem [{\citenamefont {Nicolini}()}]{Nic12}%
  \BibitemOpen
  \bibfield  {author} {\bibinfo {author} {\bibfnamefont {P.}~\bibnamefont
  {Nicolini}},\ }\href@noop {} {\ }\Eprint
  {http://arxiv.org/abs/1202.2102}{arXiv:1202.2102 [hep-th]}\BibitemShut
  {NoStop}%
\bibitem [{\citenamefont {Calcagni}\ \emph {et~al.}(2014)\citenamefont
  {Calcagni}, \citenamefont {Modesto},\ and\ \citenamefont {Nicolini}}]{CMN14}%
  \BibitemOpen
  \bibfield  {author} {\bibinfo {author} {\bibfnamefont {G.}~\bibnamefont
  {Calcagni}}, \bibinfo {author} {\bibfnamefont {L.}~\bibnamefont {Modesto}}, \
  and\ \bibinfo {author} {\bibfnamefont {P.}~\bibnamefont {Nicolini}},\ }\href
  {\doibase 10.1140/epjc/s10052-014-2999-8} {\bibfield  {journal} {\bibinfo
  {journal} {Eur. Phys. J.}\ }\textbf {\bibinfo {volume} {C74}},\ \bibinfo
  {pages} {2999} (\bibinfo {year} {2014})},\ \Eprint
  {http://arxiv.org/abs/1306.5332}{arXiv:1306.5332 [gr-qc]}\BibitemShut
  {NoStop}%
\bibitem [{\citenamefont {Nicolini}(2009)}]{Nic09}%
  \BibitemOpen
  \bibfield  {author} {\bibinfo {author} {\bibfnamefont {P.}~\bibnamefont
  {Nicolini}},\ }\href {\doibase 10.1142/S0217751X09043353} {\bibfield
  {journal} {\bibinfo  {journal} {Int. J. Mod. Phys.}\ }\textbf {\bibinfo
  {volume} {A24}},\ \bibinfo {pages} {1229} (\bibinfo {year} {2009})},\ \Eprint
  {http://arxiv.org/abs/0807.1939}{arXiv:0807.1939 [hep-th]}\BibitemShut
  {NoStop}%
\bibitem [{\citenamefont {Nicolini}\ and\ \citenamefont
  {Winstanley}(2011)}]{NiW11}%
  \BibitemOpen
  \bibfield  {author} {\bibinfo {author} {\bibfnamefont {P.}~\bibnamefont
  {Nicolini}}\ and\ \bibinfo {author} {\bibfnamefont {E.}~\bibnamefont
  {Winstanley}},\ }\href {\doibase 10.1007/JHEP11(2011)075} {\bibfield
  {journal} {\bibinfo  {journal} {JHEP}\ }\textbf {\bibinfo {volume} {11}},\
  \bibinfo {pages} {075} (\bibinfo {year} {2011})},\ \Eprint
  {http://arxiv.org/abs/1108.4419}{arXiv:1108.4419 [hep-ph]}\BibitemShut
  {NoStop}%
\bibitem [{\citenamefont {Banks}\ and\ \citenamefont {Fischler}(1999)}]{BaF99}%
  \BibitemOpen
  \bibfield  {author} {\bibinfo {author} {\bibfnamefont {T.}~\bibnamefont
  {Banks}}\ and\ \bibinfo {author} {\bibfnamefont {W.}~\bibnamefont
  {Fischler}},\ }\href@noop {} {\  (\bibinfo {year} {1999})},\ \Eprint
  {http://arxiv.org/abs/hep-th/9906038}{arXiv:hep-th/9906038
  [hep-th]}\BibitemShut {NoStop}%
\bibitem [{\citenamefont {Dimopoulos}\ and\ \citenamefont
  {Landsberg}(2001)}]{DiL01}%
  \BibitemOpen
  \bibfield  {author} {\bibinfo {author} {\bibfnamefont {S.}~\bibnamefont
  {Dimopoulos}}\ and\ \bibinfo {author} {\bibfnamefont {G.~L.}\ \bibnamefont
  {Landsberg}},\ }\href {\doibase 10.1103/PhysRevLett.87.161602} {\bibfield
  {journal} {\bibinfo  {journal} {Phys.Rev.Lett.}\ }\textbf {\bibinfo {volume}
  {87}},\ \bibinfo {pages} {161602} (\bibinfo {year} {2001})},\ \Eprint
  {http://arxiv.org/abs/hep-ph/0106295}{arXiv:hep-ph/0106295
  [hep-ph]}\BibitemShut {NoStop}%
\bibitem [{\citenamefont {Giddings}\ and\ \citenamefont
  {Thomas}(2002)}]{GiT02}%
  \BibitemOpen
  \bibfield  {author} {\bibinfo {author} {\bibfnamefont {S.~B.}\ \bibnamefont
  {Giddings}}\ and\ \bibinfo {author} {\bibfnamefont {S.~D.}\ \bibnamefont
  {Thomas}},\ }\href {\doibase 10.1103/PhysRevD.65.056010} {\bibfield
  {journal} {\bibinfo  {journal} {Phys.Rev.}\ }\textbf {\bibinfo {volume}
  {D65}},\ \bibinfo {pages} {056010} (\bibinfo {year} {2002})},\ \Eprint
  {http://arxiv.org/abs/hep-ph/0106219}{arXiv:hep-ph/0106219
  [hep-ph]}\BibitemShut {NoStop}%
\bibitem [{\citenamefont {Antoniadis}\ \emph {et~al.}(1998)\citenamefont
  {Antoniadis}, \citenamefont {Arkani-Hamed}, \citenamefont {Dimopoulos},\ and\
  \citenamefont {Dvali}}]{AAD98}%
  \BibitemOpen
  \bibfield  {author} {\bibinfo {author} {\bibfnamefont {I.}~\bibnamefont
  {Antoniadis}}, \bibinfo {author} {\bibfnamefont {N.}~\bibnamefont
  {Arkani-Hamed}}, \bibinfo {author} {\bibfnamefont {S.}~\bibnamefont
  {Dimopoulos}}, \ and\ \bibinfo {author} {\bibfnamefont {G.}~\bibnamefont
  {Dvali}},\ }\href {\doibase 10.1016/S0370-2693(98)00860-0} {\bibfield
  {journal} {\bibinfo  {journal} {Phys.Lett.}\ }\textbf {\bibinfo {volume}
  {B436}},\ \bibinfo {pages} {257} (\bibinfo {year} {1998})},\ \Eprint
  {http://arxiv.org/abs/hep-ph/9804398}{arXiv:hep-ph/9804398
  [hep-ph]}\BibitemShut {NoStop}%
\bibitem [{\citenamefont {Arkani-Hamed}\ \emph {et~al.}(1998)\citenamefont
  {Arkani-Hamed}, \citenamefont {Dimopoulos},\ and\ \citenamefont
  {Dvali}}]{ADD98}%
  \BibitemOpen
  \bibfield  {author} {\bibinfo {author} {\bibfnamefont {N.}~\bibnamefont
  {Arkani-Hamed}}, \bibinfo {author} {\bibfnamefont {S.}~\bibnamefont
  {Dimopoulos}}, \ and\ \bibinfo {author} {\bibfnamefont {G.}~\bibnamefont
  {Dvali}},\ }\href {\doibase 10.1016/S0370-2693(98)00466-3} {\bibfield
  {journal} {\bibinfo  {journal} {Phys.Lett.}\ }\textbf {\bibinfo {volume}
  {B429}},\ \bibinfo {pages} {263} (\bibinfo {year} {1998})},\ \Eprint
  {http://arxiv.org/abs/hep-ph/9803315}{arXiv:hep-ph/9803315
  [hep-ph]}\BibitemShut {NoStop}%
\bibitem [{\citenamefont {Arkani-Hamed}\ \emph {et~al.}(1999)\citenamefont
  {Arkani-Hamed}, \citenamefont {Dimopoulos},\ and\ \citenamefont
  {Dvali}}]{ADD99}%
  \BibitemOpen
  \bibfield  {author} {\bibinfo {author} {\bibfnamefont {N.}~\bibnamefont
  {Arkani-Hamed}}, \bibinfo {author} {\bibfnamefont {S.}~\bibnamefont
  {Dimopoulos}}, \ and\ \bibinfo {author} {\bibfnamefont {G.}~\bibnamefont
  {Dvali}},\ }\href {\doibase 10.1103/PhysRevD.59.086004} {\bibfield  {journal}
  {\bibinfo  {journal} {Phys.Rev.}\ }\textbf {\bibinfo {volume} {D59}},\
  \bibinfo {pages} {086004} (\bibinfo {year} {1999})},\ \Eprint
  {http://arxiv.org/abs/hep-ph/9807344}{arXiv:hep-ph/9807344
  [hep-ph]}\BibitemShut {NoStop}%
\bibitem [{\citenamefont {Randall}\ and\ \citenamefont
  {Sundrum}(1999{\natexlab{a}})}]{RaS99a}%
  \BibitemOpen
  \bibfield  {author} {\bibinfo {author} {\bibfnamefont {L.}~\bibnamefont
  {Randall}}\ and\ \bibinfo {author} {\bibfnamefont {R.}~\bibnamefont
  {Sundrum}},\ }\href {\doibase 10.1103/PhysRevLett.83.3370} {\bibfield
  {journal} {\bibinfo  {journal} {Phys.Rev.Lett.}\ }\textbf {\bibinfo {volume}
  {83}},\ \bibinfo {pages} {3370} (\bibinfo {year} {1999}{\natexlab{a}})},\
  \Eprint {http://arxiv.org/abs/hep-ph/9905221}{arXiv:hep-ph/9905221
  [hep-ph]}\BibitemShut {NoStop}%
\bibitem [{\citenamefont {Randall}\ and\ \citenamefont
  {Sundrum}(1999{\natexlab{b}})}]{RaS99b}%
  \BibitemOpen
  \bibfield  {author} {\bibinfo {author} {\bibfnamefont {L.}~\bibnamefont
  {Randall}}\ and\ \bibinfo {author} {\bibfnamefont {R.}~\bibnamefont
  {Sundrum}},\ }\href {\doibase 10.1103/PhysRevLett.83.4690} {\bibfield
  {journal} {\bibinfo  {journal} {Phys.Rev.Lett.}\ }\textbf {\bibinfo {volume}
  {83}},\ \bibinfo {pages} {4690} (\bibinfo {year} {1999}{\natexlab{b}})},\
  \Eprint {http://arxiv.org/abs/hep-th/9906064}{arXiv:hep-th/9906064
  [hep-th]}\BibitemShut {NoStop}%
\bibitem [{\citenamefont {Appelquist}\ \emph {et~al.}(2001)\citenamefont
  {Appelquist}, \citenamefont {Cheng},\ and\ \citenamefont {Dobrescu}}]{ACD01}%
  \BibitemOpen
  \bibfield  {author} {\bibinfo {author} {\bibfnamefont {T.}~\bibnamefont
  {Appelquist}}, \bibinfo {author} {\bibfnamefont {H.-C.}\ \bibnamefont
  {Cheng}}, \ and\ \bibinfo {author} {\bibfnamefont {B.~A.}\ \bibnamefont
  {Dobrescu}},\ }\href {\doibase 10.1103/PhysRevD.64.035002} {\bibfield
  {journal} {\bibinfo  {journal} {Phys. Rev.}\ }\textbf {\bibinfo {volume}
  {D64}},\ \bibinfo {pages} {035002} (\bibinfo {year} {2001})},\ \Eprint
  {http://arxiv.org/abs/hep-ph/0012100}{arXiv:hep-ph/0012100
  [hep-ph]}\BibitemShut {NoStop}%
\bibitem [{\citenamefont {Landsberg}(2002)}]{Lan02}%
  \BibitemOpen
  \bibfield  {author} {\bibinfo {author} {\bibfnamefont {G.~L.}\ \bibnamefont
  {Landsberg}},\ }in\ \href
  {http://www-library.desy.de/preparch/desy/proc/proc02-02/Proceedings/pl.7/landsberg_pr.pdf}
  {\emph {\bibinfo {booktitle} {{Supersymmetry and unification of fundamental
  interactions. Proceedings, 10th International Conference, SUSY'02, Hamburg,
  Germany, June 17-23, 2002}}}}\ (\bibinfo {year} {2002})\ pp.\ \bibinfo
  {pages} {562--577},\ \Eprint
  {http://arxiv.org/abs/hep-ph/0211043}{arXiv:hep-ph/0211043
  [hep-ph]}\BibitemShut {NoStop}%
\bibitem [{\citenamefont {Cavaglia}(2003)}]{Cav03}%
  \BibitemOpen
  \bibfield  {author} {\bibinfo {author} {\bibfnamefont {M.}~\bibnamefont
  {Cavaglia}},\ }\href {\doibase 10.1142/S0217751X03013569} {\bibfield
  {journal} {\bibinfo  {journal} {Int.J.Mod.Phys.}\ }\textbf {\bibinfo {volume}
  {A18}},\ \bibinfo {pages} {1843} (\bibinfo {year} {2003})},\ \Eprint
  {http://arxiv.org/abs/hep-ph/0210296}{arXiv:hep-ph/0210296
  [hep-ph]}\BibitemShut {NoStop}%
\bibitem [{\citenamefont {Kanti}(2004)}]{Kan04}%
  \BibitemOpen
  \bibfield  {author} {\bibinfo {author} {\bibfnamefont {P.}~\bibnamefont
  {Kanti}},\ }\href {\doibase 10.1142/S0217751X04018324} {\bibfield  {journal}
  {\bibinfo  {journal} {Int.J.Mod.Phys.}\ }\textbf {\bibinfo {volume} {A19}},\
  \bibinfo {pages} {4899} (\bibinfo {year} {2004})},\ \Eprint
  {http://arxiv.org/abs/hep-ph/0402168}{arXiv:hep-ph/0402168
  [hep-ph]}\BibitemShut {NoStop}%
\bibitem [{\citenamefont {Hossenfelder}(2004)}]{Hos04}%
  \BibitemOpen
  \bibfield  {author} {\bibinfo {author} {\bibfnamefont {S.}~\bibnamefont
  {Hossenfelder}},\ }\href@noop {} {\  (\bibinfo {year} {2004})},\ \Eprint
  {http://arxiv.org/abs/hep-ph/0412265}{arXiv:hep-ph/0412265
  [hep-ph]}\BibitemShut {NoStop}%
\bibitem [{\citenamefont {Casanova}\ and\ \citenamefont
  {Spallucci}(2006)}]{CaS06}%
  \BibitemOpen
  \bibfield  {author} {\bibinfo {author} {\bibfnamefont {A.}~\bibnamefont
  {Casanova}}\ and\ \bibinfo {author} {\bibfnamefont {E.}~\bibnamefont
  {Spallucci}},\ }\href {\doibase 10.1088/0264-9381/23/3/R01} {\bibfield
  {journal} {\bibinfo  {journal} {Class.Quant.Grav.}\ }\textbf {\bibinfo
  {volume} {23}},\ \bibinfo {pages} {R45} (\bibinfo {year} {2006})},\ \Eprint
  {http://arxiv.org/abs/hep-ph/0512063}{arXiv:hep-ph/0512063
  [hep-ph]}\BibitemShut {NoStop}%
\bibitem [{\citenamefont {Winstanley}(2007)}]{Win07}%
  \BibitemOpen
  \bibfield  {author} {\bibinfo {author} {\bibfnamefont {E.}~\bibnamefont
  {Winstanley}},\ }in\ \href
  {http://inspirehep.net/record/758640/files/arXiv:0708.2656.pdf} {\emph
  {\bibinfo {booktitle} {{Conference on Black Holes and Naked Singularities
  Milan, Italy, May 10-12, 2007}}}}\ (\bibinfo {year} {2007})\ \Eprint
  {http://arxiv.org/abs/0708.2656}{arXiv:0708.2656 [hep-th]}\BibitemShut
  {NoStop}%
\bibitem [{\citenamefont {Bleicher}\ and\ \citenamefont
  {Nicolini}(2010)}]{BlN10}%
  \BibitemOpen
  \bibfield  {author} {\bibinfo {author} {\bibfnamefont {M.}~\bibnamefont
  {Bleicher}}\ and\ \bibinfo {author} {\bibfnamefont {P.}~\bibnamefont
  {Nicolini}},\ }\href {\doibase 10.1088/1742-6596/237/1/012008} {\bibfield
  {journal} {\bibinfo  {journal} {J. Phys. Conf. Ser.}\ }\textbf {\bibinfo
  {volume} {237}},\ \bibinfo {pages} {012008} (\bibinfo {year} {2010})},\
  \Eprint {http://arxiv.org/abs/1001.2211}{arXiv:1001.2211
  [hep-ph]}\BibitemShut {NoStop}%
\bibitem [{\citenamefont {Calmet}(2010)}]{Cal10a}%
  \BibitemOpen
  \bibfield  {author} {\bibinfo {author} {\bibfnamefont {X.}~\bibnamefont
  {Calmet}},\ }\href {\doibase 10.1142/S0217732310033591} {\bibfield  {journal}
  {\bibinfo  {journal} {Mod. Phys. Lett.}\ }\textbf {\bibinfo {volume} {A25}},\
  \bibinfo {pages} {1553} (\bibinfo {year} {2010})},\ \Eprint
  {http://arxiv.org/abs/1005.1805}{arXiv:1005.1805 [hep-ph]}\BibitemShut
  {NoStop}%
\bibitem [{\citenamefont {Park}(2012)}]{Par12}%
  \BibitemOpen
  \bibfield  {author} {\bibinfo {author} {\bibfnamefont {S.~C.}\ \bibnamefont
  {Park}},\ }\href {\doibase 10.1016/j.ppnp.2012.03.004} {\bibfield  {journal}
  {\bibinfo  {journal} {Prog. Part. Nucl. Phys.}\ }\textbf {\bibinfo {volume}
  {67}},\ \bibinfo {pages} {617} (\bibinfo {year} {2012})},\ \Eprint
  {http://arxiv.org/abs/1203.4683}{arXiv:1203.4683 [hep-ph]}\BibitemShut
  {NoStop}%
\bibitem [{\citenamefont {Nicolini}\ \emph {et~al.}(2015)\citenamefont
  {Nicolini}, \citenamefont {Mureika}, \citenamefont {Spallucci}, \citenamefont
  {Winstanley},\ and\ \citenamefont {Bleicher}}]{NMS13}%
  \BibitemOpen
  \bibfield  {author} {\bibinfo {author} {\bibfnamefont {P.}~\bibnamefont
  {Nicolini}}, \bibinfo {author} {\bibfnamefont {J.}~\bibnamefont {Mureika}},
  \bibinfo {author} {\bibfnamefont {E.}~\bibnamefont {Spallucci}}, \bibinfo
  {author} {\bibfnamefont {E.}~\bibnamefont {Winstanley}}, \ and\ \bibinfo
  {author} {\bibfnamefont {M.}~\bibnamefont {Bleicher}},\ }in\ \href {\doibase
  10.1142/9789814623995_0478} {\emph {\bibinfo {booktitle} {{Proceedings, 13th
  Marcel Grossmann Meeting on Recent Developments in Theoretical and
  Experimental General Relativity, Astrophysics, and Relativistic Field
  Theories (MG13): Stockholm, Sweden, July 1-7, 2012}}}}\ (\bibinfo {year}
  {2015})\ pp.\ \bibinfo {pages} {2495--2497},\ \Eprint
  {http://arxiv.org/abs/1302.2640}{arXiv:1302.2640 [hep-th]}\BibitemShut
  {NoStop}%
\bibitem [{\citenamefont {Bleicher}\ and\ \citenamefont
  {Nicolini}(2014)}]{BlN14}%
  \BibitemOpen
  \bibfield  {author} {\bibinfo {author} {\bibfnamefont {M.}~\bibnamefont
  {Bleicher}}\ and\ \bibinfo {author} {\bibfnamefont {P.}~\bibnamefont
  {Nicolini}},\ }\href@noop {} {\bibfield  {journal} {\bibinfo  {journal}
  {Astron. Nachr.}\ }\textbf {\bibinfo {volume} {335}},\ \bibinfo {pages} {605}
  (\bibinfo {year} {2014})},\ \Eprint
  {http://arxiv.org/abs/1403.0944}{arXiv:1403.0944 [hep-th]}\BibitemShut
  {NoStop}%
\bibitem [{\citenamefont {Kanti}\ and\ \citenamefont
  {Winstanley}(2015)}]{KaW15}%
  \BibitemOpen
  \bibfield  {author} {\bibinfo {author} {\bibfnamefont {P.}~\bibnamefont
  {Kanti}}\ and\ \bibinfo {author} {\bibfnamefont {E.}~\bibnamefont
  {Winstanley}},\ }\href {\doibase 10.1007/978-3-319-10852-0_8} {\bibfield
  {journal} {\bibinfo  {journal} {Fundam. Theor. Phys.}\ }\textbf {\bibinfo
  {volume} {178}},\ \bibinfo {pages} {229} (\bibinfo {year} {2015})},\ \Eprint
  {http://arxiv.org/abs/1402.3952}{arXiv:1402.3952 [hep-th]}\BibitemShut
  {NoStop}%
\bibitem [{\citenamefont {Wondrak}\ \emph {et~al.}(2017)\citenamefont
  {Wondrak}, \citenamefont {Nicolini},\ and\ \citenamefont {Bleicher}}]{WNB17}%
  \BibitemOpen
  \bibfield  {author} {\bibinfo {author} {\bibfnamefont {M.~F.}\ \bibnamefont
  {Wondrak}}, \bibinfo {author} {\bibfnamefont {P.}~\bibnamefont {Nicolini}}, \
  and\ \bibinfo {author} {\bibfnamefont {M.}~\bibnamefont {Bleicher}},\
  }\bibfield  {booktitle} {\emph {\bibinfo {booktitle} {{Proceedings, Frontier
  Research in Astrophysics - II: Mondello, Palermo, Italy, May 23-28, 2016}}},\
  }\href {\doibase 10.22323/1.269.0082} {\bibfield  {journal} {\bibinfo
  {journal} {PoS}\ }\textbf {\bibinfo {volume} {FRAPWS2016}},\ \bibinfo {pages}
  {082} (\bibinfo {year} {2017})},\ \Eprint
  {http://arxiv.org/abs/1612.08415}{arXiv:1612.08415 [hep-ph]}\BibitemShut
  {NoStop}%
\bibitem [{\citenamefont {Wondrak}\ \emph {et~al.}(2018)\citenamefont
  {Wondrak}, \citenamefont {Bleicher},\ and\ \citenamefont {Nicolini}}]{WNB18}%
  \BibitemOpen
  \bibfield  {author} {\bibinfo {author} {\bibfnamefont {M.~F.}\ \bibnamefont
  {Wondrak}}, \bibinfo {author} {\bibfnamefont {M.}~\bibnamefont {Bleicher}}, \
  and\ \bibinfo {author} {\bibfnamefont {P.}~\bibnamefont {Nicolini}}\
  }(\bibinfo {year} {2018})\ pp.\ \bibinfo {pages} {359--373},\ \Eprint
  {http://arxiv.org/abs/1708.06763}{arXiv:1708.06763 [gr-qc]}\BibitemShut
  {NoStop}%
\bibitem [{\citenamefont {Köppel}\ \emph {et~al.}(2018)\citenamefont
  {Köppel}, \citenamefont {Knipfer}, \citenamefont {Isi}, \citenamefont
  {Mureika},\ and\ \citenamefont {Nicolini}}]{Koppel:2017rsf}%
  \BibitemOpen
  \bibfield  {author} {\bibinfo {author} {\bibfnamefont {S.}~\bibnamefont
  {Köppel}}, \bibinfo {author} {\bibfnamefont {M.}~\bibnamefont {Knipfer}},
  \bibinfo {author} {\bibfnamefont {M.}~\bibnamefont {Isi}}, \bibinfo {author}
  {\bibfnamefont {J.}~\bibnamefont {Mureika}}, \ and\ \bibinfo {author}
  {\bibfnamefont {P.}~\bibnamefont {Nicolini}},\ }in\ \href {\doibase
  10.1007/978-3-319-94256-8_16} {\emph {\bibinfo {booktitle} {{2nd Karl
  Schwarzschild Meeting on Gravitational Physics}}}},\ Vol.\ \bibinfo {volume}
  {208}\ (\bibinfo {year} {2018})\ pp.\ \bibinfo {pages} {141--147},\ \Eprint
  {http://arxiv.org/abs/1703.05222}{arXiv:1703.05222 [hep-th]}\BibitemShut
  {NoStop}%
\bibitem [{\citenamefont {Scardigli}\ and\ \citenamefont
  {Casadio}(2003)}]{ScC03}%
  \BibitemOpen
  \bibfield  {author} {\bibinfo {author} {\bibfnamefont {F.}~\bibnamefont
  {Scardigli}}\ and\ \bibinfo {author} {\bibfnamefont {R.}~\bibnamefont
  {Casadio}},\ }\href {\doibase 10.1088/0264-9381/20/18/305} {\bibfield
  {journal} {\bibinfo  {journal} {Class. Quant. Grav.}\ }\textbf {\bibinfo
  {volume} {20}},\ \bibinfo {pages} {3915} (\bibinfo {year} {2003})},\ \Eprint
  {http://arxiv.org/abs/hep-th/0307174}{arXiv:hep-th/0307174
  [hep-th]}\BibitemShut {NoStop}%
\bibitem [{\citenamefont {Maziashvili}(2013)}]{Maz13}%
  \BibitemOpen
  \bibfield  {author} {\bibinfo {author} {\bibfnamefont {M.}~\bibnamefont
  {Maziashvili}},\ }\href {\doibase 10.1088/1475-7516/2013/03/042} {\bibfield
  {journal} {\bibinfo  {journal} {JCAP}\ }\textbf {\bibinfo {volume} {1303}},\
  \bibinfo {pages} {042} (\bibinfo {year} {2013})},\ \Eprint
  {http://arxiv.org/abs/1208.5570}{arXiv:1208.5570 [hep-th]}\BibitemShut
  {NoStop}%
\bibitem [{\citenamefont {Carr}(2013)}]{Car13}%
  \BibitemOpen
  \bibfield  {author} {\bibinfo {author} {\bibfnamefont {B.}~\bibnamefont
  {Carr}},\ }\href@noop {} {\bibfield  {journal} {\bibinfo  {journal} {Mod.
  Phys. Lett. A}\ }\textbf {\bibinfo {volume} {28}},\ \bibinfo {pages}
  {1340011} (\bibinfo {year} {2013})}\BibitemShut {NoStop}%
\bibitem [{\citenamefont {Lake}\ and\ \citenamefont {Carr}(2015)}]{LaC15}%
  \BibitemOpen
  \bibfield  {author} {\bibinfo {author} {\bibfnamefont {M.~J.}\ \bibnamefont
  {Lake}}\ and\ \bibinfo {author} {\bibfnamefont {B.}~\bibnamefont {Carr}},\
  }\href {\doibase 10.1007/JHEP11(2015)105} {\bibfield  {journal} {\bibinfo
  {journal} {JHEP}\ }\textbf {\bibinfo {volume} {11}},\ \bibinfo {pages} {105}
  (\bibinfo {year} {2015})},\ \Eprint
  {http://arxiv.org/abs/1505.06994}{arXiv:1505.06994 [gr-qc]}\BibitemShut
  {NoStop}%
\bibitem [{\citenamefont {Lake}\ and\ \citenamefont {Carr}(2016)}]{LaC16}%
  \BibitemOpen
  \bibfield  {author} {\bibinfo {author} {\bibfnamefont {M.~J.}\ \bibnamefont
  {Lake}}\ and\ \bibinfo {author} {\bibfnamefont {B.}~\bibnamefont {Carr}},\
  }\href@noop {} {\  (\bibinfo {year} {2016})},\ \Eprint
  {http://arxiv.org/abs/1611.01913}{arXiv:1611.01913 [gr-qc]}\BibitemShut
  {NoStop}%
\bibitem [{\citenamefont {Carrr}(2018)}]{Carr:2017grh}%
  \BibitemOpen
  \bibfield  {author} {\bibinfo {author} {\bibfnamefont {B.~J.}\ \bibnamefont
  {Carrr}},\ }\bibfield  {booktitle} {\emph {\bibinfo {booktitle}
  {{Proceedings, 2nd Karl Schwarzschild Meeting on Gravitational Physics (KSM
  2015): Frankfurt am Main, Germany, July 20-24, 2015}}},\ }\href {\doibase
  10.1007/978-3-319-94256-8_9} {\bibfield  {journal} {\bibinfo  {journal}
  {Springer Proc. Phys.}\ }\textbf {\bibinfo {volume} {208}},\ \bibinfo {pages}
  {85} (\bibinfo {year} {2018})},\ \Eprint
  {http://arxiv.org/abs/1703.08655}{arXiv:1703.08655 [gr-qc]}\BibitemShut
  {NoStop}%
\bibitem [{\citenamefont {Maziashvili}(2012)}]{Maz12}%
  \BibitemOpen
  \bibfield  {author} {\bibinfo {author} {\bibfnamefont {M.}~\bibnamefont
  {Maziashvili}},\ }\href {\doibase 10.1103/PhysRevD.86.104066} {\bibfield
  {journal} {\bibinfo  {journal} {Phys. Rev.}\ }\textbf {\bibinfo {volume}
  {D86}},\ \bibinfo {pages} {104066} (\bibinfo {year} {2012})},\ \Eprint
  {http://arxiv.org/abs/1206.4388}{arXiv:1206.4388 [gr-qc]}\BibitemShut
  {NoStop}%
\bibitem [{\citenamefont {Dirkes}\ \emph {et~al.}(2015)\citenamefont {Dirkes},
  \citenamefont {Maziashvili},\ and\ \citenamefont {Silagadze}}]{DMS15}%
  \BibitemOpen
  \bibfield  {author} {\bibinfo {author} {\bibfnamefont {A.~R.~P.}\
  \bibnamefont {Dirkes}}, \bibinfo {author} {\bibfnamefont {M.}~\bibnamefont
  {Maziashvili}}, \ and\ \bibinfo {author} {\bibfnamefont {Z.~K.}\ \bibnamefont
  {Silagadze}},\ }\href {\doibase 10.1142/S0218271816500152} {\bibfield
  {journal} {\bibinfo  {journal} {Int. J. Mod. Phys.}\ }\textbf {\bibinfo
  {volume} {D25}},\ \bibinfo {pages} {1650015} (\bibinfo {year} {2015})},\
  \Eprint {http://arxiv.org/abs/1309.7427}{arXiv:1309.7427 [gr-qc]}\BibitemShut
  {NoStop}%
\bibitem [{\citenamefont {Maziashvili}(2015)}]{Maz15}%
  \BibitemOpen
  \bibfield  {author} {\bibinfo {author} {\bibfnamefont {M.}~\bibnamefont
  {Maziashvili}},\ }\href {\doibase 10.1103/PhysRevD.91.064040} {\bibfield
  {journal} {\bibinfo  {journal} {Phys. Rev.}\ }\textbf {\bibinfo {volume}
  {D91}},\ \bibinfo {pages} {064040} (\bibinfo {year} {2015})},\ \Eprint
  {http://arxiv.org/abs/1502.07535}{arXiv:1502.07535 [hep-th]}\BibitemShut
  {NoStop}%
\bibitem [{\citenamefont {Lake}\ and\ \citenamefont {Carr}(2018)}]{LaC18}%
  \BibitemOpen
  \bibfield  {author} {\bibinfo {author} {\bibfnamefont {M.~J.}\ \bibnamefont
  {Lake}}\ and\ \bibinfo {author} {\bibfnamefont {B.}~\bibnamefont {Carr}},\
  }\href {\doibase 10.1142/S021827181930001} {\  (\bibinfo {year} {2018}),\
  10.1142/S021827181930001},\ \Eprint
  {http://arxiv.org/abs/1808.08386}{arXiv:1808.08386 [gr-qc]}\BibitemShut
  {NoStop}%
\bibitem [{\citenamefont {Balasin}\ and\ \citenamefont
  {Nachbagauer}(1993)}]{BaN93}%
  \BibitemOpen
  \bibfield  {author} {\bibinfo {author} {\bibfnamefont {H.}~\bibnamefont
  {Balasin}}\ and\ \bibinfo {author} {\bibfnamefont {H.}~\bibnamefont
  {Nachbagauer}},\ }\href {\doibase 10.1088/0264-9381/10/11/010} {\bibfield
  {journal} {\bibinfo  {journal} {Class. Quant. Grav.}\ }\textbf {\bibinfo
  {volume} {10}},\ \bibinfo {pages} {2271} (\bibinfo {year} {1993})},\ \Eprint
  {http://arxiv.org/abs/gr-qc/9305009}{arXiv:gr-qc/9305009}\BibitemShut
  {NoStop}%
\bibitem [{\citenamefont {Balasin}\ and\ \citenamefont
  {Nachbagauer}(1994)}]{BaN94}%
  \BibitemOpen
  \bibfield  {author} {\bibinfo {author} {\bibfnamefont {H.}~\bibnamefont
  {Balasin}}\ and\ \bibinfo {author} {\bibfnamefont {H.}~\bibnamefont
  {Nachbagauer}},\ }\href {\doibase 10.1088/0264-9381/11/6/010} {\bibfield
  {journal} {\bibinfo  {journal} {Class.Quant.Grav.}\ }\textbf {\bibinfo
  {volume} {11}},\ \bibinfo {pages} {1453} (\bibinfo {year} {1994})},\ \Eprint
  {http://arxiv.org/abs/gr-qc/9312028}{arXiv:gr-qc/9312028 [gr-qc]}\BibitemShut
  {NoStop}%
\bibitem [{\citenamefont {DeBenedictis}(2008)}]{DeB08}%
  \BibitemOpen
  \bibfield  {author} {\bibinfo {author} {\bibfnamefont {A.}~\bibnamefont
  {DeBenedictis}},\ }\enquote {\bibinfo {title} {Developments in black hole
  research: classical, semi-classical, and quantum},}\ \ (\bibinfo  {publisher}
  {Nova Science Publishers},\ \bibinfo {year} {2008})\ pp.\ \bibinfo {pages}
  {371--426},\ \Eprint
  {http://arxiv.org/abs/0711.2279}{arXiv:0711.2279}\BibitemShut {NoStop}%
\bibitem [{\citenamefont {Nicolini}\ \emph {et~al.}(2006)\citenamefont
  {Nicolini}, \citenamefont {Smailagic},\ and\ \citenamefont
  {Spallucci}}]{NSS06b}%
  \BibitemOpen
  \bibfield  {author} {\bibinfo {author} {\bibfnamefont {P.}~\bibnamefont
  {Nicolini}}, \bibinfo {author} {\bibfnamefont {A.}~\bibnamefont {Smailagic}},
  \ and\ \bibinfo {author} {\bibfnamefont {E.}~\bibnamefont {Spallucci}},\
  }\href {\doibase 10.1016/j.physletb.2005.11.004} {\bibfield  {journal}
  {\bibinfo  {journal} {Phys. Lett.}\ }\textbf {\bibinfo {volume} {B632}},\
  \bibinfo {pages} {547} (\bibinfo {year} {2006})},\ \Eprint
  {http://arxiv.org/abs/gr-qc/0510112}{arXiv:gr-qc/0510112}\BibitemShut
  {NoStop}%
\bibitem [{\citenamefont {Nicolini}\ and\ \citenamefont
  {Spallucci}(2010)}]{NiS10}%
  \BibitemOpen
  \bibfield  {author} {\bibinfo {author} {\bibfnamefont {P.}~\bibnamefont
  {Nicolini}}\ and\ \bibinfo {author} {\bibfnamefont {E.}~\bibnamefont
  {Spallucci}},\ }\href {\doibase 10.1088/0264-9381/27/1/015010} {\bibfield
  {journal} {\bibinfo  {journal} {Class. Quant. Grav.}\ }\textbf {\bibinfo
  {volume} {27}},\ \bibinfo {pages} {015010} (\bibinfo {year} {2010})},\
  \Eprint {http://arxiv.org/abs/0902.4654}{arXiv:0902.4654 [gr-qc]}\BibitemShut
  {NoStop}%
\bibitem [{\citenamefont {Nicolini}\ \emph {et~al.}(2019)\citenamefont
  {Nicolini}, \citenamefont {Spallucci},\ and\ \citenamefont
  {Wondrak}}]{NSW19}%
  \BibitemOpen
  \bibfield  {author} {\bibinfo {author} {\bibfnamefont {P.}~\bibnamefont
  {Nicolini}}, \bibinfo {author} {\bibfnamefont {E.}~\bibnamefont {Spallucci}},
  \ and\ \bibinfo {author} {\bibfnamefont {M.~F.}\ \bibnamefont {Wondrak}},\
  }\href@noop {} {\  (\bibinfo {year} {2019})},\ \Eprint
  {http://arxiv.org/abs/1902.11242}{arXiv:1902.11242 [gr-qc]}\BibitemShut
  {NoStop}%
\bibitem [{\citenamefont {Batic}\ and\ \citenamefont {Nicolini}(2010)}]{BaN10}%
  \BibitemOpen
  \bibfield  {author} {\bibinfo {author} {\bibfnamefont {D.}~\bibnamefont
  {Batic}}\ and\ \bibinfo {author} {\bibfnamefont {P.}~\bibnamefont
  {Nicolini}},\ }\href {\doibase 10.1016/j.physletb.2010.07.007} {\bibfield
  {journal} {\bibinfo  {journal} {Phys. Lett.}\ }\textbf {\bibinfo {volume}
  {B692}},\ \bibinfo {pages} {32} (\bibinfo {year} {2010})},\ \Eprint
  {http://arxiv.org/abs/1001.1158}{arXiv:1001.1158 [gr-qc]}\BibitemShut
  {NoStop}%
\bibitem [{\citenamefont {Brown}\ and\ \citenamefont {Mann}(2011)}]{BrM11}%
  \BibitemOpen
  \bibfield  {author} {\bibinfo {author} {\bibfnamefont {E.}~\bibnamefont
  {Brown}}\ and\ \bibinfo {author} {\bibfnamefont {R.~B.}\ \bibnamefont
  {Mann}},\ }\href {\doibase 10.1016/j.physletb.2010.10.014} {\bibfield
  {journal} {\bibinfo  {journal} {Phys. Lett.}\ }\textbf {\bibinfo {volume}
  {B695}},\ \bibinfo {pages} {440} (\bibinfo {year} {2011})},\ \Eprint
  {http://arxiv.org/abs/1012.4787}{arXiv:1012.4787 [hep-th]}\BibitemShut
  {NoStop}%
\bibitem [{\citenamefont {Knipfer}(2014)}]{Knipfer2014}%
  \BibitemOpen
  \bibfield  {author} {\bibinfo {author} {\bibfnamefont {M.}~\bibnamefont
  {Knipfer}},\ }\href@noop {} {\enquote {\bibinfo {title} {Generalized
  uncertainty principle inspired schwarzschild black holes in extra
  dimensions},}\ } (\bibinfo {year} {2014}),\ \bibinfo {note}
  {\url{http://publikationen.ub.uni-frankfurt.de/frontdoor/index/index/docId/48519}}\BibitemShut
  {NoStop}%
\bibitem [{\citenamefont {Barriola}\ and\ \citenamefont
  {Vilenkin}(1989)}]{BaV89}%
  \BibitemOpen
  \bibfield  {author} {\bibinfo {author} {\bibfnamefont {M.}~\bibnamefont
  {Barriola}}\ and\ \bibinfo {author} {\bibfnamefont {A.}~\bibnamefont
  {Vilenkin}},\ }\href {\doibase 10.1103/PhysRevLett.63.341} {\bibfield
  {journal} {\bibinfo  {journal} {Phys. Rev. Lett.}\ }\textbf {\bibinfo
  {volume} {63}},\ \bibinfo {pages} {341} (\bibinfo {year} {1989})}\BibitemShut
  {NoStop}%
\bibitem [{\citenamefont {Frolov}\ \emph {et~al.}(1989)\citenamefont {Frolov},
  \citenamefont {Israel},\ and\ \citenamefont {Unruh}}]{FIU89}%
  \BibitemOpen
  \bibfield  {author} {\bibinfo {author} {\bibfnamefont {V.~P.}\ \bibnamefont
  {Frolov}}, \bibinfo {author} {\bibfnamefont {W.}~\bibnamefont {Israel}}, \
  and\ \bibinfo {author} {\bibfnamefont {W.~G.}\ \bibnamefont {Unruh}},\ }\href
  {\doibase 10.1103/PhysRevD.39.1084} {\bibfield  {journal} {\bibinfo
  {journal} {Phys. Rev.}\ }\textbf {\bibinfo {volume} {D39}},\ \bibinfo {pages}
  {1084} (\bibinfo {year} {1989})}\BibitemShut {NoStop}%
\bibitem [{\citenamefont {Morris}\ \emph {et~al.}(1988)\citenamefont {Morris},
  \citenamefont {Thorne},\ and\ \citenamefont {Yurtsever}}]{MTY88}%
  \BibitemOpen
  \bibfield  {author} {\bibinfo {author} {\bibfnamefont {M.~S.}\ \bibnamefont
  {Morris}}, \bibinfo {author} {\bibfnamefont {K.~S.}\ \bibnamefont {Thorne}},
  \ and\ \bibinfo {author} {\bibfnamefont {U.}~\bibnamefont {Yurtsever}},\
  }\href {\doibase 10.1103/PhysRevLett.61.1446} {\bibfield  {journal} {\bibinfo
   {journal} {Phys. Rev. Lett.}\ }\textbf {\bibinfo {volume} {61}},\ \bibinfo
  {pages} {1446} (\bibinfo {year} {1988})}\BibitemShut {NoStop}%
\bibitem [{\citenamefont {Mann}(1997)}]{Mann97}%
  \BibitemOpen
  \bibfield  {author} {\bibinfo {author} {\bibfnamefont {R.~B.}\ \bibnamefont
  {Mann}},\ }\href {\doibase 10.1088/0264-9381/14/10/018} {\bibfield  {journal}
  {\bibinfo  {journal} {Class. Quant. Grav.}\ }\textbf {\bibinfo {volume}
  {14}},\ \bibinfo {pages} {2927} (\bibinfo {year} {1997})},\ \Eprint
  {http://arxiv.org/abs/gr-qc/9705007}{arXiv:gr-qc/9705007 [gr-qc]}\BibitemShut
  {NoStop}%
\bibitem [{\citenamefont {Amati}\ \emph {et~al.}(1987)\citenamefont {Amati},
  \citenamefont {Ciafaloni},\ and\ \citenamefont {Veneziano}}]{ACV87}%
  \BibitemOpen
  \bibfield  {author} {\bibinfo {author} {\bibfnamefont {D.}~\bibnamefont
  {Amati}}, \bibinfo {author} {\bibfnamefont {M.}~\bibnamefont {Ciafaloni}}, \
  and\ \bibinfo {author} {\bibfnamefont {G.}~\bibnamefont {Veneziano}},\ }\href
  {\doibase 10.1016/0370-2693(87)90346-7} {\bibfield  {journal} {\bibinfo
  {journal} {Phys. Lett.}\ }\textbf {\bibinfo {volume} {B197}},\ \bibinfo
  {pages} {81} (\bibinfo {year} {1987})}\BibitemShut {NoStop}%
\bibitem [{\citenamefont {Amati}\ \emph {et~al.}(1988)\citenamefont {Amati},
  \citenamefont {Ciafaloni},\ and\ \citenamefont {Veneziano}}]{ACV88}%
  \BibitemOpen
  \bibfield  {author} {\bibinfo {author} {\bibfnamefont {D.}~\bibnamefont
  {Amati}}, \bibinfo {author} {\bibfnamefont {M.}~\bibnamefont {Ciafaloni}}, \
  and\ \bibinfo {author} {\bibfnamefont {G.}~\bibnamefont {Veneziano}},\ }\href
  {\doibase 10.1142/S0217751X88000710} {\bibfield  {journal} {\bibinfo
  {journal} {Int. J. Mod. Phys.}\ }\textbf {\bibinfo {volume} {A3}},\ \bibinfo
  {pages} {1615} (\bibinfo {year} {1988})}\BibitemShut {NoStop}%
\bibitem [{\citenamefont {Amati}\ \emph {et~al.}(1990)\citenamefont {Amati},
  \citenamefont {Ciafaloni},\ and\ \citenamefont {Veneziano}}]{ACV90}%
  \BibitemOpen
  \bibfield  {author} {\bibinfo {author} {\bibfnamefont {D.}~\bibnamefont
  {Amati}}, \bibinfo {author} {\bibfnamefont {M.}~\bibnamefont {Ciafaloni}}, \
  and\ \bibinfo {author} {\bibfnamefont {G.}~\bibnamefont {Veneziano}},\ }\href
  {\doibase 10.1016/0550-3213(90)90375-N} {\bibfield  {journal} {\bibinfo
  {journal} {Nucl. Phys.}\ }\textbf {\bibinfo {volume} {B347}},\ \bibinfo
  {pages} {550} (\bibinfo {year} {1990})}\BibitemShut {NoStop}%
\bibitem [{\citenamefont {Gross}\ and\ \citenamefont {Mende}(1988)}]{GrM88}%
  \BibitemOpen
  \bibfield  {author} {\bibinfo {author} {\bibfnamefont {D.~J.}\ \bibnamefont
  {Gross}}\ and\ \bibinfo {author} {\bibfnamefont {P.~F.}\ \bibnamefont
  {Mende}},\ }\href {\doibase 10.1016/0550-3213(88)90390-2} {\bibfield
  {journal} {\bibinfo  {journal} {Nucl. Phys.}\ }\textbf {\bibinfo {volume}
  {B303}},\ \bibinfo {pages} {407} (\bibinfo {year} {1988})}\BibitemShut
  {NoStop}%
\bibitem [{\citenamefont {Gross}\ and\ \citenamefont {Mende}(1987)}]{GrM87}%
  \BibitemOpen
  \bibfield  {author} {\bibinfo {author} {\bibfnamefont {D.~J.}\ \bibnamefont
  {Gross}}\ and\ \bibinfo {author} {\bibfnamefont {P.~F.}\ \bibnamefont
  {Mende}},\ }\href {\doibase 10.1016/0370-2693(87)90355-8} {\bibfield
  {journal} {\bibinfo  {journal} {Phys. Lett.}\ }\textbf {\bibinfo {volume}
  {B197}},\ \bibinfo {pages} {129} (\bibinfo {year} {1987})}\BibitemShut
  {NoStop}%
\bibitem [{\citenamefont {Nojiri}\ and\ \citenamefont
  {Odintsov}(2003)}]{NoO2003}%
  \BibitemOpen
  \bibfield  {author} {\bibinfo {author} {\bibfnamefont {S.}~\bibnamefont
  {Nojiri}}\ and\ \bibinfo {author} {\bibfnamefont {S.~D.}\ \bibnamefont
  {Odintsov}},\ }\href {\doibase 10.1103/PhysRevD.68.123512} {\bibfield
  {journal} {\bibinfo  {journal} {Phys. Rev.}\ }\textbf {\bibinfo {volume}
  {D68}},\ \bibinfo {pages} {123512} (\bibinfo {year} {2003})},\ \Eprint
  {http://arxiv.org/abs/hep-th/0307288}{arXiv:hep-th/0307288
  [hep-th]}\BibitemShut {NoStop}%
\bibitem [{\citenamefont {Olmo}(2005)}]{Olmo2005}%
  \BibitemOpen
  \bibfield  {author} {\bibinfo {author} {\bibfnamefont {G.~J.}\ \bibnamefont
  {Olmo}},\ }\href {\doibase 10.1103/PhysRevD.72.083505} {\bibfield  {journal}
  {\bibinfo  {journal} {Phys. Rev.}\ }\textbf {\bibinfo {volume} {D72}},\
  \bibinfo {pages} {083505} (\bibinfo {year} {2005})},\ \Eprint
  {http://arxiv.org/abs/gr-qc/0505135}{arXiv:gr-qc/0505135 [gr-qc]}\BibitemShut
  {NoStop}%
\bibitem [{\citenamefont {Faraoni}(2006)}]{Faraoni96}%
  \BibitemOpen
  \bibfield  {author} {\bibinfo {author} {\bibfnamefont {V.}~\bibnamefont
  {Faraoni}},\ }\href {\doibase 10.1103/PhysRevD.74.104017} {\bibfield
  {journal} {\bibinfo  {journal} {Phys. Rev.}\ }\textbf {\bibinfo {volume}
  {D74}},\ \bibinfo {pages} {104017} (\bibinfo {year} {2006})},\ \Eprint
  {http://arxiv.org/abs/astro-ph/0610734}{arXiv:astro-ph/0610734
  [astro-ph]}\BibitemShut {NoStop}%
\bibitem [{\citenamefont {Capozziello}\ \emph {et~al.}(2007)\citenamefont
  {Capozziello}, \citenamefont {Stabile},\ and\ \citenamefont
  {Troisi}}]{CST2007}%
  \BibitemOpen
  \bibfield  {author} {\bibinfo {author} {\bibfnamefont {S.}~\bibnamefont
  {Capozziello}}, \bibinfo {author} {\bibfnamefont {A.}~\bibnamefont
  {Stabile}}, \ and\ \bibinfo {author} {\bibfnamefont {A.}~\bibnamefont
  {Troisi}},\ }\href {\doibase 10.1103/PhysRevD.76.104019} {\bibfield
  {journal} {\bibinfo  {journal} {Phys. Rev.}\ }\textbf {\bibinfo {volume}
  {D76}},\ \bibinfo {pages} {104019} (\bibinfo {year} {2007})},\ \Eprint
  {http://arxiv.org/abs/0708.0723}{arXiv:0708.0723 [gr-qc]}\BibitemShut
  {NoStop}%
\bibitem [{\citenamefont {Nojiri}\ and\ \citenamefont
  {Odintsov}(2007)}]{NoO2007}%
  \BibitemOpen
  \bibfield  {author} {\bibinfo {author} {\bibfnamefont {S.}~\bibnamefont
  {Nojiri}}\ and\ \bibinfo {author} {\bibfnamefont {S.~D.}\ \bibnamefont
  {Odintsov}},\ }\href {\doibase 10.1016/j.physletb.2007.07.039} {\bibfield
  {journal} {\bibinfo  {journal} {Phys. Lett.}\ }\textbf {\bibinfo {volume}
  {B652}},\ \bibinfo {pages} {343} (\bibinfo {year} {2007})},\ \Eprint
  {http://arxiv.org/abs/0706.1378}{arXiv:0706.1378 [hep-th]}\BibitemShut
  {NoStop}%
\bibitem [{\citenamefont {Cognola}\ \emph {et~al.}(2008)\citenamefont
  {Cognola}, \citenamefont {Elizalde}, \citenamefont {Nojiri}, \citenamefont
  {Odintsov}, \citenamefont {Sebastiani},\ and\ \citenamefont
  {Zerbini}}]{CENOSZ2008}%
  \BibitemOpen
  \bibfield  {author} {\bibinfo {author} {\bibfnamefont {G.}~\bibnamefont
  {Cognola}}, \bibinfo {author} {\bibfnamefont {E.}~\bibnamefont {Elizalde}},
  \bibinfo {author} {\bibfnamefont {S.}~\bibnamefont {Nojiri}}, \bibinfo
  {author} {\bibfnamefont {S.~D.}\ \bibnamefont {Odintsov}}, \bibinfo {author}
  {\bibfnamefont {L.}~\bibnamefont {Sebastiani}}, \ and\ \bibinfo {author}
  {\bibfnamefont {S.}~\bibnamefont {Zerbini}},\ }\href {\doibase
  10.1103/PhysRevD.77.046009} {\bibfield  {journal} {\bibinfo  {journal} {Phys.
  Rev.}\ }\textbf {\bibinfo {volume} {D77}},\ \bibinfo {pages} {046009}
  (\bibinfo {year} {2008})},\ \Eprint
  {http://arxiv.org/abs/0712.4017}{arXiv:0712.4017 [hep-th]}\BibitemShut
  {NoStop}%
\bibitem [{\citenamefont {Capozziello}\ \emph {et~al.}(2010)\citenamefont
  {Capozziello}, \citenamefont {De~Laurentis},\ and\ \citenamefont
  {Faraoni}}]{CDLF2010}%
  \BibitemOpen
  \bibfield  {author} {\bibinfo {author} {\bibfnamefont {S.}~\bibnamefont
  {Capozziello}}, \bibinfo {author} {\bibfnamefont {M.}~\bibnamefont
  {De~Laurentis}}, \ and\ \bibinfo {author} {\bibfnamefont {V.}~\bibnamefont
  {Faraoni}},\ }\href {\doibase 10.2174/1874381101003010049,
  10.2174/1874381101003020049} {\bibfield  {journal} {\bibinfo  {journal} {Open
  Astron. J.}\ }\textbf {\bibinfo {volume} {3}},\ \bibinfo {pages} {49}
  (\bibinfo {year} {2010})},\ \Eprint
  {http://arxiv.org/abs/0909.4672}{arXiv:0909.4672 [gr-qc]}\BibitemShut
  {NoStop}%
\bibitem [{\citenamefont {Berry}\ and\ \citenamefont {Gair}(2011)}]{BeG2011}%
  \BibitemOpen
  \bibfield  {author} {\bibinfo {author} {\bibfnamefont {C.~P.~L.}\
  \bibnamefont {Berry}}\ and\ \bibinfo {author} {\bibfnamefont {J.~R.}\
  \bibnamefont {Gair}},\ }\href {\doibase 10.1103/PhysRevD.85.089906,
  10.1103/PhysRevD.83.104022} {\bibfield  {journal} {\bibinfo  {journal} {Phys.
  Rev.}\ }\textbf {\bibinfo {volume} {D83}},\ \bibinfo {pages} {104022}
  (\bibinfo {year} {2011})},\ \bibinfo {note} {[Erratum: Phys.
  Rev.D85,089906(2012)]},\ \Eprint
  {http://arxiv.org/abs/1104.0819}{arXiv:1104.0819 [gr-qc]}\BibitemShut
  {NoStop}%
\bibitem [{\citenamefont {Schellstede}(2016)}]{Schell2016}%
  \BibitemOpen
  \bibfield  {author} {\bibinfo {author} {\bibfnamefont {G.~O.}\ \bibnamefont
  {Schellstede}},\ }\href {\doibase 10.1007/s10714-016-2111-9} {\bibfield
  {journal} {\bibinfo  {journal} {Gen. Rel. Grav.}\ }\textbf {\bibinfo {volume}
  {48}},\ \bibinfo {pages} {118} (\bibinfo {year} {2016})}\BibitemShut
  {NoStop}%
\bibitem [{\citenamefont {Edholm}\ \emph {et~al.}(2016)\citenamefont {Edholm},
  \citenamefont {Koshelev},\ and\ \citenamefont {Mazumdar}}]{EKMaz2016}%
  \BibitemOpen
  \bibfield  {author} {\bibinfo {author} {\bibfnamefont {J.}~\bibnamefont
  {Edholm}}, \bibinfo {author} {\bibfnamefont {A.~S.}\ \bibnamefont
  {Koshelev}}, \ and\ \bibinfo {author} {\bibfnamefont {A.}~\bibnamefont
  {Mazumdar}},\ }\href {\doibase 10.1103/PhysRevD.94.104033} {\bibfield
  {journal} {\bibinfo  {journal} {Phys. Rev.}\ }\textbf {\bibinfo {volume}
  {D94}},\ \bibinfo {pages} {104033} (\bibinfo {year} {2016})},\ \Eprint
  {http://arxiv.org/abs/1604.01989}{arXiv:1604.01989 [gr-qc]}\BibitemShut
  {NoStop}%
\bibitem [{\citenamefont {Frolov}\ and\ \citenamefont
  {Zelnikov}(2016)}]{FroZ2016}%
  \BibitemOpen
  \bibfield  {author} {\bibinfo {author} {\bibfnamefont {V.~P.}\ \bibnamefont
  {Frolov}}\ and\ \bibinfo {author} {\bibfnamefont {A.}~\bibnamefont
  {Zelnikov}},\ }\href {\doibase 10.1103/PhysRevD.93.064048} {\bibfield
  {journal} {\bibinfo  {journal} {Phys. Rev.}\ }\textbf {\bibinfo {volume}
  {D93}},\ \bibinfo {pages} {064048} (\bibinfo {year} {2016})},\ \Eprint
  {http://arxiv.org/abs/1509.03336}{arXiv:1509.03336 [hep-th]}\BibitemShut
  {NoStop}%
\bibitem [{\citenamefont {Kehagias}\ and\ \citenamefont
  {Maggiore}(2014)}]{KeMa2014}%
  \BibitemOpen
  \bibfield  {author} {\bibinfo {author} {\bibfnamefont {A.}~\bibnamefont
  {Kehagias}}\ and\ \bibinfo {author} {\bibfnamefont {M.}~\bibnamefont
  {Maggiore}},\ }\href {\doibase 10.1007/JHEP08(2014)029} {\bibfield  {journal}
  {\bibinfo  {journal} {JHEP}\ }\textbf {\bibinfo {volume} {08}},\ \bibinfo
  {pages} {029} (\bibinfo {year} {2014})},\ \Eprint
  {http://arxiv.org/abs/1401.8289}{arXiv:1401.8289 [hep-th]}\BibitemShut
  {NoStop}%
\bibitem [{\citenamefont {Hawking}\ and\ \citenamefont {Hertog}(2002)}]{HaH02}%
  \BibitemOpen
  \bibfield  {author} {\bibinfo {author} {\bibfnamefont {S.~W.}\ \bibnamefont
  {Hawking}}\ and\ \bibinfo {author} {\bibfnamefont {T.}~\bibnamefont
  {Hertog}},\ }\href {\doibase 10.1103/PhysRevD.65.103515} {\bibfield
  {journal} {\bibinfo  {journal} {Phys. Rev.}\ }\textbf {\bibinfo {volume}
  {D65}},\ \bibinfo {pages} {103515} (\bibinfo {year} {2002})},\ \Eprint
  {http://arxiv.org/abs/hep-th/0107088}{arXiv:hep-th/0107088
  [hep-th]}\BibitemShut {NoStop}%
\bibitem [{\citenamefont {Perivolaropoulos}(2017)}]{Perivo16}%
  \BibitemOpen
  \bibfield  {author} {\bibinfo {author} {\bibfnamefont {L.}~\bibnamefont
  {Perivolaropoulos}},\ }\href {\doibase 10.1103/PhysRevD.95.084050} {\bibfield
   {journal} {\bibinfo  {journal} {Phys. Rev.}\ }\textbf {\bibinfo {volume}
  {D95}},\ \bibinfo {pages} {084050} (\bibinfo {year} {2017})},\ \Eprint
  {http://arxiv.org/abs/1611.07293}{arXiv:1611.07293 [gr-qc]}\BibitemShut
  {NoStop}%
\bibitem [{\citenamefont {Carr}\ and\ \citenamefont {Hawking}(1974)}]{CaH74}%
  \BibitemOpen
  \bibfield  {author} {\bibinfo {author} {\bibfnamefont {B.~J.}\ \bibnamefont
  {Carr}}\ and\ \bibinfo {author} {\bibfnamefont {S.}~\bibnamefont {Hawking}},\
  }\href@noop {} {\bibfield  {journal} {\bibinfo  {journal} {Mon. Not. Roy.
  Astron. Soc.}\ }\textbf {\bibinfo {volume} {168}},\ \bibinfo {pages} {399}
  (\bibinfo {year} {1974})}\BibitemShut {NoStop}%
\bibitem [{\citenamefont {Mann}\ and\ \citenamefont {Ross}(1995)}]{MaR95}%
  \BibitemOpen
  \bibfield  {author} {\bibinfo {author} {\bibfnamefont {R.~B.}\ \bibnamefont
  {Mann}}\ and\ \bibinfo {author} {\bibfnamefont {S.~F.}\ \bibnamefont
  {Ross}},\ }\href {\doibase 10.1103/PhysRevD.52.2254} {\bibfield  {journal}
  {\bibinfo  {journal} {Phys.Rev.}\ }\textbf {\bibinfo {volume} {D52}},\
  \bibinfo {pages} {2254} (\bibinfo {year} {1995})},\ \Eprint
  {http://arxiv.org/abs/gr-qc/9504015}{arXiv:gr-qc/9504015 [gr-qc]}\BibitemShut
  {NoStop}%
\bibitem [{\citenamefont {Bousso}\ and\ \citenamefont {Hawking}(1996)}]{BoH96}%
  \BibitemOpen
  \bibfield  {author} {\bibinfo {author} {\bibfnamefont {R.}~\bibnamefont
  {Bousso}}\ and\ \bibinfo {author} {\bibfnamefont {S.~W.}\ \bibnamefont
  {Hawking}},\ }\href {\doibase 10.1103/PhysRevD.54.6312} {\bibfield  {journal}
  {\bibinfo  {journal} {Phys.Rev.}\ }\textbf {\bibinfo {volume} {D54}},\
  \bibinfo {pages} {6312} (\bibinfo {year} {1996})},\ \Eprint
  {http://arxiv.org/abs/gr-qc/9606052}{arXiv:gr-qc/9606052 [gr-qc]}\BibitemShut
  {NoStop}%
\bibitem [{\citenamefont {Mann}\ and\ \citenamefont {Nicolini}(2011)}]{MaN11}%
  \BibitemOpen
  \bibfield  {author} {\bibinfo {author} {\bibfnamefont {R.~B.}\ \bibnamefont
  {Mann}}\ and\ \bibinfo {author} {\bibfnamefont {P.}~\bibnamefont
  {Nicolini}},\ }\href@noop {} {\bibfield  {journal} {\bibinfo  {journal}
  {Phys. Rev.}\ }\textbf {\bibinfo {volume} {D84}},\ \bibinfo {pages} {064014}
  (\bibinfo {year} {2011})},\ \Eprint
  {http://arxiv.org/abs/1102.5096}{arXiv:1102.5096 [gr-qc]}\BibitemShut
  {NoStop}%
\bibitem [{\citenamefont {Sprenger}\ \emph {et~al.}(2011)\citenamefont
  {Sprenger}, \citenamefont {Nicolini},\ and\ \citenamefont
  {Bleicher}}]{SNB11a}%
  \BibitemOpen
  \bibfield  {author} {\bibinfo {author} {\bibfnamefont {M.}~\bibnamefont
  {Sprenger}}, \bibinfo {author} {\bibfnamefont {P.}~\bibnamefont {Nicolini}},
  \ and\ \bibinfo {author} {\bibfnamefont {M.}~\bibnamefont {Bleicher}},\
  }\href@noop {} {\bibfield  {journal} {\bibinfo  {journal} {Class. Quant.
  Grav.}\ }\textbf {\bibinfo {volume} {28}},\ \bibinfo {pages} {235019}
  (\bibinfo {year} {2011})},\ \Eprint
  {http://arxiv.org/abs/1011.5225}{arXiv:1011.5225 [hep-ph]}\BibitemShut
  {NoStop}%
\bibitem [{\citenamefont {Maselli}\ \emph {et~al.}(2018)\citenamefont
  {Maselli}, \citenamefont {Pani}, \citenamefont {Cardoso}, \citenamefont
  {Abdelsalhin}, \citenamefont {Gualtieri},\ and\ \citenamefont
  {Ferrari}}]{Maselli2017}%
  \BibitemOpen
  \bibfield  {author} {\bibinfo {author} {\bibfnamefont {A.}~\bibnamefont
  {Maselli}}, \bibinfo {author} {\bibfnamefont {P.}~\bibnamefont {Pani}},
  \bibinfo {author} {\bibfnamefont {V.}~\bibnamefont {Cardoso}}, \bibinfo
  {author} {\bibfnamefont {T.}~\bibnamefont {Abdelsalhin}}, \bibinfo {author}
  {\bibfnamefont {L.}~\bibnamefont {Gualtieri}}, \ and\ \bibinfo {author}
  {\bibfnamefont {V.}~\bibnamefont {Ferrari}},\ }\href {\doibase
  10.1103/PhysRevLett.120.081101} {\bibfield  {journal} {\bibinfo  {journal}
  {Phys. Rev. Lett.}\ }\textbf {\bibinfo {volume} {120}},\ \bibinfo {pages}
  {081101} (\bibinfo {year} {2018})},\ \Eprint
  {http://arxiv.org/abs/1703.10612}{arXiv:1703.10612 [gr-qc]}\BibitemShut
  {NoStop}%
\bibitem [{\citenamefont {Addazi}\ \emph {et~al.}(2019)\citenamefont {Addazi},
  \citenamefont {Marciano},\ and\ \citenamefont {Yunes}}]{Addazi2018}%
  \BibitemOpen
  \bibfield  {author} {\bibinfo {author} {\bibfnamefont {A.}~\bibnamefont
  {Addazi}}, \bibinfo {author} {\bibfnamefont {A.}~\bibnamefont {Marciano}}, \
  and\ \bibinfo {author} {\bibfnamefont {N.}~\bibnamefont {Yunes}},\ }\href
  {\doibase 10.1103/PhysRevLett.122.081301} {\bibfield  {journal} {\bibinfo
  {journal} {Phys. Rev. Lett.}\ }\textbf {\bibinfo {volume} {122}},\ \bibinfo
  {pages} {081301} (\bibinfo {year} {2019})},\ \Eprint
  {http://arxiv.org/abs/1810.10417}{arXiv:1810.10417 [gr-qc]}\BibitemShut
  {NoStop}%
\bibitem [{\citenamefont {Sirunyan}\ \emph {et~al.}(2018)\citenamefont
  {Sirunyan} \emph {et~al.}}]{Sirunyan:2018xwt}%
  \BibitemOpen
  \bibfield  {author} {\bibinfo {author} {\bibfnamefont {A.~M.}\ \bibnamefont
  {Sirunyan}} \emph {et~al.} (\bibinfo {collaboration} {CMS}),\ }\href
  {\doibase 10.1007/JHEP11(2018)042} {\bibfield  {journal} {\bibinfo  {journal}
  {JHEP}\ }\textbf {\bibinfo {volume} {11}},\ \bibinfo {pages} {042} (\bibinfo
  {year} {2018})},\ \Eprint {http://arxiv.org/abs/1805.06013}{arXiv:1805.06013
  [hep-ex]}\BibitemShut {NoStop}%
\bibitem [{\citenamefont {Mureika}\ \emph {et~al.}(2012)\citenamefont
  {Mureika}, \citenamefont {Nicolini},\ and\ \citenamefont
  {Spallucci}}]{MNS12}%
  \BibitemOpen
  \bibfield  {author} {\bibinfo {author} {\bibfnamefont {J.}~\bibnamefont
  {Mureika}}, \bibinfo {author} {\bibfnamefont {P.}~\bibnamefont {Nicolini}}, \
  and\ \bibinfo {author} {\bibfnamefont {E.}~\bibnamefont {Spallucci}},\
  }\href@noop {} {\bibfield  {journal} {\bibinfo  {journal} {Phys.Rev.}\
  }\textbf {\bibinfo {volume} {D85}},\ \bibinfo {pages} {106007} (\bibinfo
  {year} {2012})},\ \Eprint {http://arxiv.org/abs/1111.5830}{arXiv:1111.5830
  [hep-ph]}\BibitemShut {NoStop}%
\bibitem [{\citenamefont {Niikura}\ \emph {et~al.}(2019)\citenamefont {Niikura}
  \emph {et~al.}}]{Niikura:2017zjd}%
  \BibitemOpen
  \bibfield  {author} {\bibinfo {author} {\bibfnamefont {H.}~\bibnamefont
  {Niikura}} \emph {et~al.},\ }\href {\doibase 10.1038/s41550-019-0723-1}
  {\bibfield  {journal} {\bibinfo  {journal} {Nat. Astron.}\ }\textbf {\bibinfo
  {volume} {3}},\ \bibinfo {pages} {524} (\bibinfo {year} {2019})},\ \Eprint
  {http://arxiv.org/abs/1701.02151}{arXiv:1701.02151 [astro-ph.CO]}\BibitemShut
  {NoStop}%
\bibitem [{\citenamefont {Carr}\ \emph {et~al.}()\citenamefont {Carr},
  \citenamefont {Mentzer}, \citenamefont {Mureika},\ and\ \citenamefont
  {Nicolini}}]{CMMN}%
  \BibitemOpen
  \bibfield  {author} {\bibinfo {author} {\bibfnamefont {B.}~\bibnamefont
  {Carr}}, \bibinfo {author} {\bibfnamefont {H.}~\bibnamefont {Mentzer}},
  \bibinfo {author} {\bibfnamefont {J.}~\bibnamefont {Mureika}}, \ and\
  \bibinfo {author} {\bibfnamefont {P.}~\bibnamefont {Nicolini}},\ }\href@noop
  {} {\enquote {\bibinfo {title} {Self-complete and {GUP}-modified charged and
  spinning black holes},}\ }\bibinfo {note} {In preparation}\BibitemShut
  {NoStop}%
\bibitem [{\citenamefont {Szapudi}\ \emph {et~al.}(2005)\citenamefont
  {Szapudi}, \citenamefont {Pan}, \citenamefont {Prunet},\ and\ \citenamefont
  {Budavari}}]{Szapudi:2005vq}%
  \BibitemOpen
  \bibfield  {author} {\bibinfo {author} {\bibfnamefont {I.}~\bibnamefont
  {Szapudi}}, \bibinfo {author} {\bibfnamefont {J.}~\bibnamefont {Pan}},
  \bibinfo {author} {\bibfnamefont {S.}~\bibnamefont {Prunet}}, \ and\ \bibinfo
  {author} {\bibfnamefont {T.}~\bibnamefont {Budavari}},\ }\href {\doibase
  10.1086/496971} {\bibfield  {journal} {\bibinfo  {journal} {Astrophys. J.}\
  }\textbf {\bibinfo {volume} {631}},\ \bibinfo {pages} {L1} (\bibinfo {year}
  {2005})},\ \Eprint
  {http://arxiv.org/abs/astro-ph/0505389}{arXiv:astro-ph/0505389
  [astro-ph]}\BibitemShut {NoStop}%
\bibitem [{\citenamefont {Ogata}(2005)}]{Ogata2005}%
  \BibitemOpen
  \bibfield  {author} {\bibinfo {author} {\bibfnamefont {H.}~\bibnamefont
  {Ogata}},\ }\href {\doibase 10.2977/prims/1145474602} {\bibfield  {journal}
  {\bibinfo  {journal} {Publications of the Research Institute for Mathematical
  Sciences}\ }\textbf {\bibinfo {volume} {41}} (\bibinfo {year} {2005}),\
  10.2977/prims/1145474602}\BibitemShut {NoStop}%
\end{thebibliography}

%

\begin{appendix}
\section{Numerical integration of energy densities}\label{apx:integration}
Within this work, $n$-dimensional Fourier transforms of radially symmetric kernels 
$f(\vec x)=f(|\vec x|)$,
\begin{equation} \label{eq:ftrafo}
\mathcal{F} (\vec{p} )=\int f(\vec{x} )e^{-i\vec{p} \cdot \vec{x} }\d^{n} x
\end{equation}
appear several times. For algebraically complex $f(\vec x)$, it might be hard or impossible to find an analytic solution to the integral \eqref{eq:ftrafo}, as it appears in Sections \ref{sec:Cas-Scar} and \ref{sec:ourGUP}. In order to compute \eqref{eq:ftrafo} numerically, it is useful to rewrite the Fourier transform as a Hankel transform. The relationship is well known in literature and given by 
\begin{equation} \label{eq:hankeltrafo}
|\vec{p}|^{\frac {n-2}{2}}\mathcal{F}(|\vec{p}|)=(2\pi )^{n/2}\int _{0}^{\infty }r^{\frac {n-2}{2}}f(r)J_{\frac {n-2}{2}}(|\vec{p}| r)\ r\, \d r,
\end{equation}
if the function $f(\vec{x})$ is spherically symmetric and $r\equiv|\vec{x}|$.
Here $J_\alpha(z)$ is the Bessel function of first kind. This Bessel function is oscillatory, and there is a rich literature to perform accurate numerical
Hankel transformations which can be adopted in order to solve \eqref{eq:hankeltrafo},
such as \cite{Szapudi:2005vq,Ogata2005}.

In the following, a simple and robust scheme is proposed to
integrate \eqref{eq:hankeltrafo} for sufficiently convergent integrands. The intention is to solve the energy density $\mathbb{T}_0^0$ in equation \eqref{eq:t00newgup}, given by
\begin{align*}
\mathbb{T}^0_0(\vec{x}) = -\rho(\vec{x})=-
\frac{M}{(2\pi)^{n}} \int \frac{\diff^n p}{1 + 
(\sqrt{\beta} |\vec{p}|)^{n-1} }e^{i \vec{x}\cdot\vec{p}} \,.
\end{align*}
Accordingly to \eqref{eq:ftrafo} and \eqref{eq:hankeltrafo}, the energy density can be written as
\begin{align}
\rho(r) &=  \frac{M}{(2\pi)^{n/2}} \frac{1}{r^{\frac{n-2}{2}}} \int_0^\infty \diff |\vec{p}|\, |\vec{p}|^{\frac{n}{2}}
    \frac{1}{1+(\sqrt{\beta} |\vec{p}|)^{n-1}} J_{\frac{n-2}{2}}(r |\vec{p}|)\ .
    \label{eq:source}
\end{align}
For arbitrary values of $n$, the above integral cannot be solved analytically. It is, however, possible to perform a numerical integration. By introducing the dimensionless variables $z=r/\sqrt{\beta}$ and $q=\sqrt{\beta}|\vec{p}|$, the above integral reads:
\begin{align}
\rho(z) 
&= \frac{M \beta^{-n/2}}{(2\pi)^{n/2}} \frac{1}{z^{\frac{n-2}{2}}} 
\int_0^\infty 
\diff q\, 
q^{\frac{n}{2}}
\frac{1}{1+q^{n-1}} J_{\frac{n-2}{2}}(zq) \,.
\label{eq:rho-ndim-integral}
\end{align}
For small arguments the Bessel function behaves as
\begin{equation}
J_{\alpha }(z)\approx {\frac {1}{\Gamma (\alpha +1)}}\left({\frac {z}{2}}\right)^{\alpha }.
\end{equation} 
This means the integral is well defined at the lower bound. For large arguments the Bessel function can be written as
\begin{equation}
J_{\alpha }(z)\approx \frac{1}{|z|}.
\end{equation} 
This guarantees the expected convergence of the integral for $q\to\infty$. On these grounds, the numerical evaluation is possible by integrating from zero-crossing (\textit{i.e.} the $z$ where $J_\alpha(z)=0$) to zero-crossing in order to stabilize the integration and to ensure convergence. For numerical purposes, the density can be approximated as
\begin{align}
\rho(z) 
&\approx \frac{M \beta^{-n/2}}{(2\pi)^{n/2}} \frac{1}{z^{\frac{n-2}{2}}} 
\sum_{k=0}^K
\sum_{i=0}^{I\Delta q}
\Delta q~
q_{i,k}^{\frac{n}{2}}
\frac{1}{1+q_{i,k}^{n-1}} J_{\frac{n-2}{2}}(zq_{i,k})
\,.
\label{eq:rho-ndim-integral-discrete}
\end{align}
Here, $K\in \mathbb{N}$ are the number of zero crossings taken into account, and $I\in \mathbb{N}$ are the total number of integration support points, each given by 
$q_{i,k}=j_k + i\Delta q$, with $j_k$ the coordinate of the $k$th root of
$J_{(n-2)/2}(zq)$.
Clearly, with $I,K\to\infty$ and $\Delta q\to 0$, the continous
integral \eqref{eq:rho-ndim-integral} is recovered. We checked converge with
different grid sizes
$I,K \in \{10^2, 10^3, 10^4\}$. 
For the actual numerical integration, a standard Gaussian quadrature rule is applied.
The function values
$\rho_{\beta,n}(z)$ are then available on a discrete sample set $\{z_i\}\subset 
\mathbb R$ with arbitrary resolution and coverage.
With this numerical approach, one can also integrate the mass distribution as in~\eqref{eq:mass-ndim}, 
\begin{equation}
\mathcal{M}(r) = \int_{B_r}\diff^n x\,\rho(\vec{x}) = 
M A_{n-1}\int_0^r\diff r\,r^{n-1}\rho(r)\,,
\label{eq:modGUPmass}
\end{equation}
where $A_{n-1}=2\pi^{n/2}/\Gamma(n/2)$ is the surface of the $r$-ball $B_r$ in $n$ dimensions.
Again, this integral is carried out
numerically as a cumulative sum in a straightforward manner. 
From the matter distribution, one obtains the metric coefficients \eqref{eq:tangherlini} and can derive the related Hawking temperature. Once again, all final results (metric coefficient, matter distribution, temperature) obtained with this method are (only) fully discrete.

\end{appendix}
\end{document}